\documentclass[12pt, draftclsnofoot, onecolumn]{IEEEtran}
\pdfoutput=1
\usepackage{graphicx}
\usepackage{float}
\usepackage{caption}
\usepackage{multicol}
\usepackage{mathrsfs}
\usepackage{amsmath}
\usepackage{amssymb}
\usepackage{amsthm}
\usepackage{multicol}
\usepackage{float}
\usepackage{placeins}
\usepackage{epstopdf}
\usepackage{multirow}	
\usepackage{subfigure}
\floatstyle{ruled}
\newfloat{algorithm}{tbp}{loa}
\providecommand{\algorithmname}{Algorithm}
\floatname{algorithm}{\protect\algorithmname}

\theoremstyle{remark}
\newtheorem{theorem}{Theorem}
\newtheorem{lemma}{Lemma}
\newtheorem{definition}{Definition}
\theoremstyle{remark}
\newtheorem{example}{Example}
\title{Optimal Index Coding with Min-Max Probability of Error over Fading Channels}
\begin{document}
\author{Anoop Thomas, Kavitha R., Chandramouli A.,  and B. Sundar Rajan,~\IEEEmembership{Fellow,~IEEE}
}
\maketitle

\begin{abstract}
An index coding scheme in which the source (transmitter) transmits  binary symbols over a wireless fading channel is considered. Index codes with the transmitter using minimum number of transmissions are known as optimal index codes. Different optimal index codes give different performances in terms of probability of error in a fading environment and this also varies from receiver to receiver. In this paper we deal with optimal index codes which minimizes the maximum probability of error among all the receivers. We identify a criterion for optimal index codes that minimizes the maximum probability of error among all the receivers. For a special class of index coding problems, we give an algorithm to identify optimal index codes which minimize the  maximum error probability. We illustrate our techniques and claims with simulation results leading to conclude that a careful choice among the optimal index codes  will give a considerable gain in fading channels.

\end{abstract}

\section{Introduction}
\label{sec:Introduction}
The problem of index coding with side information was introduced by Birk and Kol \cite{ISCO} in which a central server (source/transmitter) has to transmit a set of data blocks to a set of caching clients (receivers). The clients may receive only a part of the data which the central server transmits. The receivers inform the server about the data blocks which they possess through a backward channel. The server has to make use of this additional information and find a way to satisfy each client using minimum number of transmissions. This problem of finding a code which uses minimum number of transmissions is the index coding problem.

Bar-Yossef \textit{et al.} \cite{ICSI} studied a type of index coding problem in which each receiver demands only one single message and the number of receivers equals number of messages. A side information graph was used to characterize the side information possessed by the receivers. It was found that the length of the optimal linear index code is equal to the minrank of the side information graph of the index coding problem. Also  few classes of index coding problems in which linear index codes are optimal were identified. However Lubetzky and Stav \cite{NLBL} showed that, in general, non-linear index codes are better than linear codes. 

Ong and Ho \cite{OMIC} classify the index coding problem depending on the demands and the side information possessed by the receivers. An index coding problem is unicast if the demand sets of the receivers are disjoint. It is referred to as single unicast if it is unicast and the size of each of the demand set is one. If the side information possessed by the receivers are disjoint then the problem is referred to as uniprior index coding problem. A uniprior index coding problem in which the size of the side information is one at all receivers is referred to as single uniprior problem. All other types of index coding problems are referred to as multicast/multiprior problems. It is  proved that for single uniprior index coding problems, linear index codes are sufficient to get optimality in terms of minimum number of transmissions. 

In this paper, we consider the scenario in which the binary symbols are transmitted in a fading channel and hence are subject to channel errors. We assume a fading channel between the source and the receivers along with additive white Gaussian noise (AWGN) at the receivers. Each of the transmitted symbol goes through a Rayleigh fading channel.  To the best of our knowledge, this is the first work that considers the performance of index coding in a fading environment. We use the following decoding procedure. A receiver decodes each of the transmitted symbol first and then uses these decoded symbols to obtain the message demanded by the receiver. Simulation curves showing Bit Error Probability (BEP) as a function of SNR are provided. We observe that the BEP performance at each receiver depends on the optimal index code used. We derive a condition on the optimal index codes which minimizes the maximum probability of error among all the receivers. For a special class of index coding problems, we give an algorithm to identify an optimal index code which gives the best performance in terms of minimal maximum  error probability across all the receivers.

The problem of index coding with erroneous transmissions was studied by Dau et al \cite{ECIC}. The problem of finding the minimum length index code which enables all receivers to correct a specific number of errors is addressed. Error-correction is achieved by using extra transmissions. In this paper, we consider only errors due to a wireless fading channel and among the optimal index codes we are identifying the code which minimizes the maximal error probability across the receivers. Since the number of transmissions remain same, we do not pay in bandwidth.

The rest of the manuscript is organized as follows. Section \ref{sec:Model} introduces the system model and necessary notations. In Section \ref{sec:Analysis} we present a criterion for an index code to  minimize the maximum probability of error. In Section \ref{sec:MinPEforSingleUniprior} we give an algorithm to identify an optimal index code which minimizes the maximum probability of error for single uniprior problems. In Section \ref{sec:Simulation} we show the simulation results. We summarize the results in Section \ref{sec:Conclusion}, and also discuss some open problems. 

\section{Model}
\label{sec:Model}
In index coding problems  there is a unique source $S$ having a set of $n$ messages  $X= \lbrace x_{1},x_{2},\ldots,x_{n} \rbrace $  and  a set of $m$ receivers $\mathcal{R}=\lbrace R_{1},R_{2},\ldots,R_{m} \rbrace  $. Each message $x_{i} \in X$ belongs to the finite field $\mathbb{F}_{2}$. Each $R_{i} \in \mathcal{R}$ is specified by the tuple $(\mathcal{W}_{i},\mathcal{K}_i)$, where $\mathcal{W}_{i} \subseteq X$ are the messages demanded by $R_{i}$ and $\mathcal{K}_{i} \subseteq X \setminus \mathcal{W}_{i}$ is the information known at the receiver. An index coding problem is completely specified by $(X,\mathcal{R})$ and we refer the index coding problem as $\mathcal{I}(X,\mathcal{R})$. 

The set $\lbrace 1,2,\ldots,n\rbrace$ is denoted by $\lceil n \rfloor$. An index code for an index coding problem is defined as:
\begin{definition}
\label{def:indexCode}
An \textit{index code} over $\mathbb{F}_{2}$ for an instance of the index coding problem $\mathcal{I}(X,\mathcal{R})$, is an encoding function  $\mathfrak{C}:\mathbb{F}_{2}^{n} \rightarrow \mathbb{F}_{2}^N$ such that for each receiver $R_{i}$, $i \in \left\lceil m\right\rfloor$, there exists a decoding function $\mathfrak{D}_{i}:\mathbb{F}_{2}^{N} \times \mathbb{F}_{2}^{|\mathcal{K}_{i}|}\rightarrow \mathbb{F}_{2}^{\mathcal{W}_{i}} $ satisfying $ \mathfrak{D}_{i}(\mathfrak{C}(X),\mathcal{K}_{i})=\mathcal{W}_{i} , \forall \; X \in \mathbb{F}_{2}^{n}$. The parameter $N$ is called the \textit{length} of the index code.
\end{definition}
An index code is said to be \textit{linear} if the encoding function $\mathfrak{C}$ is  linear over  $\mathbb{F}_{2}$. A linear index code can be described as $\mathfrak{C}(x)=xL,\forall \; x \in \mathbb{F}_{2}^{n} $ where $L$ is an $n\times N$ matrix over $\mathbb{F}_{q}$. The matrix $L$ is called the matrix corresponding to the linear index code $\mathfrak{C}$. The code $\mathfrak{C}$ is referred to as the linear index code based on $L$.

Consider an index coding problem $\mathcal{I}(X,\mathcal{R})$ with index code $\mathfrak{C}$, such that $\mathfrak{C}(X)=\lbrace c_{1},c_{2},\ldots, c_{N} \rbrace$. The source has to transmit the index code over a fading channel. Let $\mathcal{S}$ denote the constellation  used by the source. Let $\nu : \mathbb{F}_{2} \rightarrow \mathcal{S}$ denote the mapping of bits to the channel symbol used at the source. Let $\nu(\mathfrak{C}(X))=s_{X}$, denote the sequence of channel symbols transmitted by the source. Assuming quasi-static fading,  the received symbol sequence  at receiver $R_{j}$ corresponding to the transmission of $s_{X}$ is given by $y_{j}=h_{j}s_{X}+n_{j}$ where $h_{j}$ is the fading coefficient associated with the link from source to receiver $R_{j}$. The additive noise $n_{j}$ is assumed to be a sequence of noise samples distributed as $\mathcal{C}\mathcal{N}(0,1)$, which denotes circularly symmetric complex Gaussian random variable with variance one. Coherent detection is assumed at the receivers. In our model, the receiver decodes $\mathfrak{C}(X)$  and then tries to find the demanded message $x_{i} \in \mathcal{W}_{i}$ using the decoded index code. In this paper we will see that different optimal index codes give rise to different performance in terms of probability of error.

We recall few of the relevant standard definitions in graph theory. A \textit{graph} is a pair $G=(V,E)$ of sets where the elements of $V$ are the vertices of graph and the elements of $E$ are its edges. The vertex set of a graph is referred to as $V(G)$, its edge set as $E(G)$. Two vertices $v_1,v_2$ of $G$ are \textit{adjacent} if $v_1v_2$ is an edge of $G$. An \textit{arc} is a directed edge. For an arc $v_1v_2$, vertex $v_1$ is the tail of the arc and vertex $v_2$ is the head of the arc. If all the vertices of $G$ are pairwise adjacent then $G$ is \textit{complete}. Consider a graph $G'=(V',E')$. If $V' \subseteq V$ and $E' \subseteq E$, then $G'$ is a \textit{subgraph} of $G$ written as $G' \subseteq G$. A subgraph $G'$ is a \textit{spanning subgraph} if $V' = V$. A \textit{path} is a non-empty graph $P=(V,E)$ of the form $V=\{v_0,v_1,\ldots,v_k\}$, $E=\{v_{0}v_{1},v_{1}v_{2},\ldots,v_{k-1}v_{k}\}$ where the $v_i$ are all distinct. If $P=v_{0}v_{1}\ldots v_{k-1}$ is a path and $k \geq 3$, then a cycle is a path with an additional edge $v_{k-1}v_{0}$. A graph is \textit{acyclic} if it does not contain any cycle. The number of edges of a path is its \textit{length}. The \textit{distance} $d_{G}(x,y)$ in $G$ of two vertices $x,y$ is the length of a shortest $x$-$y$ path in $G$. The greatest distance between any two vertices in $G$ is the \textit{diameter} of $G$. A graph $G$ is called \textit{connected} if any two of its vertices are linked by a path in $G$. A \textit{tree} is a connected acyclic graph. A \textit{spanning tree} is a tree which spans the graph. For two graphs $G_1=(V_1,E_1)$ and $G_2=(V_2,E_2)$, $G_1 \cup G_2 := (V_1 \cup V_2,E_1 \cup E_2)$, $G_1 \cap G_2 := (V_1 \cap V_2, E_1 \cap E_2) $ and $G_1 \backslash G_2 := (V_1 \backslash V_2, E_1 \backslash E_2)$. 
\\

\section{A Criterion for Minimum Maximum Probability of Error}
\label{sec:Analysis}

In this section we identify a condition that is required to minimize the maximum probability of error for decoding a message across all the receivers. Since the transmissions are over a fading channel each transmitted symbol has a probability of error. Let the probability of error of each transmitted symbol (denoted by $t_x$) be $p$. Let us consider an index code $\mathfrak{C}$ of length $N$ for an index coding problem $\mathcal{I}(X,\mathcal{R})$. Consider a receiver $R_{i} \in \mathcal{R}$, which uses $c$ of the $N$ transmissions to recover a message $x_{i} \in \mathcal{W}_{i}$. We try to find the probability of error in decoding the message $x_{i}$.
 Let the decoded message be $\widehat{x_{i}}$. The probability of error in decoding the message $x_{i}$ is 

\small{ 
\begin{flalign}
\begin{split}
Pr( \widehat{x_{i}} \neq x_{i}) &= Pr( 1 \text{ $t_x$ in error } \cup 3 \text{ $t_x$ in error }  \cup \ldots c \text{ $t_x$ in error }) \\
& = \underset{i \text{ odd},i \leq c}{\sum} Pr ( i \text{ $t_x$ in error } ) 
= \underset{i \text{ odd},i \leq c}{\sum} \left(\begin{array}{c}
c\\
i
\end{array}\right)p^{i}(1-p)^{c-i}.
\end{split}
\label{eq:ProbError}
\end{flalign}
}
\normalsize
We show that the probability of error in decoding a message decreases if receiver uses less number of transmissions to decode that message.

\begin{lemma}
The probability of error in decoding a message at a particular receiver decreases with a decrease in the number of transmissions used to decode the message among the total of $N$ transmissions. 
\begin{proof}
This lemma can be proved by showing that the expression obtained for probability of error in  \eqref{eq:ProbError} is an increasing function on $c$ which is the number of transmissions used to decode the message. We have 
\small {
\begin{flalign*}
\begin{split}
\underset{i \text{ odd},i \leq c}{\sum} \left(\begin{array}{c}
c\\
i 
\end{array}\right)p^{i}(1-p)^{c-i}  &= \frac{(p+(1-p))^{c} - ((1-p)-p)^{c}}{2} = \frac{1-(1-2p)^{c}}{2}.
\end{split}
\end{flalign*}
}
\normalsize
Consider,
\small{
\begin{flalign*}
\begin{split}
\frac{1-(1-2p)^{c+1}}{2}-  \frac{1-(1-2p)^{c}}{2} &= \frac{(1-2p)^{n}(1-(1-2p))}{2} = (1-2p)^{n}p.
\end{split}
\end{flalign*}
}
\normalsize
As $c$ increases the difference remains positive as long as $p < 0.5$. As probability of transmitted symbol to be in error is less than $0.5$, the lemma is proved. 
\end{proof}
\end{lemma}

We have considered only decoding of one message at a particular receiver. However a receiver may have multiple demands. Also there are many receivers to be considered. So we try to bound the maximum error probability. To achieve this we try to identify those optimal index codes which will reduce the maximum number of transmissions used by any receiver to decode any of its demanded message. Such optimal index codes perform better than other optimal index codes of the same number of transmissions. Such index codes are not only bandwidth optimal (since the minimum number of transmissions are used)  but are also optimal in the sense of minimum maximum probability of error.

\begin{example}
Consider a single uniprior index coding problem $\mathcal{I}(X,\mathcal{R})$ with $X=\lbrace x_{1},x_{2},\ldots,x_{9}\rbrace$ and $\mathcal{R}=\lbrace R_{1},R_{2},\ldots,R_{9} \rbrace$. Each receiver $R_{i} \in \mathcal{R}$, knows $x_{i}$ and demands $x_{i+2}$ where $+$ denotes modulo 9 addition. In addition to the above demands, receiver $R_{1}$ and $R_{2}$ also demands $x_{2}$ and $x_{3}$ respectively. The length of the optimal linear code for this problem is eight. In this example we consider four optimal linear codes and show that the number of transmissions used in decoding the demands at receivers depends on the code. 

\begin{table*}
\centering
\tiny
\begin{tabular}{cc}
$L_{1} = \left[\begin{array}{ccccccccc}
1 & 1 & 1 & 1 & 1 & 1 & 1 & 1\\
1 & 0 & 0 & 0 & 0 & 0 & 0 & 0\\
0 & 1 & 0 & 0 & 0 & 0 & 0 & 0\\
0 & 0 & 1 & 0 & 0 & 0 & 0 & 0\\
0 & 0 & 0 & 1 & 0 & 0 & 0 & 0\\
0 & 0 & 0 & 0 & 1 & 0 & 0 & 0\\
0 & 0 & 0 & 0 & 0 & 1 & 0 & 0\\
0 & 0 & 0 & 0 & 0 & 0 & 1 & 0\\
0 & 0 & 0 & 0 & 0 & 0 & 0 & 1\\
\end{array}\right]$ & $L_{2} = \left[\begin{array}{ccccccccc}
1 & 1 & 0 & 1 & 0 & 0 & 0 & 0\\
0 & 0 & 0 & 0 & 1 & 1 & 0 & 0\\
0 & 0 & 0 & 1 & 1 & 0 & 0 & 0\\
0 & 0 & 1 & 0 & 0 & 0 & 0 & 0\\
0 & 0 & 0 & 0 & 0 & 0 & 0 & 1\\
0 & 1 & 1 & 0 & 0 & 0 & 0 & 0\\
0 & 0 & 0 & 0 & 0 & 0 & 1 & 1\\
1 & 0 & 0 & 0 & 0 & 0 & 0 & 0\\
0 & 0 & 0 & 0 & 0 & 1 & 1 & 0\\
\end{array}\right]$ \\
\\
$L_{3} = \left[\begin{array}{ccccccccc}
1 & 1 & 0 & 0 & 0 & 0 & 0 & 0\\
1 & 0 & 1 & 0 & 0 & 0 & 0 & 0\\
0 & 1 & 0 & 1 & 0 & 0 & 0 & 0\\
0 & 0 & 1 & 0 & 1 & 0 & 0 & 0\\
0 & 0 & 0 & 1 & 0 & 1 & 0 & 0\\
0 & 0 & 0 & 0 & 1 & 0 & 1 & 0\\
0 & 0 & 0 & 0 & 0 & 1 & 0 & 1\\
0 & 0 & 0 & 0 & 0 & 0 & 1 & 0\\
0 & 0 & 0 & 0 & 0 & 0 & 0 & 1\\
\end{array}\right]$ ~ &
$L_{4} = \left[\begin{array}{ccccccccc}
1 & 0 & 0 & 0 & 0 & 0 & 0 & 0\\
0 & 0 & 0 & 1 & 1 & 0 & 0 & 0\\
0 & 0 & 0 & 0 & 1 & 1 & 0 & 0\\
0 & 0 & 1 & 1 & 0 & 0 & 0 & 0\\
0 & 0 & 0 & 0 & 0 & 1 & 1 & 0\\
0 & 1 & 1 & 0 & 0 & 0 & 0 & 0\\
0 & 0 & 0 & 0 & 0 & 0 & 1 & 1\\
1 & 1 & 0 & 0 & 0 & 0 & 0 & 0\\
0 & 0 & 0 & 0 & 0 & 0 & 0 & 1\\
\end{array}\right]$

\end{tabular}
\caption{\small Matrices describing codes $\mathfrak{C}_{1}$,$\mathfrak{C}_{2}$,$\mathfrak{C}_{3}$ and $\mathfrak{C}_{4}$ of Example \ref{eg:NoofTransmissions}}
\label{Tab:MatricesComparisonExample}
\end{table*}
 
Consider codes $\mathfrak{C}_{1},\mathfrak{C}_{2},\mathfrak{C}_{3}$ and $\mathfrak{C}_{4}$ represented by the matrices $L_{1},L_{2},L_{3}$ and $L_{4}$ respectively. The matrices representing the codes are given in Table \ref{Tab:MatricesComparisonExample}. The number of transmissions required by each receiver in decoding its demand for each of the codes is given in Table \ref{Tab:NoOfTransmissions}. Since receivers $R_{1}$ and $R_{2}$ have two demands, two entries are given in its column each corresponding to one of its demands. The maximum number of transmissions used for each code is underlined. From the table we can observe that the maximum number of transmissions required by a receiver in decoding its demands is four for the codes $\mathfrak{C}_{2}$ and $\mathfrak{C}_{3}$. For the code $\mathfrak{C}_{4}$, the maximum number of transmissions used to decode the message is five. However for code $\mathfrak{C}_{1}$, the maximum number is two. Among the four codes considered, code $\mathfrak{C}_{1}$  gives minimum maximum error probability across the receivers. In this section we give an algorithm to identify such codes which gives minimum maximum error probability across the receivers. 

\begin{table}
\centering{}
\scriptsize
\begin{tabular}[h]{|c|c|c|c|c|c|c|c|c|c|c|}
\hline
Codes & $R_{1}$ & $R_{2}$ & $R_{3}$ & $R_{4}$ & $R_{5}$ & $R_{6}$ & $R_{7}$ & $R_{8}$ & $R_{9}$\\
\hline
$\mathfrak{C}_{1}$ & $1,1$ & $\underline{2},\underline{2}$ & $\underline{2}$ & $\underline{2}$ & $\underline{2}$ & $\underline{2}$ & $\underline{2}$ & $1$ & $\underline{2}$ \\
\hline 
$\mathfrak{C}_{2}$ & $2,1$ & $1,\underline{4}$ & $\underline{4}$ & $1$ & $1$ & $2$ & $1$ & $1$ & $1$ \\ 
\hline
$\mathfrak{C}_{3}$ & $1,1$ & $2,1$ & $1$ & $1$ & $1$ & $1$ & $1$ & $1$ & $\underline{4}$ \\ 
\hline
$\mathfrak{C}_{4}$ & $4,\underline{5}$ & $1,1$ & $1$ & $1$ & $1$ & $1$ & $1$ & $1$ & $4$ \\ 
\hline
\end{tabular}
\caption{\small Number of transmissions used at receivers to decode its demands for codes $\mathfrak{C}_{1},\mathfrak{C}_{2},\mathfrak{C}_{3}$ and $\mathfrak{C}_{4}$.}
\label{Tab:NoOfTransmissions}
\end{table}

\label{eg:NoofTransmissions}
\end{example}

This motivates us to find those index codes which are not only bandwidth optimal but also which gives minimum maximum probability of error. There are no algorithms to find the optimal solution to the general index coding problem, however for the binary single uniprior index coding problem, it was shown that scalar linear index codes are optimal. In the next session we identify index codes for binary single uniprior index coding problems which are not only bandwidth optimal but also optimal in terms of minimizing the maximum error probability.

\section{Bandwidth optimal index code which minimizes the maximum probability of error}
\label{sec:MinPEforSingleUniprior}

In Section \ref{sec:Analysis}, we derived a condition for minimizing the maximum probability of error. The index code should be such that the maximum number of transmissions used by any receiver to decode any of its demands should be as less as possible. In this section, we identify such index codes for single uniprior index coding problems. Recall that in a single uniprior problem each receiver $R_{i}$ demands a set of messages $W_i $ and knows only one message $x_{i}$. There  are several linear solutions which are optimal in terms of least bandwidth for this problem but among them we try to identify the index code which minimizes the maximum number of transmissions that is required by any receiver in decoding its desired messages. We illustrate the problem with the following example.

The single uniprior problem can be represented by information flow graph $G$ of $m$ vertices each representing a receiver, with directed edge from vertex $i$ to vertex $j$ if and only if node $j$ wants $x_{i}$. Note that in a single uniprior problem the number of receivers is equal to the number of messages. This is because each receiver knows only one message and the message known to each receiver is different. So $n > m$ implies that there are some messages which does not form part of side information of any of the receivers. Such messages have to be transmitted directly and we can reduce that to an index coding problem where $n = m$. Ong and Ho have proved that all single uniprior problems have bandwidth optimal linear solutions. The Algorithm \ref{alg:pruningalgorithm} (Pruning algorithm), which takes information flow graph as input was proposed. The output of Algorithm \ref{alg:pruningalgorithm} is $G'$ which is a set of non-trivial strongly connected components each represented by $G'_{sub,i}$ and a collection of arcs. The benefit is that a coding scheme satisfying $G'$ will satisfy the original index coding problem $G$ as well. We propose Algorithm \ref{alg:optimalic} for the single uniprior problem which finds the bandwidth optimal index code that minimizes the maximum probability of error.

\begin{algorithm}
\small
Initialization: $G'=(V',E')\leftarrow G=(V,A)$ \\
1)Iteration
\begin{description}

\item [{while}] there exists a vertex $i\in V'$ with

\begin{description}
\item [{(i)}] more than one outgoing arc, and
\item [{(ii)}] an outgoing arc that does not belong to any cycle {[}denote
any such arc by (i,j){]}
\end{description}
\item [{do}]~

\begin{list}{}
\item remove from $G'$, all outgoing arcs of vertex $i$ except for the
arc $(i,j)$;
\end{list}
\item [{end}]~\end{description}

2) label each non-trivial strongly connected component in $G'$
as $G'_{sub,i}$, $i \in \{1,2,\ldots,N_{sub}\}$;

\caption{The Pruning Algorithm}
\label{alg:pruningalgorithm}

\end{algorithm}

\begin{algorithm}
\small
\begin{enumerate}
\item Perform the pruning algorithm on the information flow graph of the single uniprior problem and obtain the sets $G'$ and $G'_{sub,i},i \in \{1,2,\ldots,N_{sub}\}$.
\item For each $G'_{sub,i}$ perform the following:
\begin{itemize}
\item Form a complete graph on vertices of $G'_{sub,i}$.
\item Identify the spanning tree $T$, which has the minimum maximum distance between $(i,j)$ for all $(i,j) \in E(G'_{sub,i})$.
\item For each edge $(i,j)$ of $T$, transmit $x_{i} \oplus x_{j}$.
\end{itemize}
\item For each edge $(i,j)$ of $G' \backslash G'_{sub}$, transmit $x_{i}$.
\end{enumerate}
\caption{}
\label{alg:optimalic}
\end{algorithm}

The first step of Algorithm \ref{alg:optimalic} is the pruning algorithm which gives $G'$ and its connected components $G'_{sub,i}$. The number of such connected components in $G'$ is $N_{sub}$. Algorithm \ref{alg:optimalic} operates on each of the connected components $G'_{sub,i}$. A complete graph is formed on the vertices of $G'_{sub,i}$. Recall that in a complete graph all the vertices are pairwise adjacent. Consider a spanning tree $T_{i}$ of the complete graph on vertices of $G'_{sub,i}$. Consider an edge $(i,j) \in E(G'_{sub,i})$. An edge $(i,j) \in E(G'_{sub,i})$ indicates that vertex $j$ demands the message $x_{i}$. In the spanning tree $T_{i}$, there will be a unique path between the vertices $i$ and $j$. Algorithm \ref{alg:optimalic} computes the distance of that unique path. This is done for all edges $(i,j) \in E(G'_{sub,i})$ and the maximum distance is observed. This is repeated for different spanning tress and among the spanning tress the one which has the minimum maximum distance is identified by the algorithm. Let $T$ be the spanning tree identified by the algorithm. From $T$ we obtain the index code as follows. For each edge $(i,j)$ of $T$, transmit $x_{i} \oplus x_{j}$. There will be few demands which correspond to arcs in $G' \setminus G'{sub}$ where $G'_{sub}$ is the union of all connected components $G'_{sub,i}$. For each arc $(i,j) \in G' \setminus G'_{sub}, x_{i}$  is transmitted.

\begin{theorem}
For every single uniprior index coding problem, the Algorithm \ref{alg:optimalic} gives a bandwidth optimal index code which minimizes the maximum probability of error. Moreover the number of transmissions used by any receiver in decoding any of its message is at most two for the index code obtained from Algorithm \ref{alg:optimalic}.
\begin{proof}
First we prove that Algorithm \ref{alg:optimalic} gives a valid index code. Symbols transmitted in third step of algorithm are messages itself and any receiver demanding those messages gets satisfied. All receiver nodes in $T$ are able to decode the message of every other vertex in $T$ in the following way. Consider two vertices $i$ and $j$ with vertex $j$ demanding $x_i$. Since $T$ is a spanning tree there exists a unique path between any pair of its vertices. Consider that unique path $P=(i,k_1,k_2,\ldots,j)$ between $i$ and $j$. Receiver $j$ can obtain $x_{i} \oplus x_{j}$ by performing XOR operation on all the transmitted symbols corresponding to the edges in the path $P$. Now we prove the optimality in bandwidth. The number of edges of every spanning tree is $V(G'_{sub,i})-1$. For each $G'_{sub,i}$ we transmit $V(G'_{sub,i})-1$ symbols. The total number of transmissions for our index code is equal to $\overset{N_{sub}}{\underset{i=1}{\sum}}(V(G'_{sub,i})-1)+ |E(G' \backslash G'_{sub})|$. The index code of Algorithm  \ref{alg:optimalic} uses the same number of transmissions as the bandwidth optimal index code \cite{OMIC}. Observe that for every connected graph $G_{conn}$ representing a single uniprior problem, the source cannot achieve optimal bandwidth if it transmits any of the message directly. Let us assume that the source transmits $x_i$. Note that message $x_{i}$ is the side information of one of the receivers say $j$. So to satisfy the demands of receiver $j$ the source has to transmit its want-set directly. Thus to satisfy all the receivers, the source needs to transmit $|V(G_{conn})|$ symbols where as the optimal number of transmissions is $|V(G_{conn})-1|$. Hence for any connected component $G'_{sub,i}$, the  source cannot transmit the messages directly. Finally, observe that the number of transmissions used by the receiver to decode the desired message is equal to the distance between the vertices in the corresponding spanning tree. So the spanning tree which minimizes the maximum distance for all the demands of the index coding problem gives the index code which minimizes the maximum probability of error. There exists spanning trees for a complete graph with diameter two, so every receiver can decode any of its desired message using at most two transmissions.
\end{proof}  
\end{theorem}

Algorithm \ref{alg:optimalic} identifies an index code which minimizes the maximum number of transmissions required by any receiver to decode its demanded message. Note that the spanning tree identified in step 2 of the algorithm need not be unique. Hence there are multiple index codes which offers the same minimum maximum number of transmissions. Among these we could find those index codes which reduces the total number of transmissions used by all the receivers. This could be achieved by modifying the Step $2$ of Algorithm \ref{alg:optimalic}. Identify the set of spanning trees which has the minimum maximum distance between $(i,j)$ for all $(i,j) \in E(G'_{sub,i})$. Among these spanning trees we can compute the total distance between all edges $(i,j) \in E(G'_{sub,i})$ and identify the spanning tree $T_{i}$ which minimizes the overall sum. For each edge $(i,j) \in T_{i}$ transmit $x_{i}+x_{j}$. This will give the index code which minimizes the total number of transmissions used in decoding all the messages at all the receivers.   

In the remainder of this section we show few examples which illustrate the use of the algorithm. The simulation results showing the improved performance at receivers is given in Section \ref{sec:Simulation}.

\begin{example}
\label{eg:3userproblem}
In this example we consider a single uniprior index coding problem having three receivers. The index coding problem has a message set $X= \lbrace x_{1},x_{2},x_{3} \rbrace$ and the set of receivers $\mathcal{R}= \lbrace R_{1},R_{2},R_{3} \rbrace$. Receiver $R_{1}$ demands messages $x_{2}$ and $x_{3}$. Receiver $R_{2}$ demands $x_{1}$ and receiver $R_{3}$ demands $x_{1}$ and $x_{2}$. The information flow graph $G$ for this problem is given in Figure \ref{fig:3userIFGraph}. 
\begin{figure}
\centering{}\includegraphics[scale=1]{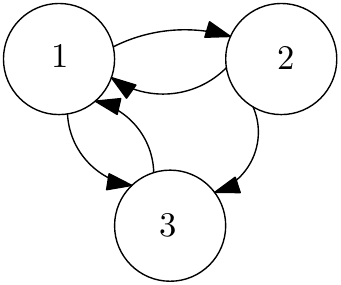}
\caption{\small Information flow graph $G$ of Example \ref{eg:3userproblem}.}
\label{fig:3userIFGraph}
\end{figure}
For this index coding problem, length of the optimal index code is two. Total number of optimal linear index codes is three. The list of optimal index codes are as follows:
\begin{itemize}
\item Code $\mathfrak{C}_{1}$ which transmits $ \lbrace x_{1}+x_{2},x_{1}+x_{3}\rbrace .$
\item Code $\mathfrak{C}_{2}$ which transmits $ \lbrace x_{1}+x_{2},x_{2}+x_{3}\rbrace .$
\item Code $\mathfrak{C}_{3}$ which transmits $\lbrace x_{1}+x_{3},x_{2}+x_{3}\rbrace .$
\end{itemize}
The number of transmissions used by each of the receivers in decoding its demanded message for the codes above is given in Table \ref{Tab:3userTableNoOfTransmissions}. From Table \ref{Tab:3userTableNoOfTransmissions}, we can infer that for all the optimal index codes, the maximum number of transmissions used by any receiver is two. So for this specific instance of index coding problem, any index code which is optimal in terms of bandwidth is optimal in terms of minimum maximum error probability also. 

\begin{table*}
\scriptsize
\centering{}
\begin{tabular}{|c|c|c|c|c|c|c|c|}
\hline
\multirow{2}{*}{Code} & \multirow{2}{*}{Encoding} &  \multicolumn{2}{|c|}{$R_{1}$}& $R_{2}$ & \multicolumn{2}{|c|}{$R_{3}$}\\
\cline{3-7}
& &  $x_{2} \in \mathcal{W}_{1}$ & $x_{3} \in \mathcal{W}_{1}$ & $x_{1} \in \mathcal{W}_{2}$ &$x_{1} \in \mathcal{W}_{3}$ & $x_{2} \in \mathcal{W}_{3}$ \\
\hline
$\mathfrak{C}_{1}$ & $x_{1}+x_{2},x_{1}+x_{3}$ & $1$ & $1$ & $1$ & $1$ & $2$\\
\hline
$\mathfrak{C}_{2}$ & $x_{2}+x_{1},x_{2}+x_{3}$ & $1$ & $2$ & $1$ & $1$ & $2$\\
\hline
$\mathfrak{C}_{3}$ & $x_{3}+x_{1},x_{3}+x_{2}$ & $1$ & $2$ & $2$ & $1$ & $1$\\
\hline
\end{tabular}
\caption{\small Comparison of optimal length linear codes for Example \ref{eg:3userproblem}. Each row in the table gives code and the corresponding number of transmissions the receiver uses in decoding its demanded messages.}
\label{Tab:3userTableNoOfTransmissions}
\end{table*}

\end{example}

\begin{example}
\label{eg:4userproblem}
Consider a single uniprior index coding problem with four messages $x_{1},x_{2},x_{3},x_{4}$ and four receivers $R_{1},R_{2},R_{3},R_{4}$. Each receiver $R_{i}$ knows $x_{i}$ and wants $x_{i+1}$ where $+$ denotes modulo 4 addition. The information flow graph $G$ for this problem is given in Figure \ref{fig:4userproblemIFGraph}. The optimal length of the index code for this index coding problem is three. We list out all possible optimal length linear index codes by an exhaustive search. Total number of optimal length linear index codes for this problem is $28$. We list out all possible index codes in Table \ref{Tab:4userTable}. There are many index codes in which the maximum number of transmissions used by a receiver is three. However there are twelve index codes in which the maximum number of transmissions used is two. The output of Algorithm \ref{alg:optimalic} belongs to the category of index codes which allows any receiver to decode its wanted message with the help of at most any two of the 3 transmissions. Observe that out of the 28 codes, 12 of them are good in terms of minimum-maximum error probability. Among the 12, there is one code which is the best in terms of minimizing the error probabilities of all the receivers as well. Our algorithm may not give that one. Algorithm \ref{alg:optimalic} ensures that the code which it outputs will belong to this group of 12 codes whose worst case error probabilities are same. Note that the number of codes which perform better in terms of minimizing the maximum error probability is less than 50\% of the total number of optimal length index codes. For a similar problem involving five receivers we were able to identify the total number of optimal length index codes as  840 and out of which at least 480 codes does not satisfy the minimum maximum error probability criterion. Hence we conclude that arbitrarily choosing an optimal length index code could result in using an index code which performs badly in terms of minimizing the maximum probability of error.

\begin{figure}
\centering{}\includegraphics[scale=1]{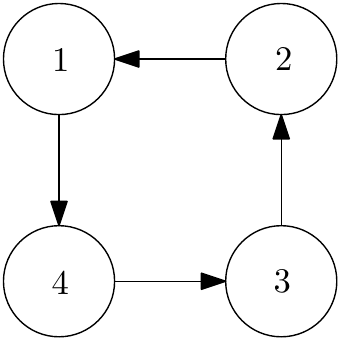}
\caption{\small Information flow graph $G$ of Example \ref{eg:4userproblem}.}
\label{fig:4userproblemIFGraph}
\end{figure}
\begin{table*}
\scriptsize
\centering{}
\begin{tabular}{|c|c|c|c|c|c|c|}
\hline
\multirow{2}{*}{Code} & \multirow{2}{*}{Encoding} & {$R_{1}$}& $R_{2}$ & {$R_{3}$} & {$R_{4}$}\\
\cline{3-6}
& &  $\mathcal{W}_{1}=\{x_2\}$ & $\mathcal{W}_{2}=\{x_3\}$ &$\mathcal{W}_{3}=\{x_4\}$ & $\mathcal{W}_{4}=\{x_1\}$\\
\hline
$\mathfrak{C}_{1}$ & $x_{1}+x_{2},x_{2}+x_{3},x_{3}+x_{4}$ &1 & 1 & 1 & 3 \\
\hline
$\mathfrak{C}_{2}$ & $x_{1}+x_{2},x_{2}+x_{3},x_2+x_4$ &1 & 1 & 2 & 2 \\
\hline
$\mathfrak{C}_{3}$ & $x_{1}+x_{2},x_{2}+x_{3},x_1+x_2+x_3+x_4$ &1 & 1 & 2 & 2 \\
\hline
$\mathfrak{C}_{4}$ & $x_{1}+x_{2},x_{2}+x_{3},x_1+x_4$ &1 & 1 & 3 & 1 \\
\hline
$\mathfrak{C}_{5}$ & $x_{1}+x_{2},x_{3}+x_{4},x_1+x_3$ & 1 & 2 & 1 & 2 \\
\hline
$\mathfrak{C}_{6}$ & $x_{1}+x_{2},x_{3}+x_{4},x_2+x_4$ &1 & 2 &1& 2 \\
\hline
$\mathfrak{C}_{7}$ & $x_{1}+x_{2},x_{3}+x_{4},x_1+x_4$ &1 &3 &1 & 1 \\
\hline
$\mathfrak{C}_{8}$  & $x_{1}+x_{2},x_{1}+x_{3},x_2+x_4$ &1 & 2 & 3 & 2 \\
\hline
$\mathfrak{C}_{9}$ & $x_{1}+x_{2}, x_{1}+x_{3},x_1+x_2+x_3+x_4$ &1 & 2 & 2& 3 \\
\hline
$\mathfrak{C}_{10}$ & $x_{1}+x_{2}, x_{1}+x_{3},x_1+x_4$ &1 & 2 & 2 &1 \\
\hline
$\mathfrak{C}_{11}$ & $x_{1}+x_{2},x_{2}+x_{4},x_1+x_2+x_3+x_4$ &1 &3 & 2 & 2 \\
\hline
$\mathfrak{C}_{12}$ & $x_{1}+x_{2},x_1+x_2+x_3+x_4,x_1+x_4$ &1 & 2 &2 &1 \\
\hline
$\mathfrak{C}_{13}$  & $x_{2}+x_{3},x_3+x_4,x_1+x_3$ &2 & 1 & 1 & 2 \\
\hline
$\mathfrak{C}_{14}$ & $x_{2}+x_{3},x_3+x_4,x_1+x_2+x_3+x_4$ &2 & 1 &1 & 2 \\
\hline
$\mathfrak{C}_{15}$ & $x_{2}+x_{3},x_3+x_4,x_1+x_4$ &3 & 1 & 1 & 1 \\
\hline
$\mathfrak{C}_{16}$ & $x_{2}+x_{3},x_1+x_3,x_2+x_4$ &2 &1 &2 &3 \\
\hline
$\mathfrak{C}_{17}$ & $x_{2}+x_{3},x_1+x_3,x_1+x_2+x_3+x_4$ &2 & 1 &3 &2 \\
\hline
$\mathfrak{C}_{18}$ & $x_{2}+x_{3},x_1+x_3,x_1+x_4$ &2 & 1 &2 &1 \\
\hline
$\mathfrak{C}_{19}$  & $x_{2}+x_{3},x_2+x_4,x_1+x_2+x_3+x_4$ &3 & 1& 2 & 2 \\
\hline
$\mathfrak{C}_{20}$  & $x_{2}+x_{3},x_2+x_4,x_1+x_4$ &2 &1 &2 & 1\\
\hline
$\mathfrak{C}_{21}$ & $x_{3}+x_{4},x_1+x_3,x_2+x_4$ &3 &2 & 1 & 2 \\
\hline
$\mathfrak{C}_{22}$  & $x_{3}+x_{4},x_1+x_3,x_1+x_2+x_3+x_4$ &2 & 3& 1 & 2 \\
\hline
$\mathfrak{C}_{23}$ & $x_{1}+x_{3},x_2+x_4,x_1+x_4$ &2 & 3 & 2 & 1 \\
\hline
$\mathfrak{C}_{24}$ & $x_{1}+x_{3},x_1+x_2+x_3+x_4,x_1+x_4$ &3 &2 & 2 & 1 \\
\hline
$\mathfrak{C}_{25}$ & $x_{2}+x_{4},x_1+x_2+x_3+x_4,x_1+x_4$ &2 & 2 &3 & 1 \\
\hline
$\mathfrak{C}_{26}$ & $x_{3}+x_{4},x_2+x_4,x_1+x_2+x_3+x_4$ &2 &2 & 1 & 3 \\
\hline
$\mathfrak{C}_{27}$  & $x_{3}+x_{4},x_2+x_4,x_1+x_4$ &2 & 2 & 1 &1 \\
\hline
$\mathfrak{C}_{28}$  & $x_{3}+x_{4},x_1+x_2+x_3+x_4,x_1+x_4$ &2 & 2 & 1 & 1 \\
\hline
\end{tabular}
\caption{\small Comparison of optimal length linear codes for Example \ref{eg:4userproblem}. Each row in the table gives code and the corresponding number of transmissions the receiver uses in decoding its demanded messages.}
\label{Tab:4userTable}
\end{table*}
\end{example}

\begin{example}
\label{example:theorem1}
Consider a single uniprior index coding problem with four messages $x_{1},x_{2},x_{3},x_{4}$ and four receivers $R_{1},R_{2},R_3 \text{ and } R_4$. Each receiver $R_{i}$ knows $x_{i}$. The want-sets for the receivers are as follows: $\mathcal{W}_{1}= \lbrace x_{2},x_{4} \rbrace ,\mathcal{W}_{2}=\lbrace x_{3} \rbrace, \mathcal{W}_{3}=\lbrace x_{1} \rbrace \text{ and } \mathcal{W}_{4}=\lbrace x_{2},x_{3} \rbrace.   
$
\begin{figure}[htbp]
\centering{}
\subfigure[Information flow graph $G$]{
\includegraphics[scale=1]{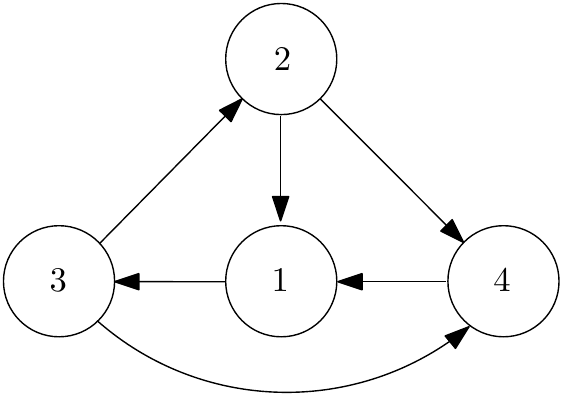}
\label{fig:IFGraphExampleTheorem}
}
\subfigure[Spanning tree $T$]{
\includegraphics[scale=1]{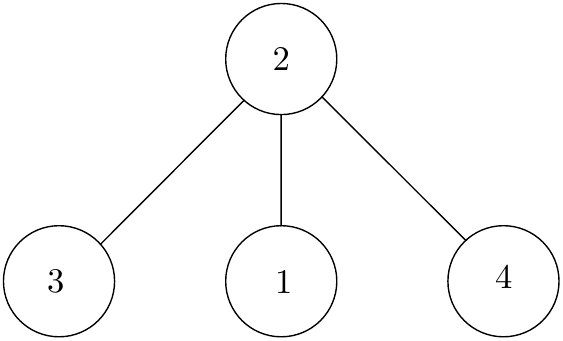}
\label{fig:SpanTreeExampleTheorem}
}
\caption{\small Information flow graph $G$ and Spanning tree $T$ of Example \ref{example:theorem1}.}
\end{figure}
The information flow graph $G$ of the problem is given in Figure ~\ref{fig:IFGraphExampleTheorem}. Note that the side information flow graph is a strongly connected graph. Hence the output of the pruning algorithm is $G$ itself. We perform Algorithm \ref{alg:optimalic} and the spanning tree obtained is given in Figure \ref{fig:SpanTreeExampleTheorem}. The index code which minimizes the maximum probability of error is $\lbrace c_{1}, c_{2},c_{3} \rbrace$ where $c_{1} =x_{2} \oplus x_{3},c_{2}=x_{2} \oplus x_{1}$ and $c_{3}= x_{2} \oplus x_{4}$. This enables all the receivers to decode any of its demands by using at most two transmissions. At receiver $R_{1}, x_{2}$ can be obtained by performing $x_{1} \oplus c_{2}$ and $x_{4}$ can be obtained by performing $x_{1} \oplus c_{2} \oplus c_{3}$. The decoding procedure used by receivers is given in Table \ref{Tab:DecExampleTheorem1}.

\scriptsize{
\begin{table}
\centering{}
\small
\begin{tabular}[h]{|c|c|c|}
\hline
Receivers & Demands & Decoding procedure \\
\hline
\multirow{2}{*}{$R_{1}$} & $x_{2}$ &  $x_{1} \oplus c_{2}$ \\ 
\cline{2-3}
&  $x_{4}$  & ~~~~~~$ x_{1} \oplus c_{2} \oplus c_{3} $ \\
\hline
$R_{2}$ & $x_{3}$ &  $x_{2} \oplus c_{1}$ \\ 
\hline
$R_{3}$ & $x_{1}$ &  ~~~~~~$x_{3} \oplus c_{2} \oplus c_{1}$ \\ 
\hline
\multirow{2}{*}{$R_{4}$} & $x_{2}$ &  $x_{4} \oplus c_{3}$ \\
\cline{2-3} 
&  $x_{3}$  & ~~~~~~$ x_{4} \oplus c_{3} \oplus c_{1} $ \\
\hline
\end{tabular}
\caption{\small Decoding procedure for Example \ref{example:theorem1}.}
\label{Tab:DecExampleTheorem1}
\end{table}
}
\normalsize
\end{example} 

\begin{example}
\label{example:theorem2}
Consider a single uniprior problem with five messages $x_{1},x_{2},x_{3},x_{4},x_{5}$ and five receivers $R_{1},R_{2},R_{3},R_{4},R_{5}$. Each $R_{i}$  knows $ x_{i} $ and wants $ x_{i+1} $ and $ x_{i+2} $ where $+$ denotes modulo 5 addition. The information  flow graph $G_{2}$ is given in Figure \ref{fig:IFExampleTheorem2}. The graph is strongly connected and all the edges are parts of some cycle. We perform Algorithm \ref{alg:optimalic} on $G_{2}$ and the spanning tree which minimizes the maximum distance is given in Figure \ref{fig:SpanTreeExampleTheorem2}.
\begin{figure}
\centering{}\subfigure[Information flow graph $G_{2}$ of Example \ref{example:theorem2}]{
\includegraphics[scale=1]{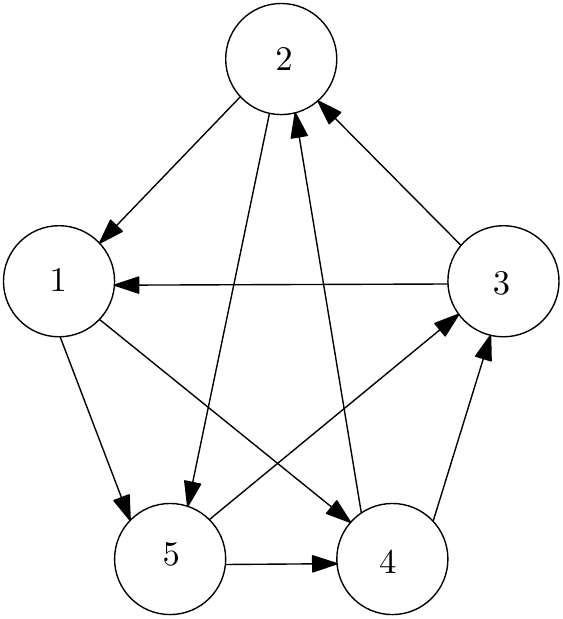}
\label{fig:IFExampleTheorem2}
}
~~~
\subfigure[Spanning tree $T$ obtained from Algorithm \ref{alg:optimalic} for Example \ref{example:theorem2}.]{
\includegraphics[scale=1]{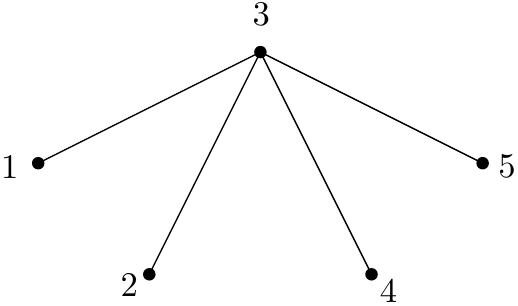}
\label{fig:SpanTreeExampleTheorem2}
}
\caption{\small Information flow graph $G$ and Spanning tree $T$ of Example \ref{example:theorem1}.}
\end{figure}

The index code which minimizes the maximum probability of error is $\lbrace c_{1},c_{2},c_{3},c_{4} \rbrace$ where $c_{1}= x_{1} \oplus x_{3}, c_{2}=x_{2} \oplus x_{3}, c_{3}=x_{3} \oplus x_{4}$ and $c_{4}= x_{3} \oplus x_{5}$. The decoding procedure at receivers is given in Table \ref{Tab:DecExampleTheorem2}. From the table we can observe that any receiver would take at most two transmissions to decode any of its messages. We also observe that for any $n$ (number of receivers), we will get a similar solution and number of transmissions required to decode any particular demanded message would be at most two. 

\begin{table}
\centering{}
\scriptsize
\scalebox{1}{
\begin{tabular}[h]{|c|c|c|}
\hline
Receivers & Demands & Decoding procedure \\
\hline
\multirow{2}{*}{$R_{1}$} & $x_{2}$ &  ~~~~~~$x_{1} \oplus c_{1} \oplus c_{2}$ \\ 
\cline{2-3}
&  $x_{3}$  & $ x_{1} \oplus c_{1}  $ \\
\hline
\multirow{2}{*}{$R_{2}$} & $x_{3}$ &  $x_{2} \oplus c_{2} $ \\ 
\cline{2-3}
&  $x_{4}$  & ~~~~~~$ x_{2} \oplus c_{2} \oplus c_{3}  $ \\
\hline
\multirow{2}{*}{$R_{3}$} & $x_{4}$ &  $x_{3} \oplus c_{3} $ \\ 
\cline{2-3}
&  $x_{5}$  & $ x_{3} \oplus c_{4}  $ \\
\hline
\multirow{2}{*}{$R_{4}$} & $x_{5}$ &  ~~~~~~$x_{4} \oplus c_{3} \oplus c_{4}$ \\ 
\cline{2-3}
&  $x_{1}$  & ~~~~~~$ x_{4} \oplus c_{3} \oplus c_{1}  $ \\
\hline
\multirow{2}{*}{$R_{5}$} & $x_{1}$ &  ~~~~~~$x_{5} \oplus c_{4} \oplus c_{1}$ \\
\cline{2-3}
&  $x_{2}$  & ~~~~~~$ x_{5} \oplus c_{4} \oplus c_{2}  $ \\
\hline
\end{tabular}
}
\caption{\small Decoding procedure for Example \ref{example:theorem2}.}
\label{Tab:DecExampleTheorem2}
\end{table}

\end{example}   

\begin{example}
Consider the index coding problem of Example \ref{eg:NoofTransmissions}. The information flow graph $G$ of this problem is given in Figure \ref{fig:IFgraph1}. To obtain the index code which gives minimum maximum probability of error across all receivers, we perform Algorithm \ref{alg:optimalic}. The spanning tree obtained from Algorithm \ref{alg:optimalic} is given in Figure \ref{fig:SpanTreeExample1}. The index code which minimizes the maximum probability of error is $\mathfrak{C}_{1}$ described by matrix $L_{1}$ given in Example \ref{eg:NoofTransmissions}. Length of the code is eight and can be represented as $\lbrace c_{1},c_{2},\ldots,c_{8}\rbrace$ where $c_{i}=x_{1} \oplus x_{i+1}$. The decoding procedure at receivers for the code $\mathfrak{C}_{1}$ is given in Table \ref{Tab:DecNoofTransmissions}. It is evident from the table that for code $\mathfrak{C}_{1}$, the maximum number of transmissions required to decode any demanded message across all receivers is two.

\begin{figure}
\centering
\subfigure[Information flow graph $G$ of Example \ref{eg:Theorem9receivers}.]{
\centering{
\includegraphics[scale=0.8]{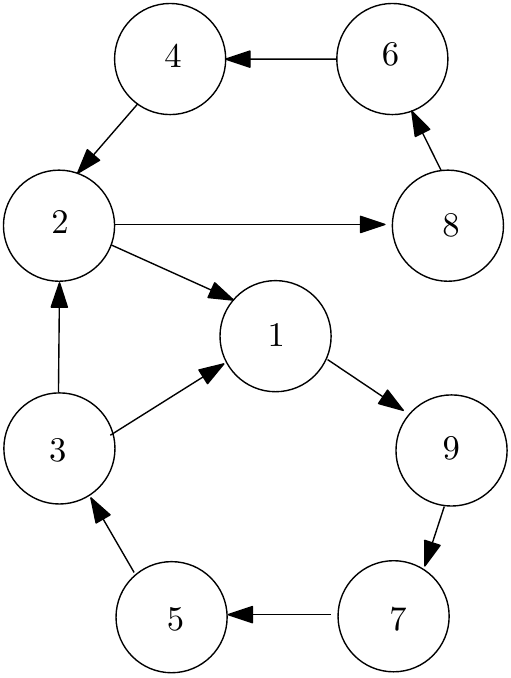}
\label{fig:IFgraph1}
}
}
~~
\subfigure[Spanning tree $T$ obtained from Algorithm \ref{alg:optimalic} for Example \ref{eg:Theorem9receivers}.]{
\centering{
\includegraphics[scale=1.2]{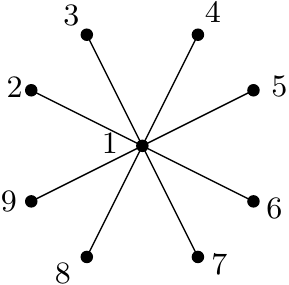}
\label{fig:SpanTreeExample1}
}
}
\caption{\small Information flow graph $G$ and Spanning tree $T$ of Example \ref{eg:Theorem9receivers}.}
\end{figure}

\label{eg:Theorem9receivers}
\end{example}  

\begin{table}
\centering{}
\scriptsize
\scalebox{1}{
\begin{tabular}[h]{|c|c|c|}
\hline
Receivers & Demands & Decoding procedure \\
\hline
\multirow{2}{*}{$R_{1}$} & $x_{2}$ &  $x_{1} \oplus c_{1}$ \\ 
\cline{2-3}
&  $x_{3}$  & $ x_{1} \oplus c_{2}  $ \\
\hline
\multirow{2}{*}{$R_{2}$} & $x_{3}$ &  ~~~~~~$x_{2} \oplus c_{1} \oplus c_{2} $ \\ 
\cline{2-3}
&  $x_{4}$  & ~~~~~~$ x_{2} \oplus c_{1} \oplus c_{3}  $ \\
\hline
$R_{3}$ & $x_{5}$ &  ~~~~~~$x_{3} \oplus c_{2} \oplus c_{4} $ \\ 
\hline
$R_{4}$ & $x_{6}$ &  ~~~~~~$x_{4} \oplus c_{3} \oplus c_{5}$ \\ 
\hline
$R_{5}$ & $x_{7}$ &  ~~~~~~$x_{5} \oplus c_{4} \oplus c_{6}$ \\ 
\hline
$R_{6}$ & $x_{8}$ &  ~~~~~~$x_{6} \oplus c_{5} \oplus c_{7}$ \\ 
\hline
$R_{7}$ & $x_{9}$ &  ~~~~~~$x_{7} \oplus c_{6} \oplus c_{8}$ \\ 
\hline
$R_{8}$ & $x_{1}$ &  $x_{8} \oplus c_{7}$ \\ 
\hline
$R_{9}$ & $x_{2}$ &  ~~~~~~$x_{9} \oplus c_{8} \oplus c_{1}$ \\ 
\hline
\end{tabular}
}
\caption{\small Decoding procedure for Example \ref{eg:Theorem9receivers}.}
\label{Tab:DecNoofTransmissions}
\end{table}

\section{Simulation Results}
\label{sec:Simulation}

In this section we give simulation results  which show that the choice of the optimal index codes matters. We show that optimal index codes which use lesser number of transmissions to decode the messages perform better than those using more number of transmissions. We consider the index coding problem in Example \ref{example1} below and  observe an improvement in the performance by choosing index code obtained from Algorithm \ref{alg:optimalic} over another arbitrary optimal index code. This shows the significance of  optimal index codes which use small number of transmissions to decode the messages at the receivers. 

\begin{example}
\label{example1}
Consider a single uniprior index coding problem $\mathcal{I}(X,\mathcal{R})$ with $X=\lbrace x_{1},x_{2},\ldots,x_{7} \rbrace$ and $\mathcal{R}= \lbrace R_{1},R_{2},\ldots,R_{7} \rbrace$. Each receiver $R_{i} \in \mathcal{R}$, knows $x_{i}$ and has a want-set $\mathcal{W}_{i}=X \setminus \lbrace x_{i} \rbrace$. We consider two index codes for the problem and show by simulation the improvement in using the index code obtained from Algorithm \ref{alg:optimalic}. 

Let $\mathfrak{C}_{1}$ be the linear index code obtained from the proposed Algorithm \ref{alg:optimalic}. We use code $\mathfrak{C}_{2}$, another valid index code of optimal bandwidth for performance comparison. Codes $\mathfrak{C}_{1}$ and $\mathfrak{C}_{2}$ are  described by the matrices $L_{1}$ and $L_{2}$ respectively. The matrices are given below.

\scriptsize{
\[ L_{1}=
\left[\begin{array}{cccccc}
1 & 1 & 1 & 1 & 1 & 1\\
1 & 0 & 0 & 0 & 0 & 0\\
0 & 1 & 0 & 0 & 0 & 0\\
0 & 0 & 1 & 0 & 0 & 0\\
0 & 0 & 0 & 1 & 0 & 0\\
0 & 0 & 0 & 0 & 1 & 0\\
0 & 0 & 0 & 0 & 0 & 1
\end{array}\right],
L_{2}= \left[\begin{array}{cccccc}
1 & 0 & 0 & 0 & 0 & 0\\
1 & 1 & 0 & 0 & 0 & 0\\
0 & 1 & 1 & 0 & 0 & 0\\
0 & 0 & 1 & 1 & 0 & 0\\
0 & 0 & 0 & 1 & 1 & 0\\
0 & 0 & 0 & 0 & 1 & 1\\
0 & 0 & 0 & 0 & 0 & 1
\end{array}\right]
.\] 
}
\normalsize
Consider receiver $R_{1}$. For code $\mathfrak{C}_{1}$, receiver $R_{1}$ uses only one transmission for decoding any of its demands. However for code $\mathfrak{C}_{2}$, receiver $R_{1}$ uses more than one transmission for decoding all the demands. For example in order to decode message $x_{4} \in \mathcal{W}_{1}$, receiver $R_{1}$ has to make use of three transmissions. 

In the simulation, the source uses symmetric $4$-PSK signal set which is equivalent to two binary transmissions. The mapping from bits to complex symbols is assumed to be Gray Mapping. We first consider the scenario in which the fading is Rayleigh and the fading coefficient $h_{j}$ of the channel between source and receiver $R_{j}$ is $\mathcal{C}\mathcal{N}(0,1)$. The SNR Vs. BEP curves for all the receivers for code $\mathfrak{C}_{1}$ is plotted in Fig. \ref{fig:SimExample1RayleighAll1}. From Fig \ref{fig:SimExample1RayleighAll1}, we can observe that maximum error probability occurs at receiver $R_{7}$. Similar plot for all the receivers while using code $\mathfrak{C}_{2}$ is shown in Fig \ref{fig:SimExample1RayleighAll2}. From Fig. \ref{fig:SimExample1RayleighAll2} we can observe that for code $\mathfrak{C}_{2}$ maximum error probability occurs at receiver $R_{7}$. We compare the performance of both the codes at receiver $R_{7}$ in Fig. \ref{fig:SimExampleReceiver7}. We can observe from Fig. \ref{fig:SimExampleReceiver7} that the maximum probability of error across receivers is less for code $\mathfrak{C}_{1}$ compared to code $\mathfrak{C}_{2}$. 
\begin{figure*}[htbp]
\centering{}
\includegraphics[width=15cm,height=9cm]{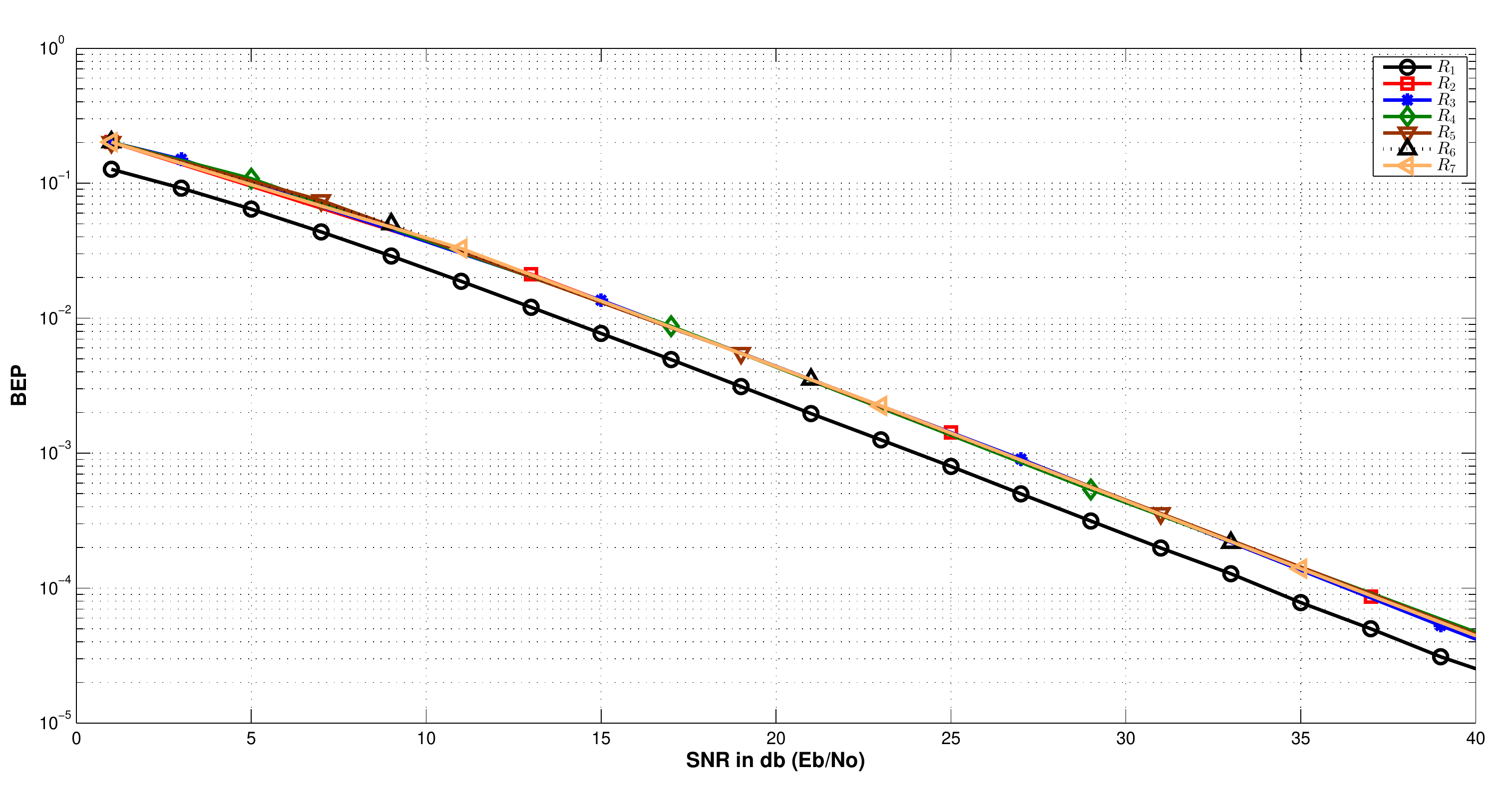}
\caption{\small SNR Vs BEP for code $\mathfrak{C}_{1}$ for Rayleigh fading scenario, at all receivers of Example \ref{example1}. }
\label{fig:SimExample1RayleighAll1}
\end{figure*}
\begin{figure*}
\centering{}
\includegraphics[width=15cm,height=9cm]{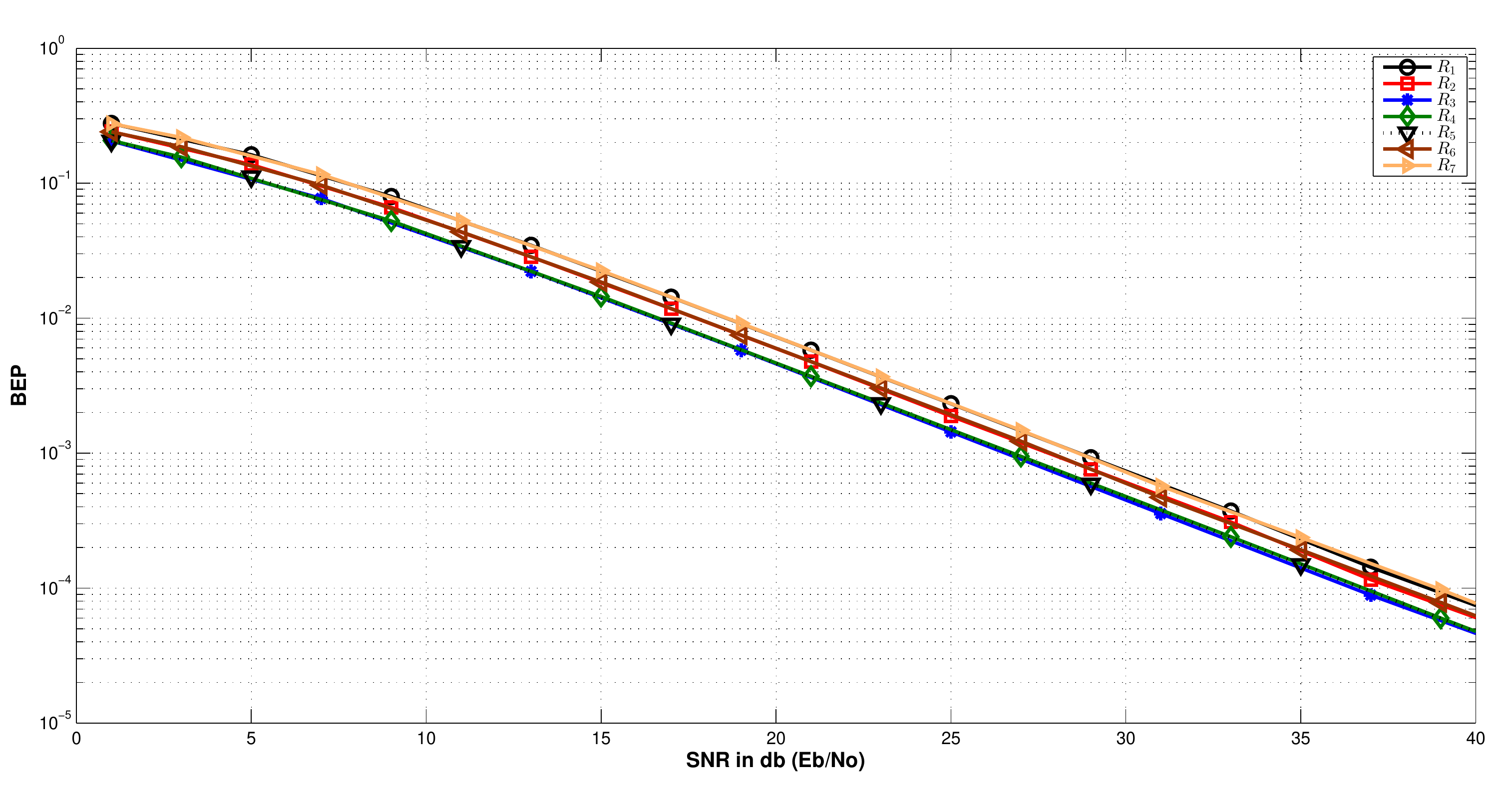}
\caption{\small SNR Vs BEP for code $\mathfrak{C}_{2}$ for Rayleigh fading scenario, at all receivers of Example \ref{example1}. }
\label{fig:SimExample1RayleighAll2}
\end{figure*} 
\begin{figure*}
\centering{}
\includegraphics[scale=0.5]{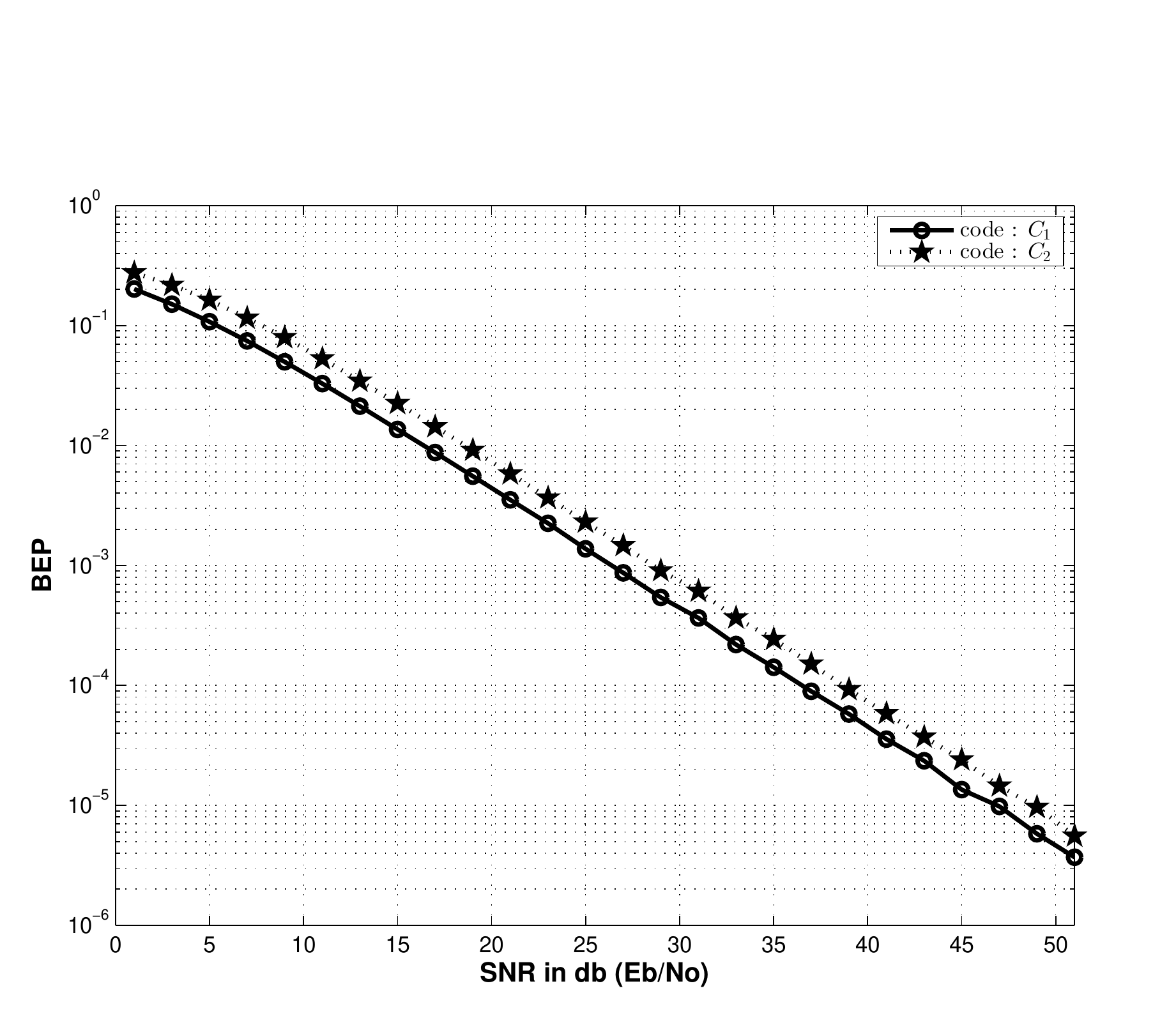}
\caption{\small SNR Vs BEP for codes $\mathfrak{C}_{1}$ and $\mathfrak{C}_{2}$ for Rayleigh fading scenario, at receiver $R_{7}$ of Example \ref{example1}. }
\label{fig:SimExampleReceiver7}
\end{figure*}

\begin{figure}
\centering
\begin{minipage}{0.45\textwidth}
\centering
\includegraphics[width=3in,height=2.25in]{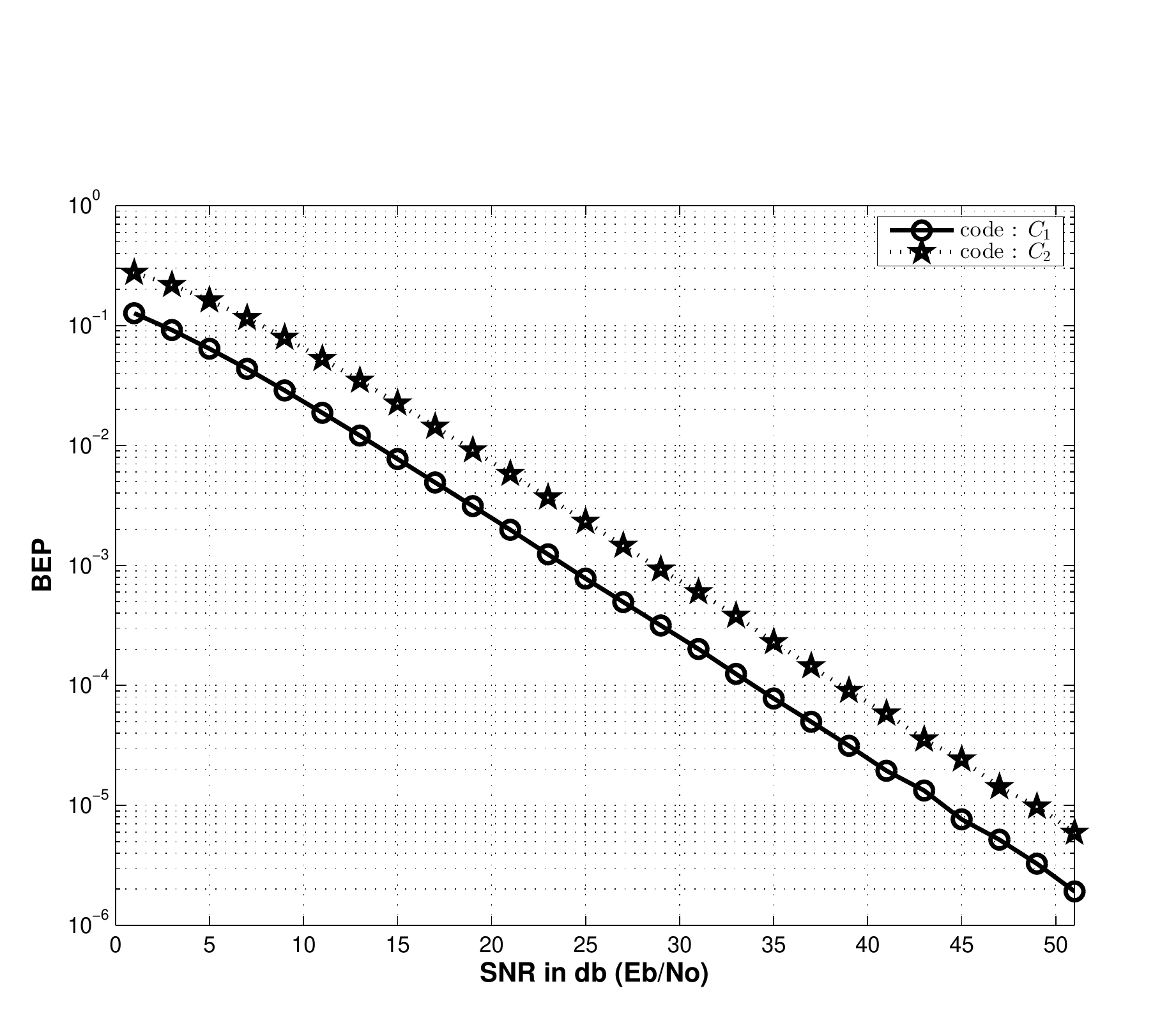}
\caption{\scriptsize SNR Vs BEP for codes $\mathfrak{C}_{1}$ and $\mathfrak{C}_{2}$ for Rayleigh fading scenario, at receiver $R_{1}$ of Example \ref{example1}. }
\label{fig:SimExampleReceiver1}
\end{minipage}\hfill
\begin{minipage}{0.45\textwidth}
\centering
\includegraphics[width=3in,height=2.25in]{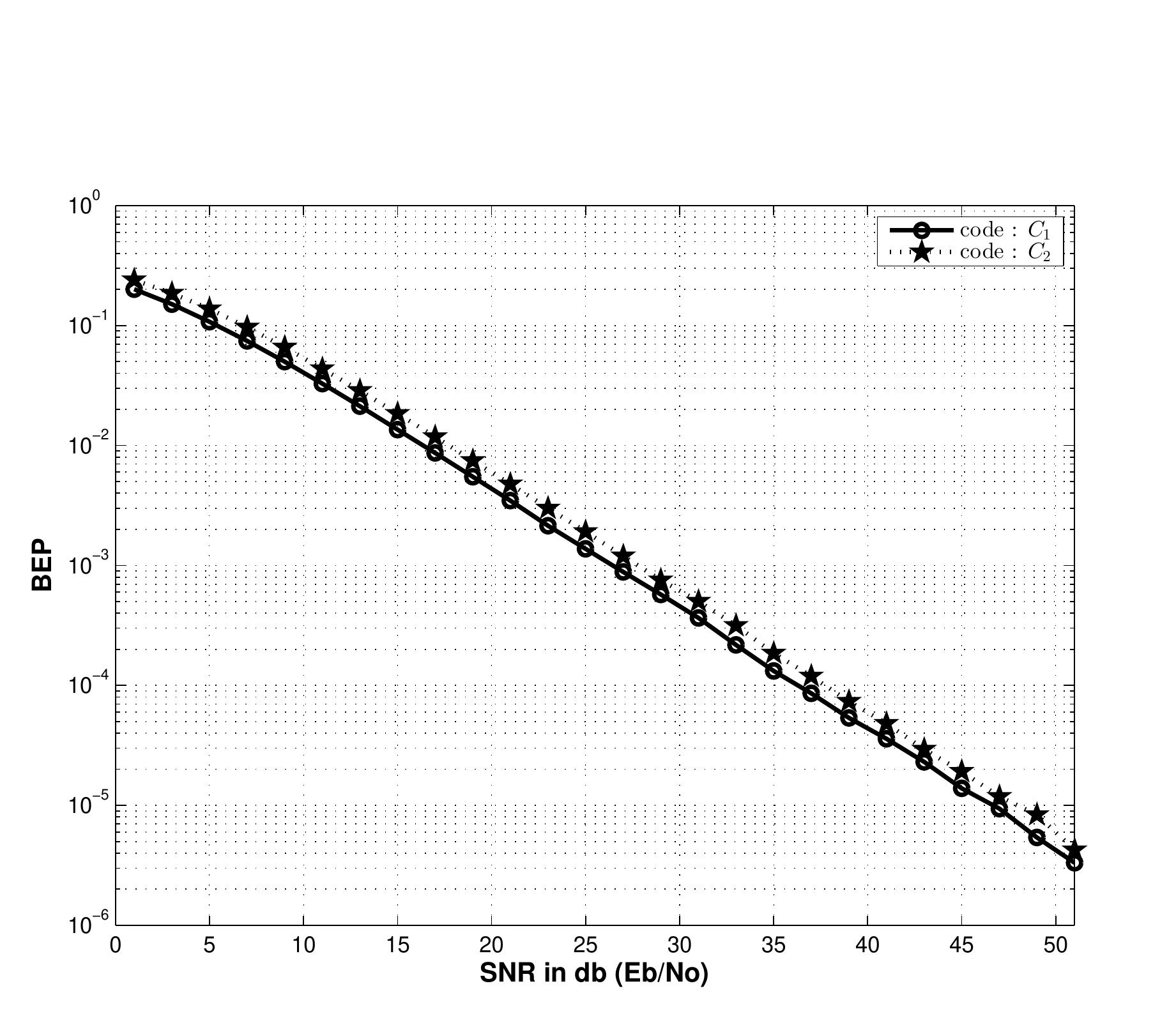}
\caption{\scriptsize SNR Vs BEP for codes $\mathfrak{C}_{1}$ and $\mathfrak{C}_{2}$ for Rayleigh fading scenario, at receiver $R_{2}$ of Example \ref{example1}. }
\label{fig:SimExampleReceiver2}
\end{minipage}
\end{figure}

\begin{figure}
\centering
\begin{minipage}{0.45\textwidth}
\centering
\includegraphics[width=3in,height=2.25in]{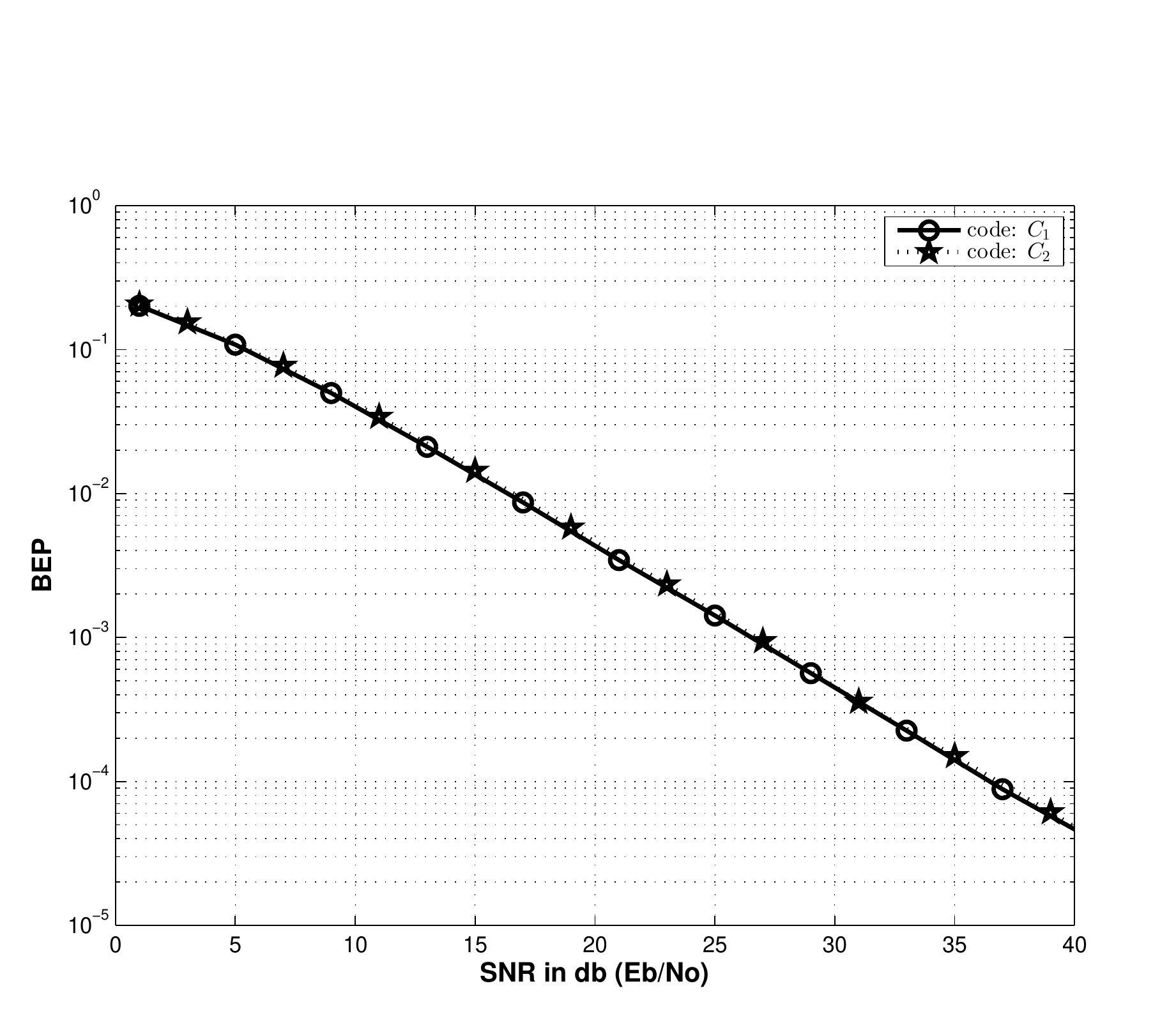}
\caption{\scriptsize SNR Vs BEP for codes $\mathfrak{C}_{1}$ and $\mathfrak{C}_{2}$ for Rayleigh fading scenario, at receiver $R_{3}$ of Example \ref{example1}. }
\label{fig:SimExampleReceiver3}
\end{minipage}\hfill
\begin{minipage}{0.45\textwidth}
\centering
\includegraphics[width=3in,height=2.25in]{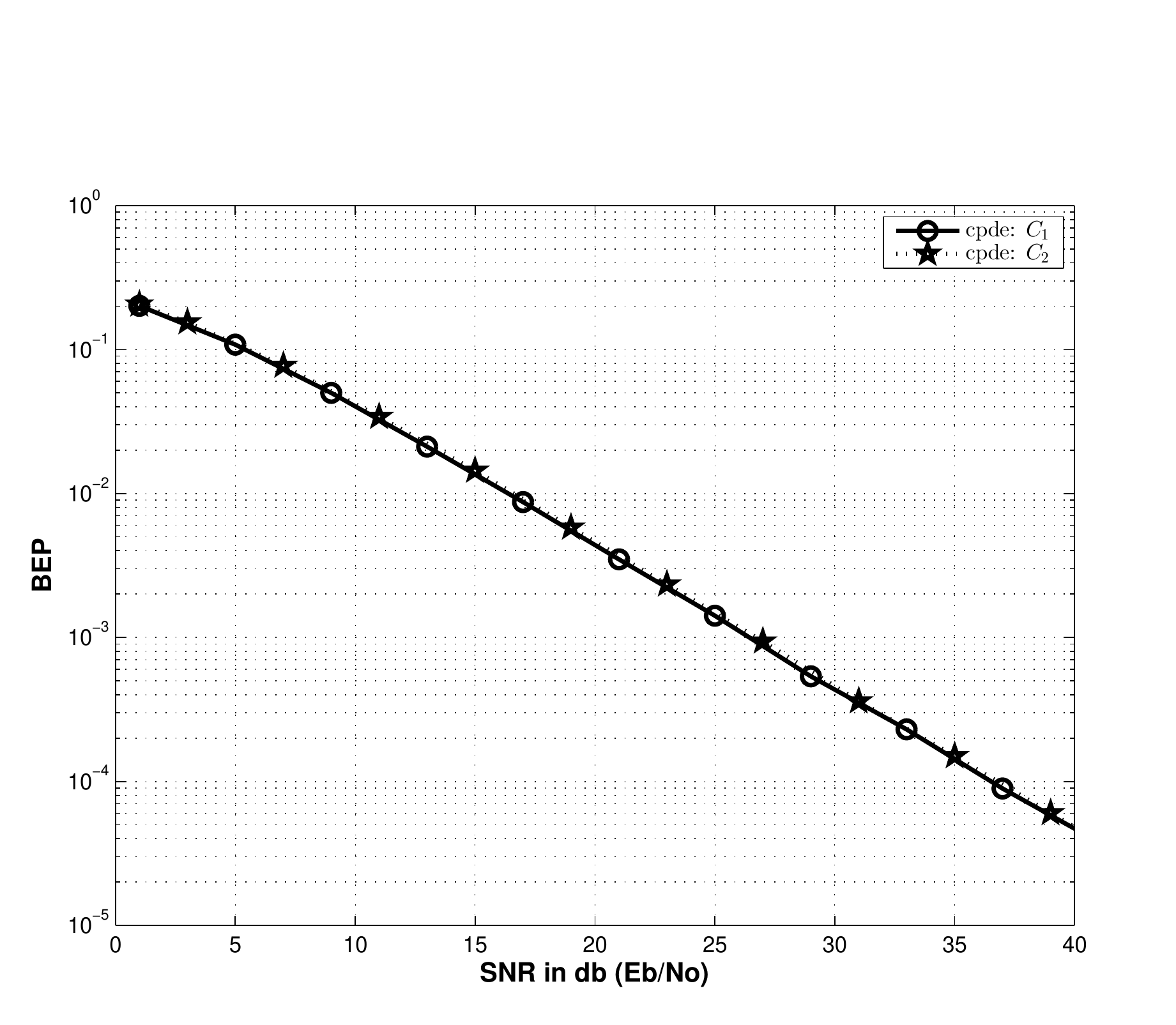}
\caption{\scriptsize SNR Vs BEP for codes $\mathfrak{C}_{1}$ and $\mathfrak{C}_{2}$ for Rayleigh fading scenario, at receiver $R_{4}$ of Example \ref{example1}. }
\label{fig:SimExampleReceiver4}
\end{minipage}
\end{figure}
\begin{figure}
\centering
\begin{minipage}{0.45\textwidth}
\centering
\includegraphics[width=3in,height=2.25in]{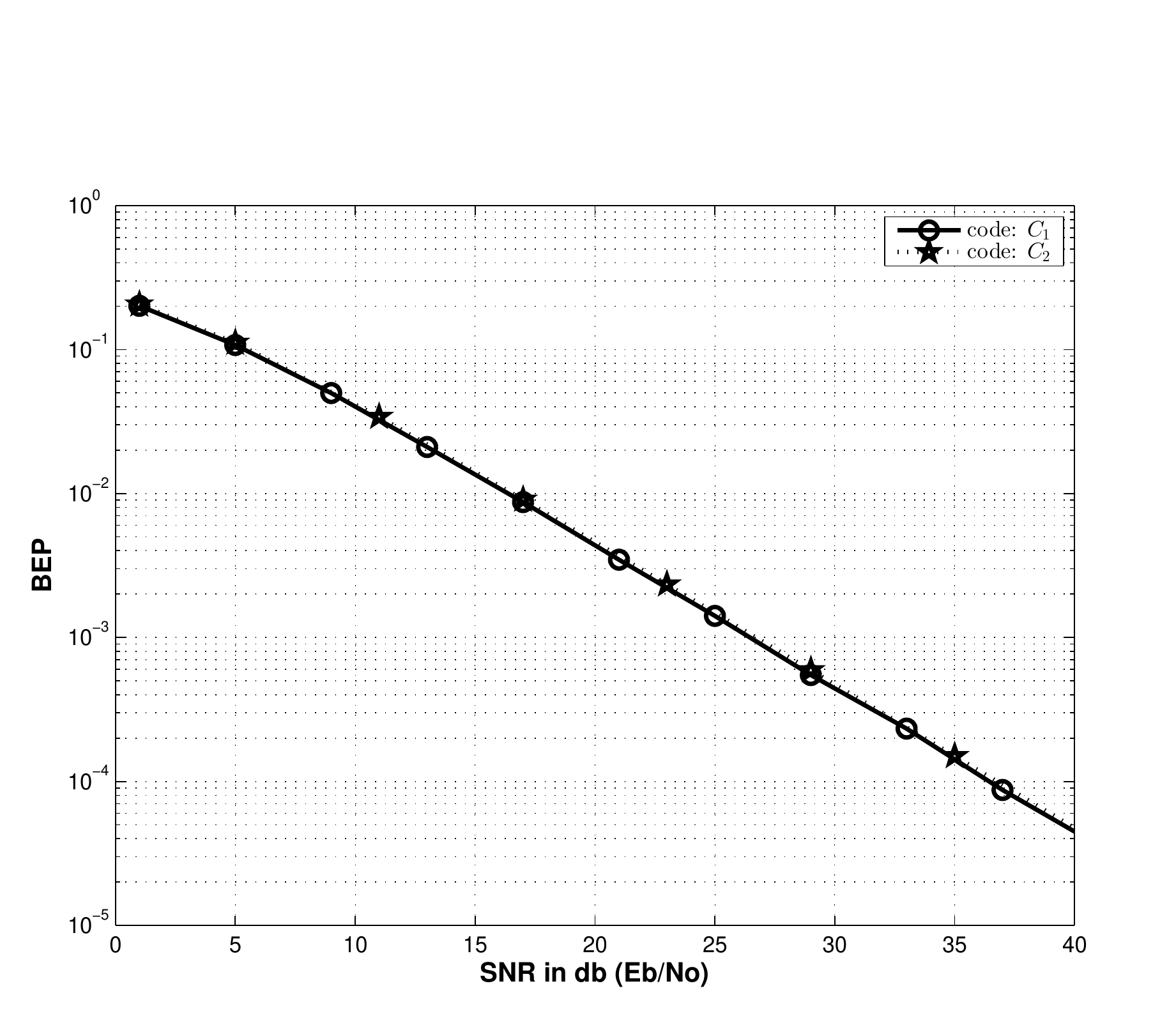}
\caption{\scriptsize SNR Vs BEP for codes $\mathfrak{C}_{1}$ and $\mathfrak{C}_{2}$ for Rayleigh fading scenario, at receiver $R_{5}$ of Example \ref{example1}. }
\label{fig:SimExampleReceiver5}
\end{minipage}\hfill
\begin{minipage}{0.45\textwidth}
\centering
\includegraphics[width=3in,height=2.25in]{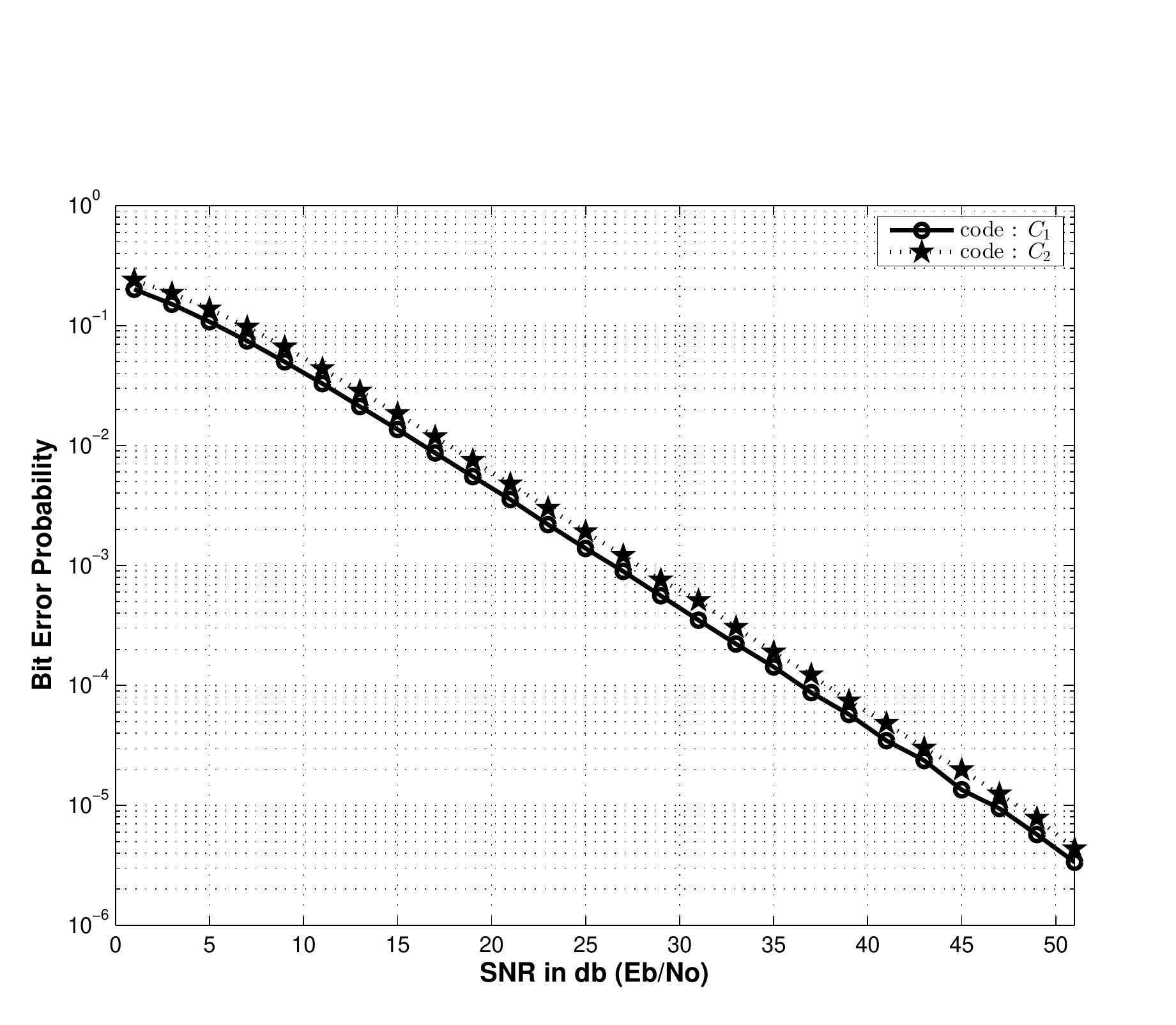}
\caption{\scriptsize SNR Vs BEP for codes $\mathfrak{C}_{1}$ and $\mathfrak{C}_{2}$ for Rayleigh fading scenario, at receiver $R_{6}$ of Example \ref{example1}. }
\label{fig:SimExampleReceiver6}
\end{minipage}
\end{figure}

The SNR Vs. BEP curves for codes $\mathfrak{C}_{1}$ and $\mathfrak{C}_{2}$ for remaining receivers are shown in Fig. \ref{fig:SimExampleReceiver1} - Fig. \ref{fig:SimExampleReceiver6}. Fig. \ref{fig:SimExampleReceiver1} shows the SNR Vs. BEP at receiver $R_{1}$. From Fig. \ref{fig:SimExampleReceiver1}, we can clearly see that code $\mathfrak{C}_{1}$ shows a better performance of around $4.5$dB compared to code $\mathfrak{C}_{2}$. Similar increase in performance was observed at all other receivers. We can observe that in all receivers Code $\mathfrak{C}_{1}$ performs at least as good as code $\mathfrak{C}_{2}$. So in terms of reducing the probability of error, Code $\mathfrak{C}_{1}$ performs better than Code $\mathfrak{C}_{2}$.

We also consider the scenario in which the channel between source and receiver $R_{j}$ is a Rician fading channel. The fading coefficient $h_{j}$  is Rician with a Rician factor 2. The source uses $4$-PSK signal set along with Gray mapping.  The SNR Vs. BEP curves for all receivers while using code $\mathfrak{C}_{1}$ and code $\mathfrak{C}_{2}$ is given in Fig. \ref{fig:SimExample1RicianAll1} and Fig. \ref{fig:SimExample1RicianAll2} respectively. We observe that maximum error probability occurs at receiver $R_{7}$ for both the codes $\mathfrak{C}_{1}$ and $\mathfrak{C}_{2}$. The SNR Vs. BEP curves for both the codes at receiver $R_{7}$ is shown in Fig. \ref{fig:SimExampleRicianR7}. From Fig. \ref{fig:SimExampleRicianR7} we observe that maximum error probability for code $\mathfrak{C}_{1}$ is lesser than for code $\mathfrak{C}_{2}$. The SNR Vs. BEP plots for both the codes at other receivers are given in Fig. \ref{fig:SimExampleRicianR1} - Fig. \ref{fig:SimExampleRicianR6}. It is evident from the plots that code $\mathfrak{C}_{1}$ performs better than code $\mathfrak{C}_{2}$. Though at some receivers it matches the performance, improvement is evident at receivers $R_{1}$ and $R_{7}$. From the simulation results we can conclude that in both Rayleigh and Rician fading models, code $\mathfrak{C}_{1}$ performs better than code $\mathfrak{C}_{2}$ in terms of reducing the probability of error. 

\begin{figure}
\centering{}
\includegraphics[width=15cm,height=9cm]{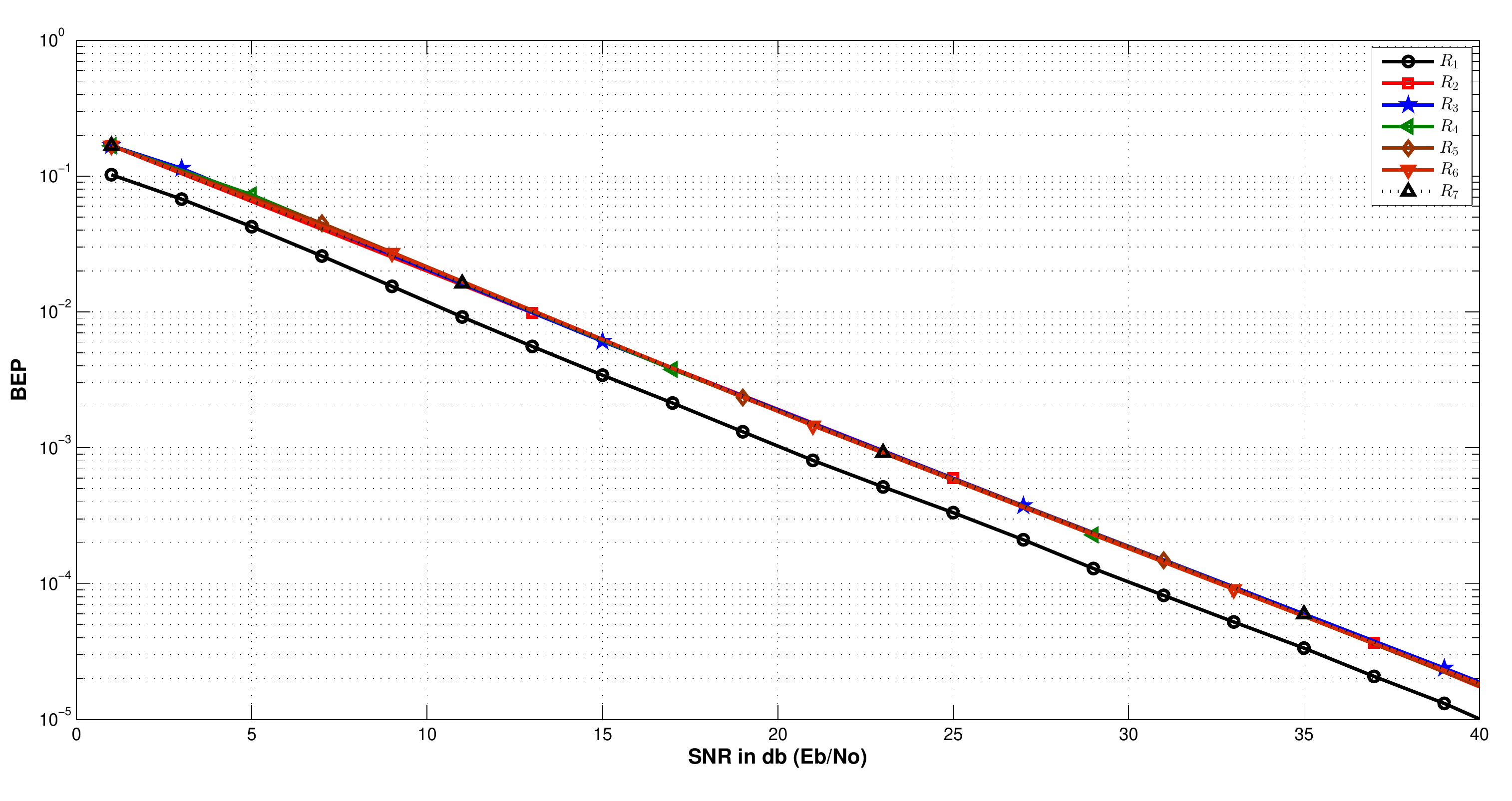}
\caption{\small SNR Vs BEP for code $\mathfrak{C}_{1}$ for Rician fading scenario, at all receivers of Example \ref{example1}. }
\label{fig:SimExample1RicianAll1}
\end{figure} 
\begin{figure}
\centering{}
\includegraphics[width=15cm,height=9cm]{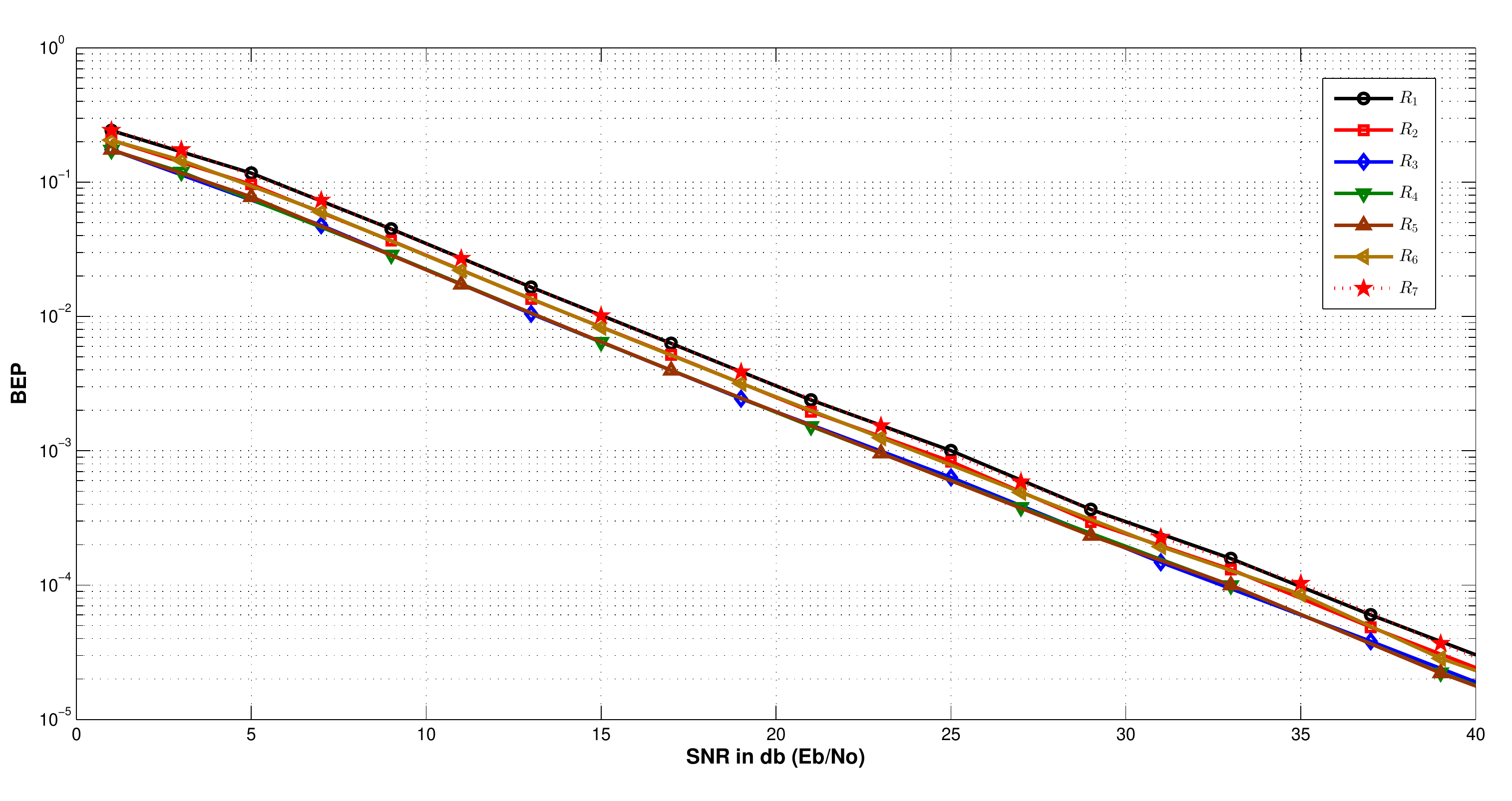}
\caption{\small SNR Vs BEP for code $\mathfrak{C}_{2}$ for Rician fading scenario, at all receivers of Example \ref{example1}. }
\label{fig:SimExample1RicianAll2}
\end{figure} 


\begin{figure}
\centering{}
\includegraphics[scale=0.5]{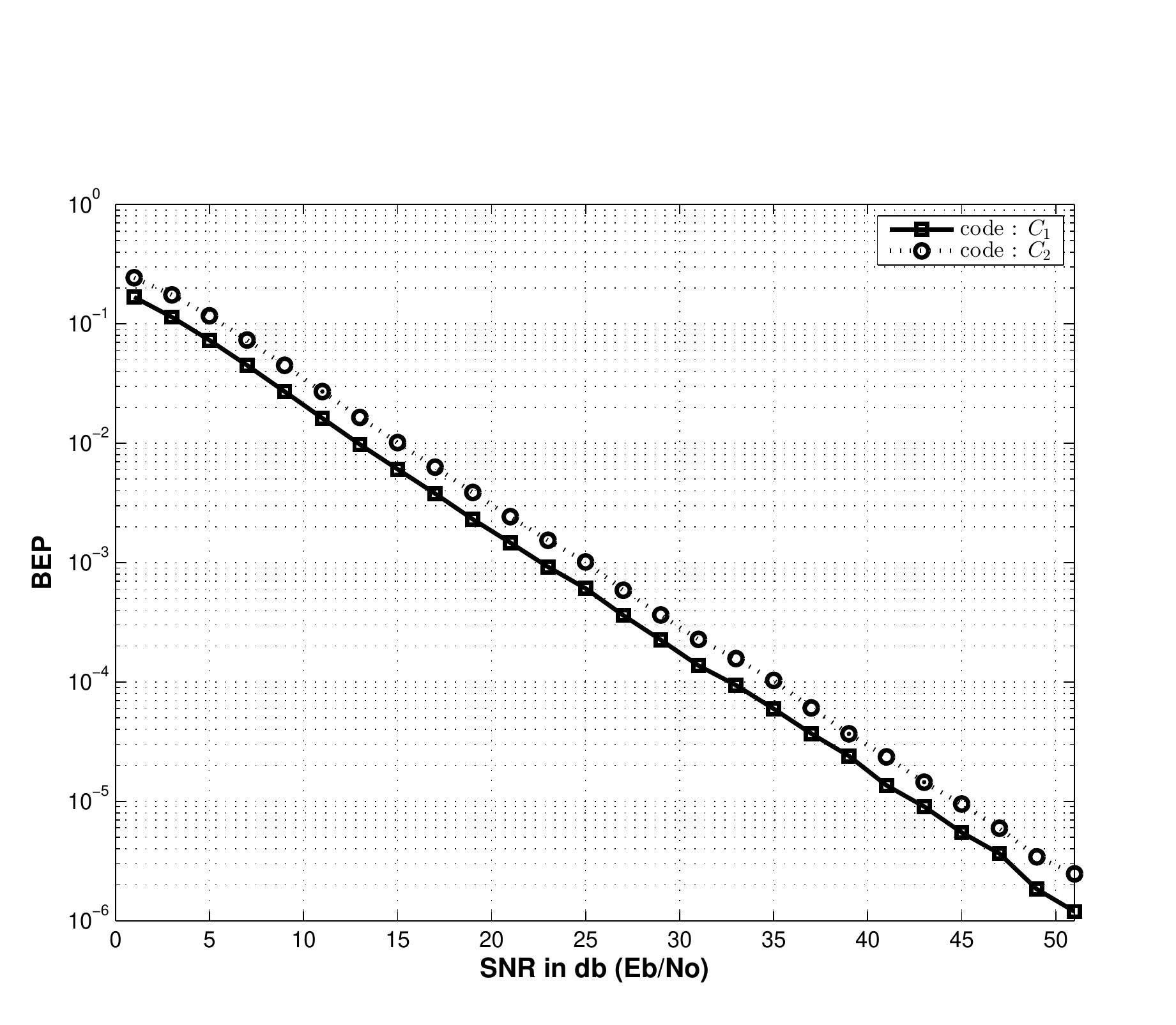}
\caption{\small SNR Vs BEP for codes $\mathfrak{C}_{1}$ and $\mathfrak{C}_{2}$ for Rician fading scenario, at receiver $R_{7}$ of Example \ref{example1}. }
\label{fig:SimExampleRicianR7}
\end{figure} 

\begin{figure}
\centering
\begin{minipage}{0.45\textwidth}
\centering
\includegraphics[scale=0.4]{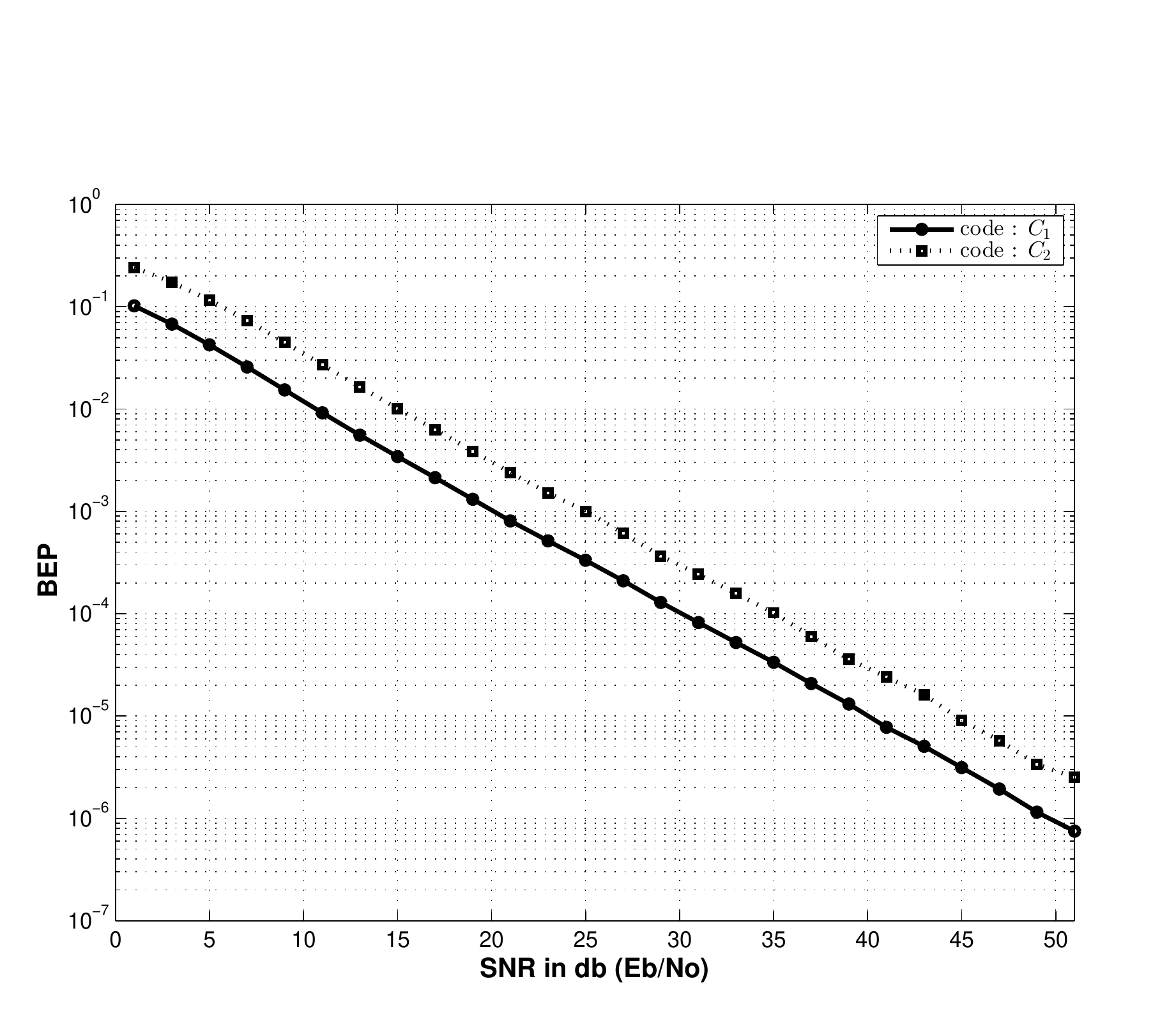}
\caption{\scriptsize SNR Vs BEP for codes $\mathfrak{C}_{1}$ and $\mathfrak{C}_{2}$ for Rician fading scenario, at receiver $R_{1}$ of Example \ref{example1}. }
\label{fig:SimExampleRicianR1}
\end{minipage}\hfill
\begin{minipage}{0.45\textwidth}
\centering
\includegraphics[scale=0.4]{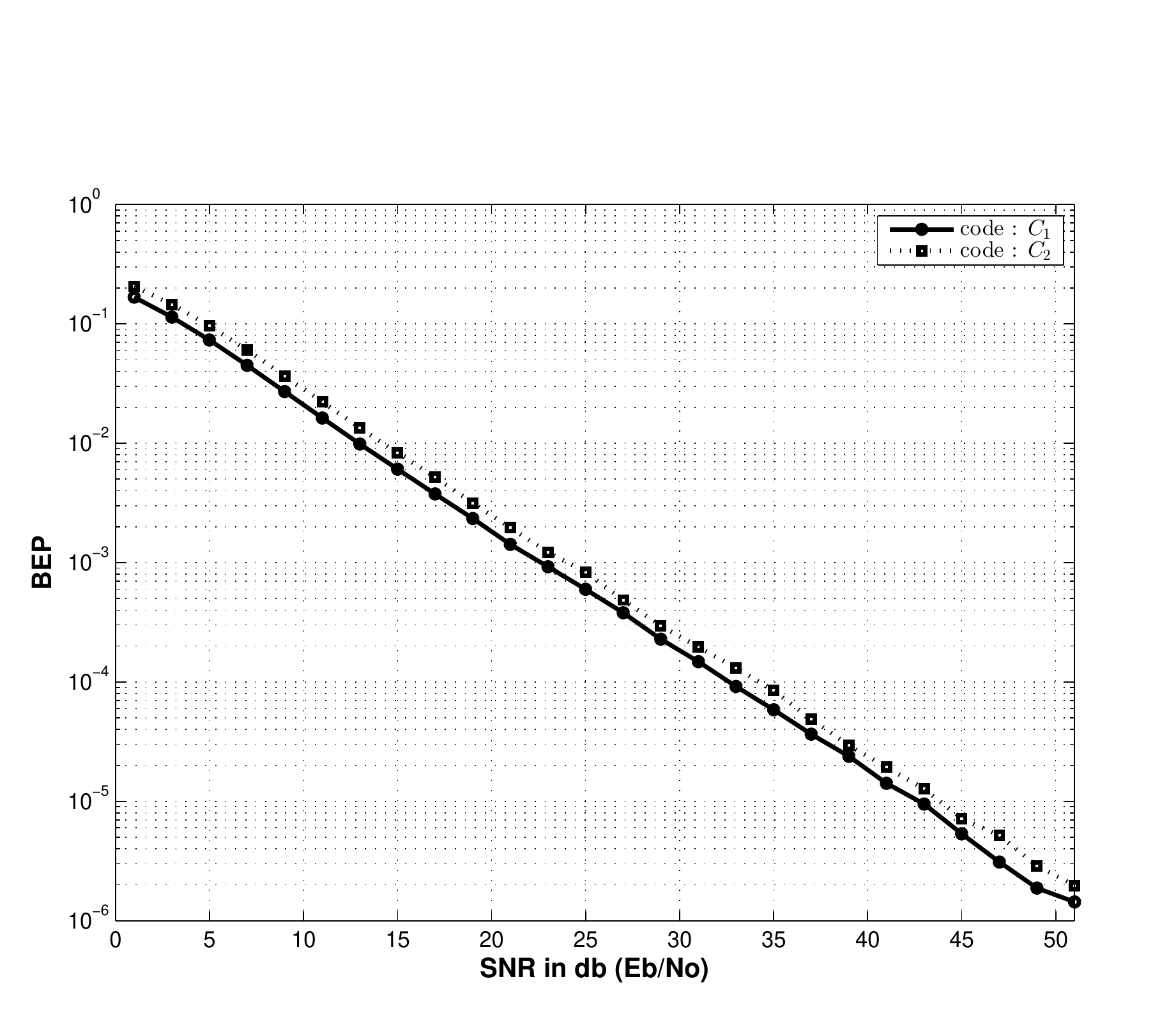}
\caption{\scriptsize SNR Vs BEP for codes $\mathfrak{C}_{1}$ and $\mathfrak{C}_{2}$ for Rician fading scenario, at receiver $R_{2}$ of Example \ref{example1}. }
\label{fig:SimExampleRicianR2}
\end{minipage}
\end{figure}

%
\begin{figure}
\centering
\begin{minipage}{0.45\textwidth}
\centering
\includegraphics[width=3in,height=2.25in]{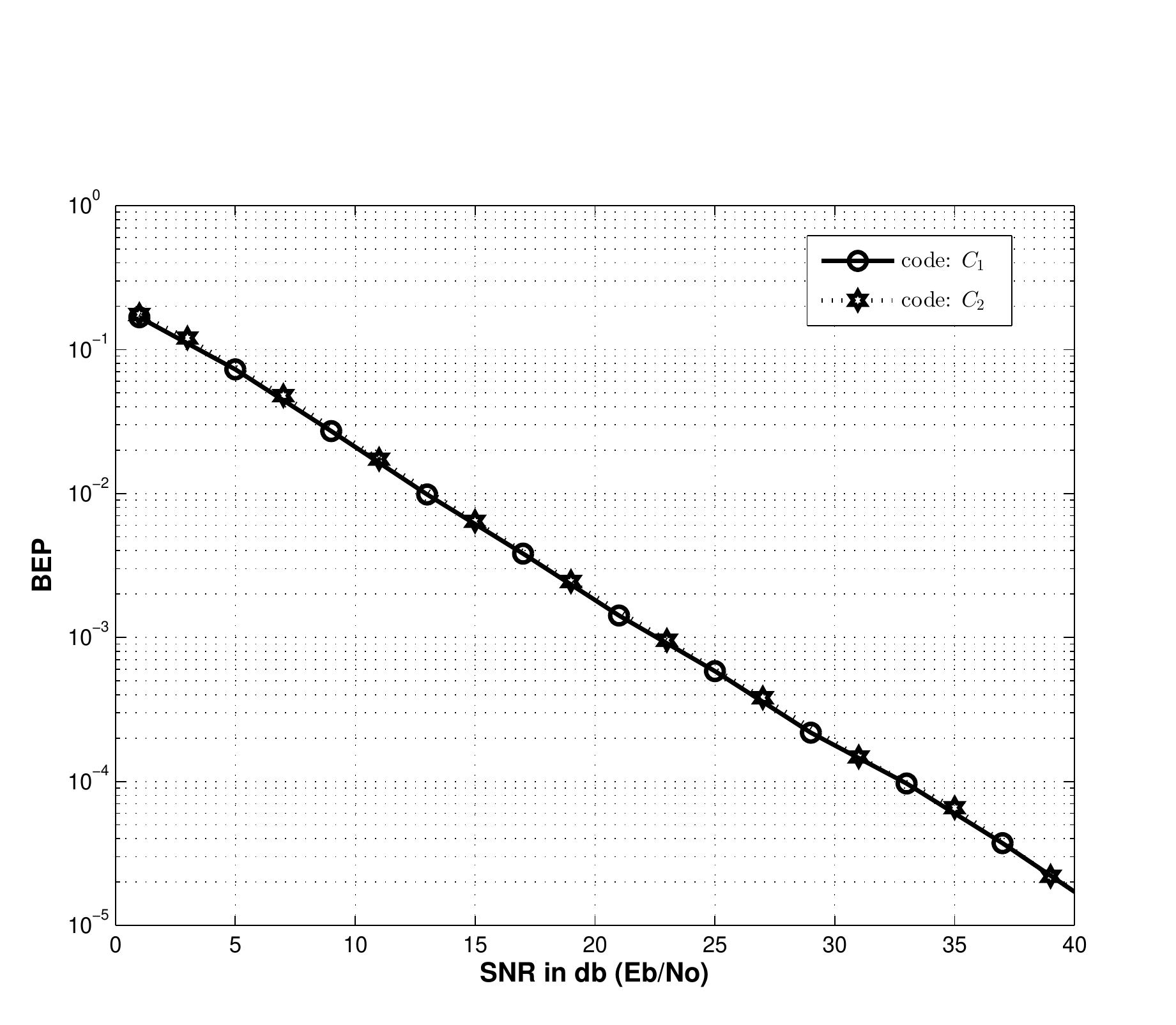}
\caption{\scriptsize SNR Vs BEP for codes $\mathfrak{C}_{1}$ and $\mathfrak{C}_{2}$ for Rician fading scenario, at receiver $R_{3}$ of Example \ref{example1}. }
\label{fig:SimExampleRicianR3}
\end{minipage}\hfill
\begin{minipage}{0.45\textwidth}
\centering
\includegraphics[width=3in,height=2.25in]{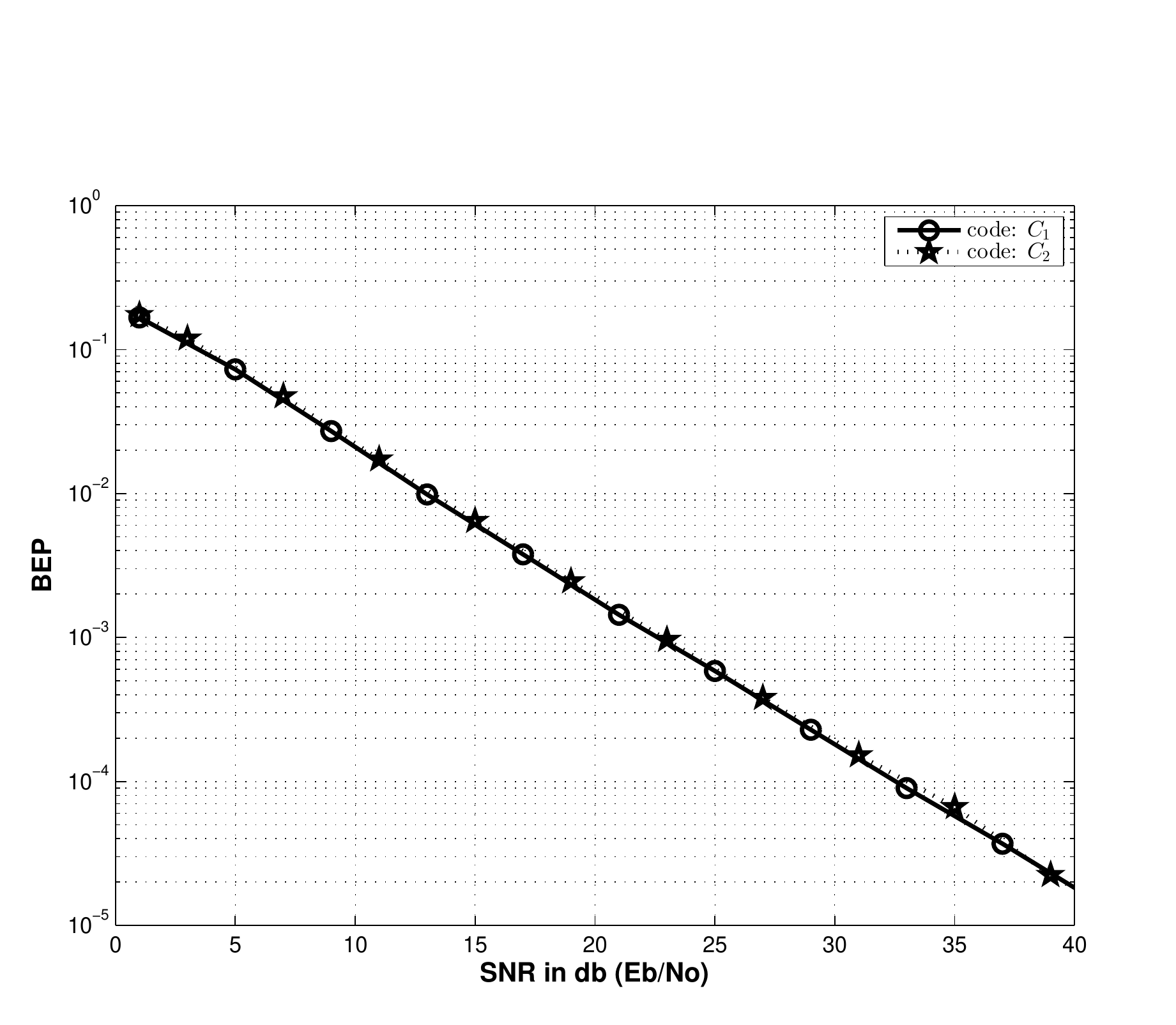}
\caption{\scriptsize SNR Vs BEP for codes $\mathfrak{C}_{1}$ and $\mathfrak{C}_{2}$ for Rician fading scenario, at receiver $R_{4}$ of Example \ref{example1}. }
\label{fig:SimExampleRicianR4}
\end{minipage}
\end{figure}
%
\begin{figure}
\centering
\begin{minipage}{0.45\textwidth}
\centering
\includegraphics[width=3in,height=2.25in]{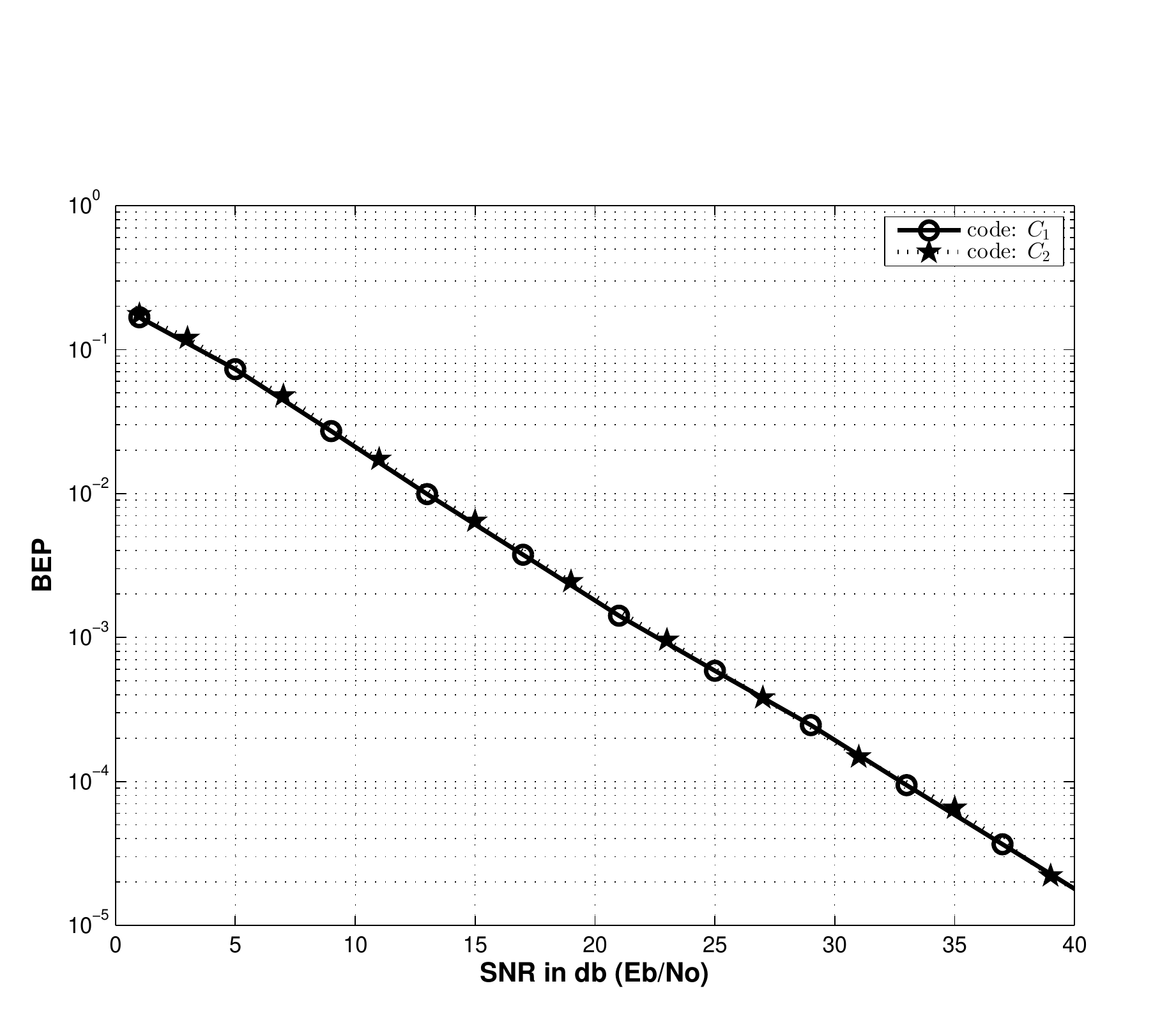}
\caption{\scriptsize SNR Vs BEP for codes $\mathfrak{C}_{1}$ and $\mathfrak{C}_{2}$ for Rician fading scenario, at receiver $R_{5}$ of Example \ref{example1}. }
\label{fig:SimExampleRicianR5}
\end{minipage}\hfill
\begin{minipage}{0.45\textwidth}
\centering
\includegraphics[width=3in,height=2.25in]{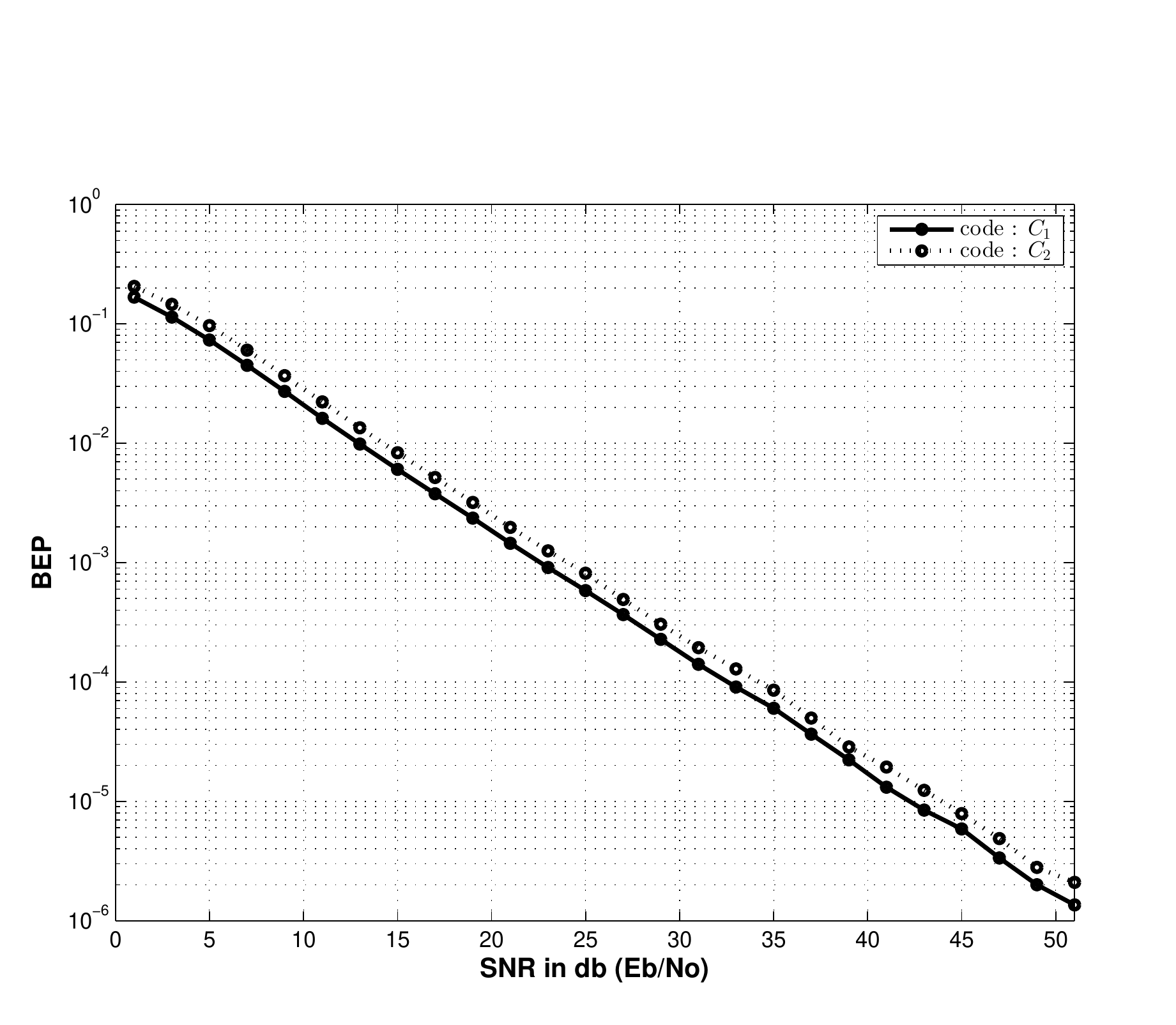}
\caption{\scriptsize SNR Vs BEP for codes $\mathfrak{C}_{1}$ and $\mathfrak{C}_{2}$ for Rician fading scenario, at receiver $R_{6}$ of Example \ref{example1}. }
\label{fig:SimExampleRicianR6}
\end{minipage}
\end{figure}
%


We also consider the scenario in which the source uses 8-PSK signal set for transmission. The mapping from bits to complex symbol is assumed to be Gray Mapping. Rayleigh fading scenario is considered. The SNR Vs. BEP curves for all receivers while using code $\mathfrak{C}_{1}$ and $\mathfrak{C}_{2}$ are given in Fig. \ref{fig:SimExample1Rayleigh8PSK_C1} and Fig. \ref{fig:SimExample1Rayleigh8PSK_C2} respectively. From Fig. \ref{fig:SimExample1Rayleigh8PSK_C1} and Fig. \ref{fig:SimExample1Rayleigh8PSK_C1}, we observe that the maximum error probability across different receivers occurs at receiver $R_{1}$ for code $\mathfrak{C}_{1}$ and at receiver $R_{2}$ for code $\mathfrak{C}_{2}$. Note that while using 8-PSK signal set the error probabilities of the transmissions depends on the mapping. The SNR Vs. BEP curves at receiver $R_{1}$ for code $\mathfrak{C}_{1}$ and at receiver $R_{2}$ for code $\mathfrak{C}_{2}$ is plotted in Fig. \ref{fig:SimExample1Rayleigh8PSKMax}. From Fig. \ref{fig:SimExample1Rayleigh8PSKMax}, it is evident that the maximum probability of error is less for $\mathfrak{C}_{1}$.  Thus code $\mathfrak{C}_{1}$ performs better than code $\mathfrak{C}_{2}$ in terms of reducing the maximum probability of error.

\begin{figure*}
\centering{}
\includegraphics[width=15cm,height=9cm]{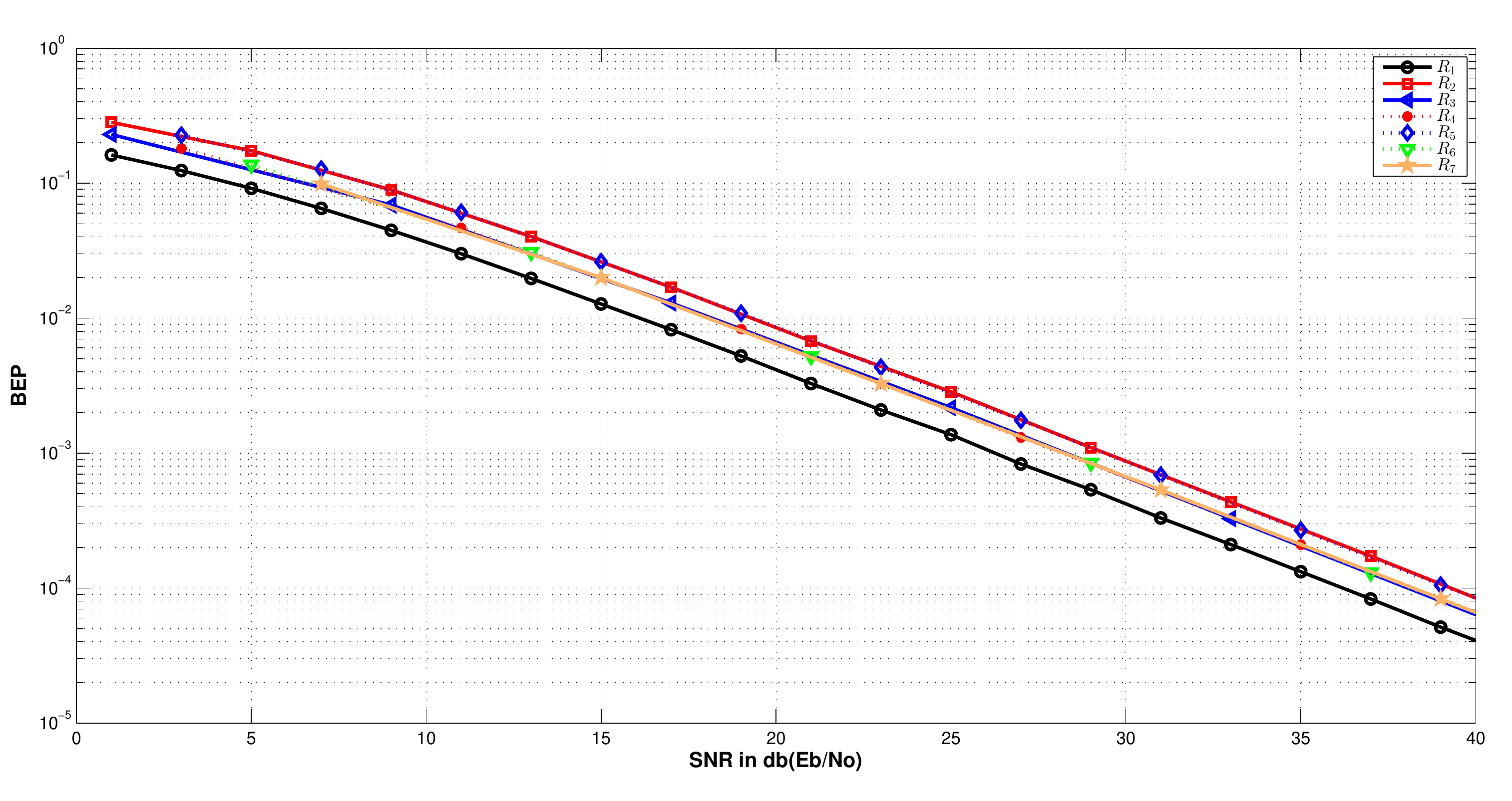}
\caption{\small SNR Vs BEP for code $\mathfrak{C}_{1}$ for Rayleigh fading scenario using 8-PSK modulation, at all receivers of Example \ref{example1}. }
\label{fig:SimExample1Rayleigh8PSK_C1}
\end{figure*} 

\begin{figure*}
\centering{}
\includegraphics[width=15cm,height=9cm]{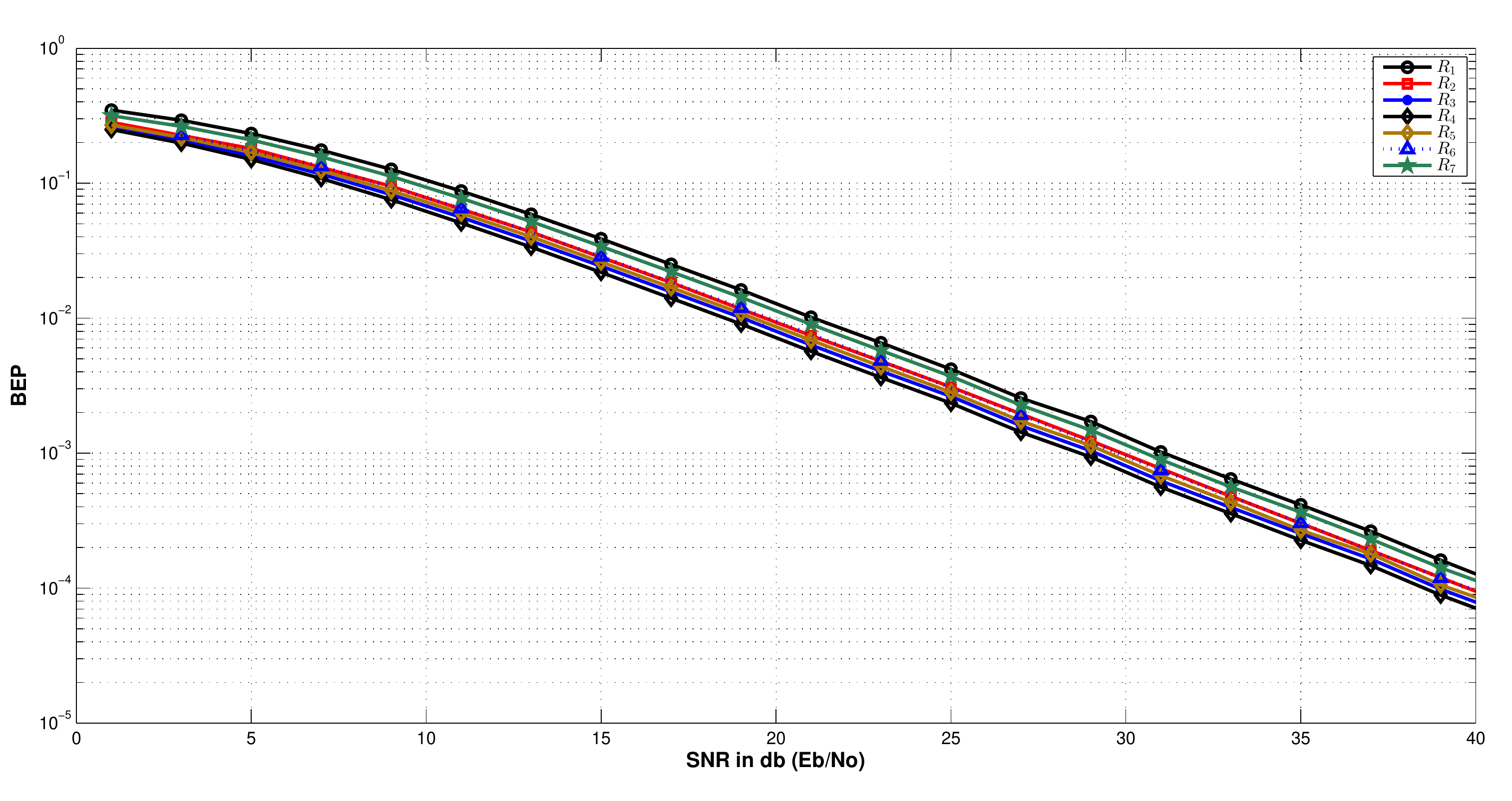}
\caption{\small SNR Vs BEP for code $\mathfrak{C}_{2}$ for Rayleigh fading scenario using 8-PSK modulation, at all receivers of Example \ref{example1}. }
\label{fig:SimExample1Rayleigh8PSK_C2}
\end{figure*} 

\begin{figure}
\centering{}
\includegraphics[scale=0.5]{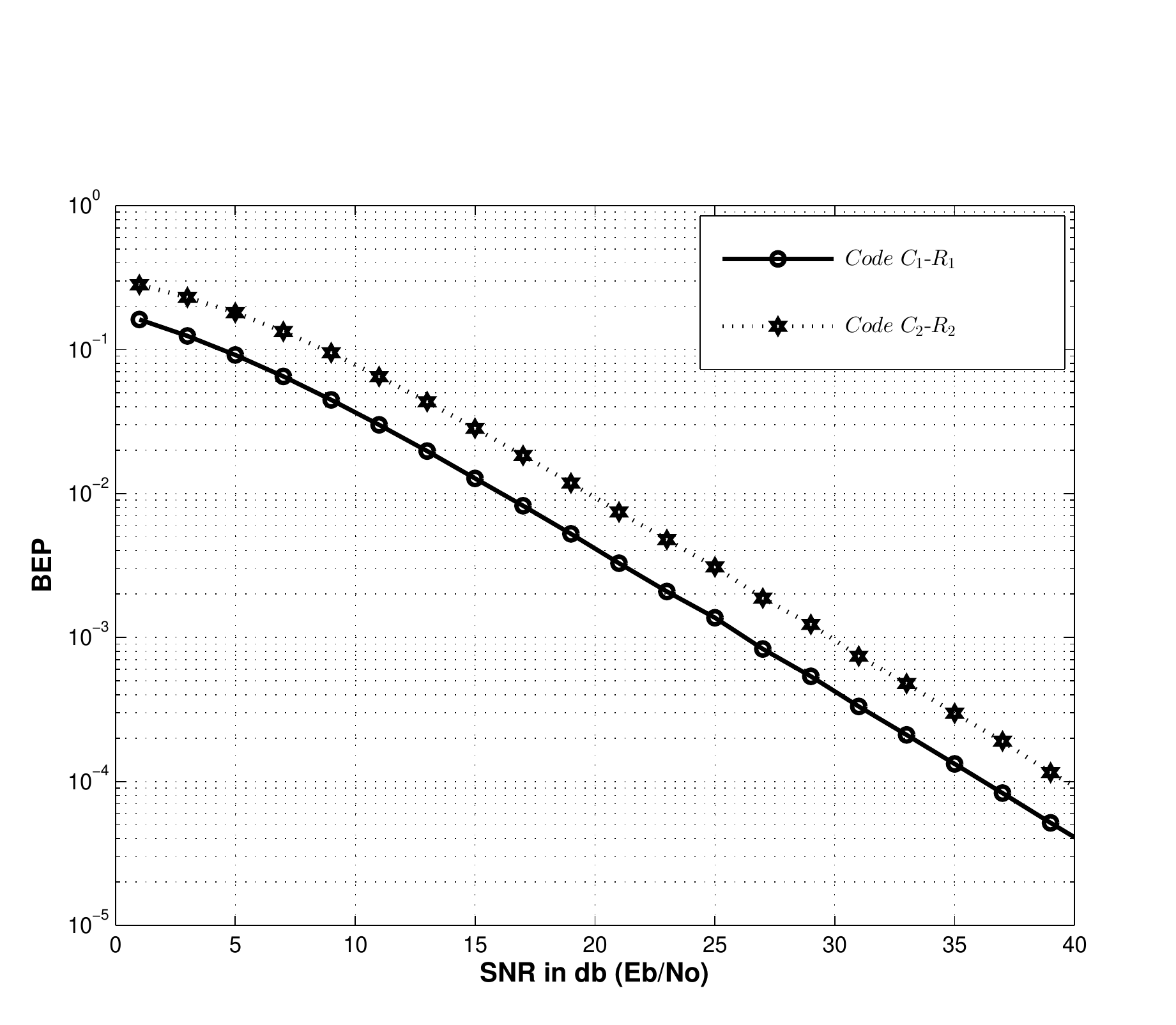}
\caption{\small SNR Vs BEP for Rayleigh fading scenario using 8PSK signal set at receiver $R_{1}$ for code $\mathfrak{C}_{1}$ and at receiver $R_{2}$ for code $\mathfrak{C}_{2}$  of Example \ref{example1}.}
\label{fig:SimExample1Rayleigh8PSKMax}
\end{figure} 

\end{example} 

\begin{example}
\label{eg:simulation7receiversF3}

Consider the index coding problem $\mathcal{I}(X,\mathcal{R})$ of Example \ref{example1}, with $x_{i} \in \mathbb{F}_{3}$. We consider two index codes for the problem and the simulation results are given below. Consider two index codes $\mathfrak{C}_{1}$ and $\mathfrak{C}_{2}$ described by the matrices $L_{1}$ and $L_{2}$ of Example \ref{example1}. However note that the matrices are over $\mathbb{F}_{3}$. In the simulation source uses ternary PSK. Both Rayleigh and Rician fading scenario were considered. In both the cases maximum probability of error occured at reciver $R_{7}$. The SNR Vs. BEP curve at reciever $R_{7}$ for both codes $\mathfrak{C}_{1}$ and $\mathfrak{C}_{2}$ and for both Rayleigh and Rician are given in Fig. \ref{fig:SimExampleF3U7Max}. From Fig. \ref{fig:SimExampleF3U7Max}, we can observe that code $\mathfrak{C}_{1}$ performs better than code $\mathfrak{C}_{2}$ in terms of maximum probability of error. 

\begin{figure}
\centering{}
\includegraphics[scale=0.5]{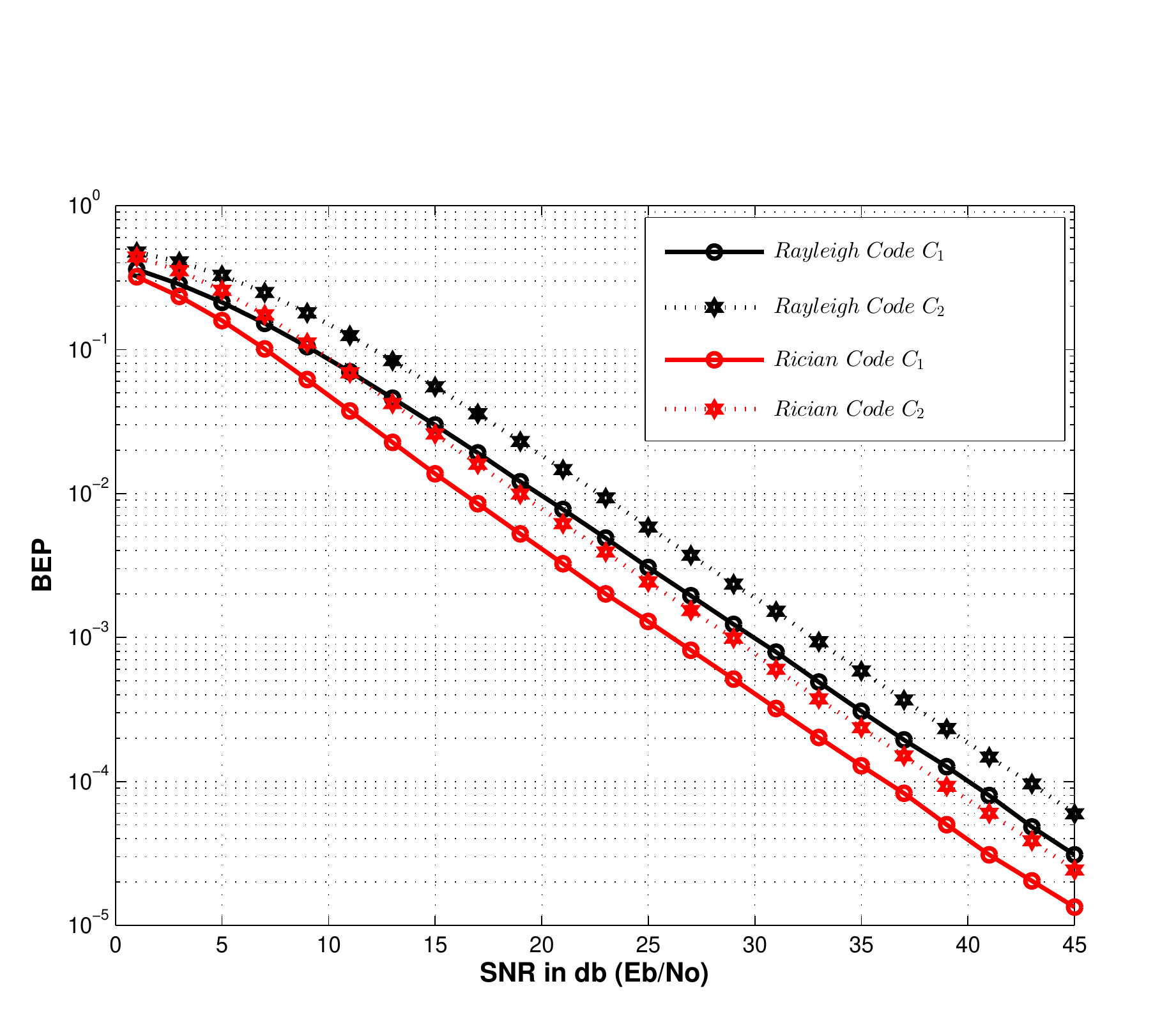}
\caption{\small SNR Vs BEP for both Rayleigh and Rician fading scenario using ternary PSK signal set at receiver $R_{7}$ for both code $\mathfrak{C}_{1}$ and code $\mathfrak{C}_{2}$  of Example \ref{example1}.}
\label{fig:SimExampleF3U7Max}
\end{figure} 

\end{example}

\begin{example}
\label{eg:simulation9receivers}
In this example we consider the index coding problem in Example \ref{eg:NoofTransmissions}. We compare the performance of codes $\mathfrak{C}_{1}$ and $\mathfrak{C}_{2}$ of Example \ref{eg:NoofTransmissions}. The matrices describing code $\mathfrak{C}_{1}$ and code $\mathfrak{C}_{2}$ are

\footnotesize{
\[
L_{1} = \left[\begin{array}{ccccccccc}
1 & 1 & 1 & 1 & 1 & 1 & 1 & 1\\
1 & 0 & 0 & 0 & 0 & 0 & 0 & 0\\
0 & 1 & 0 & 0 & 0 & 0 & 0 & 0\\
0 & 0 & 1 & 0 & 0 & 0 & 0 & 0\\
0 & 0 & 0 & 1 & 0 & 0 & 0 & 0\\
0 & 0 & 0 & 0 & 1 & 0 & 0 & 0\\
0 & 0 & 0 & 0 & 0 & 1 & 0 & 0\\
0 & 0 & 0 & 0 & 0 & 0 & 1 & 0\\
0 & 0 & 0 & 0 & 0 & 0 & 0 & 1\\
\end{array}\right]L_{2} = \left[\begin{array}{ccccccccc}
1 & 1 & 0 & 1 & 0 & 0 & 0 & 0\\
0 & 0 & 0 & 0 & 1 & 1 & 0 & 0\\
0 & 0 & 0 & 1 & 1 & 0 & 0 & 0\\
0 & 0 & 1 & 0 & 0 & 0 & 0 & 0\\
0 & 0 & 0 & 0 & 0 & 0 & 0 & 1\\
0 & 1 & 1 & 0 & 0 & 0 & 0 & 0\\
0 & 0 & 0 & 0 & 0 & 0 & 1 & 1\\
1 & 0 & 0 & 0 & 0 & 0 & 0 & 0\\
0 & 0 & 0 & 0 & 0 & 1 & 1 & 0\\
\end{array}\right]
\]
}
\normalsize 
respectively.
The source uses symmetric $4$-PSK signal set for transmission. The mapping used from bits to complex symbols is Gray mapping. Rayleigh fading scenario is considered first in which the fading coefficient  $h_{j}$ of the channel between the source and receiver $R_{j}$ is $\mathcal{C}\mathcal{N}(0,1)$. The simulation curves showing, SNR Vs. BEP for all the receivers while using code $\mathfrak{C}_{1}$ is given in Fig \ref{fig:SimExample2RayleighAll1}. From Fig. \ref{fig:SimExample2RayleighAll1}, we can observe that maximum error probability occurs at all receivers except $R_{1},R_{2}$ and $R_{8}$. The SNR Vs. BEP curves for all the receivers while using code $\mathfrak{C}_{2}$ is given in Fig. \ref{fig:SimExample2RayleighAll2}. From Fig \ref{fig:SimExample2RayleighAll2} we can observe that maximum error probability of error occurs at receiver $R_{3}$. In Fig. \ref{fig:SimExample2RayleighR3} we compare these maximum error probabilities by showing the SNR Vs. BEP curves for both the codes at receiver $R_{3}$. From Fig. \ref{fig:SimExample2RayleighR3} we are able to observe a gain of 2dB at Receiver $R_3$ by using code $\mathfrak{C}_{1}$ over code $\mathfrak{C}_{2}$.

\begin{figure*}
\centering{}
\includegraphics[width=15cm,height=9cm]{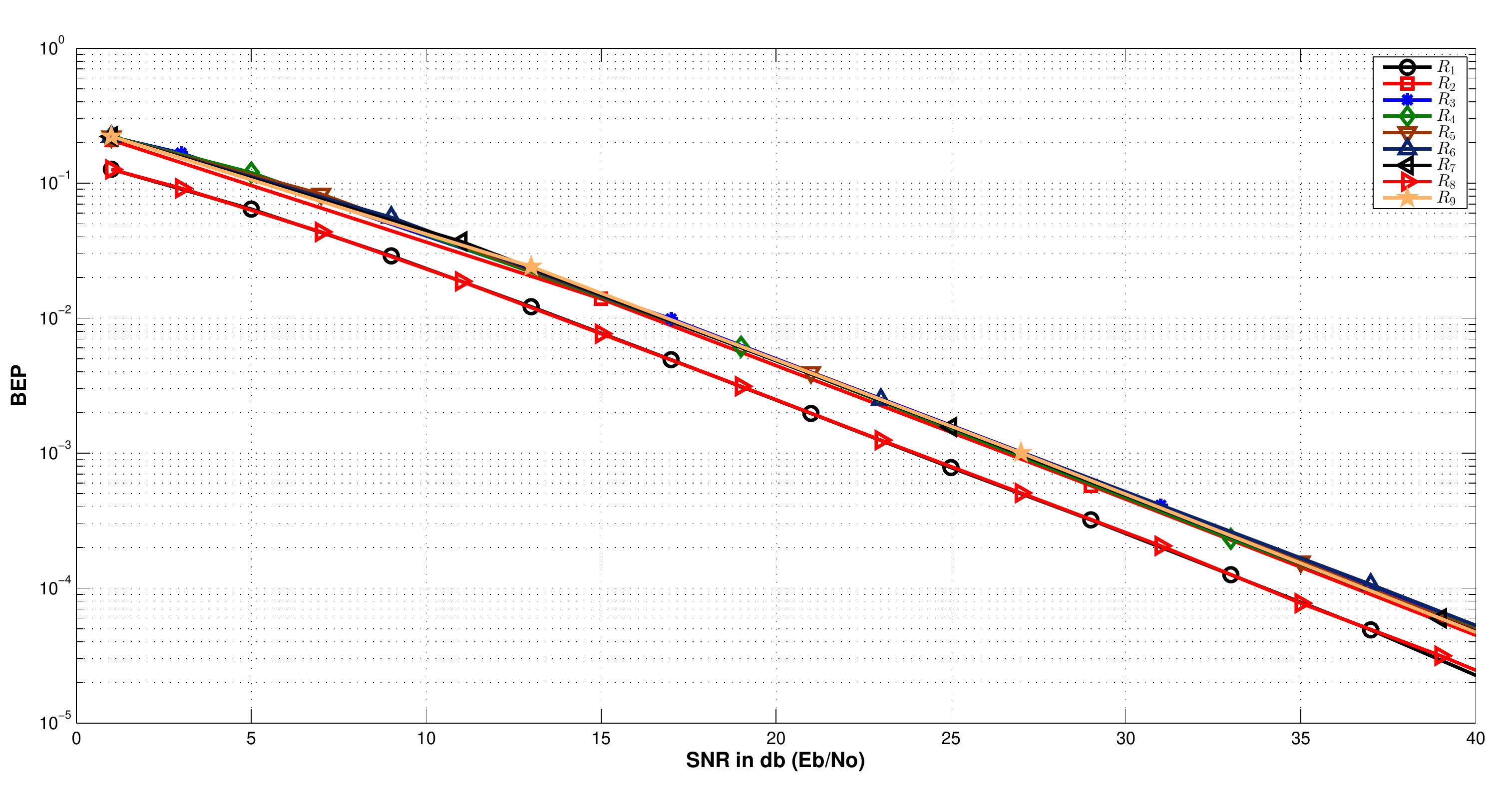}
\caption{\small SNR Vs BEP for code $\mathfrak{C}_{1}$ for Rayleigh fading scenario, at all receivers of Example \ref{eg:simulation9receivers}. }
\label{fig:SimExample2RayleighAll1}
\end{figure*}
\begin{figure*}
\centering{}
\includegraphics[width=15cm,height=9cm]{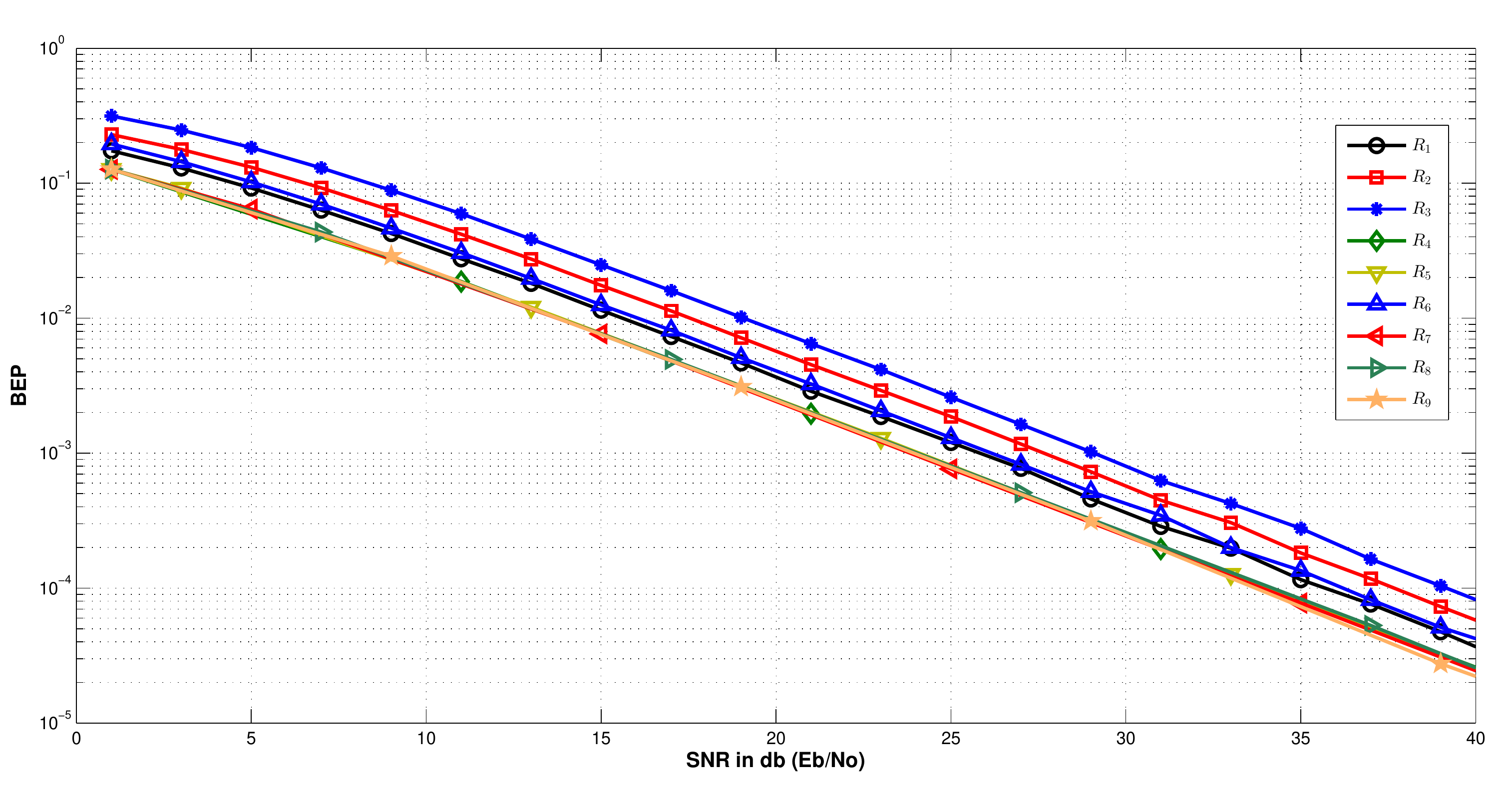}
\caption{\small SNR Vs BEP for code $\mathfrak{C}_{2}$ for Rayleigh fading scenario, at all receivers of Example \ref{eg:simulation9receivers}. }
\label{fig:SimExample2RayleighAll2}
\end{figure*}
\begin{figure}
\centering{}
\includegraphics[width=3.in,height=2.25in]{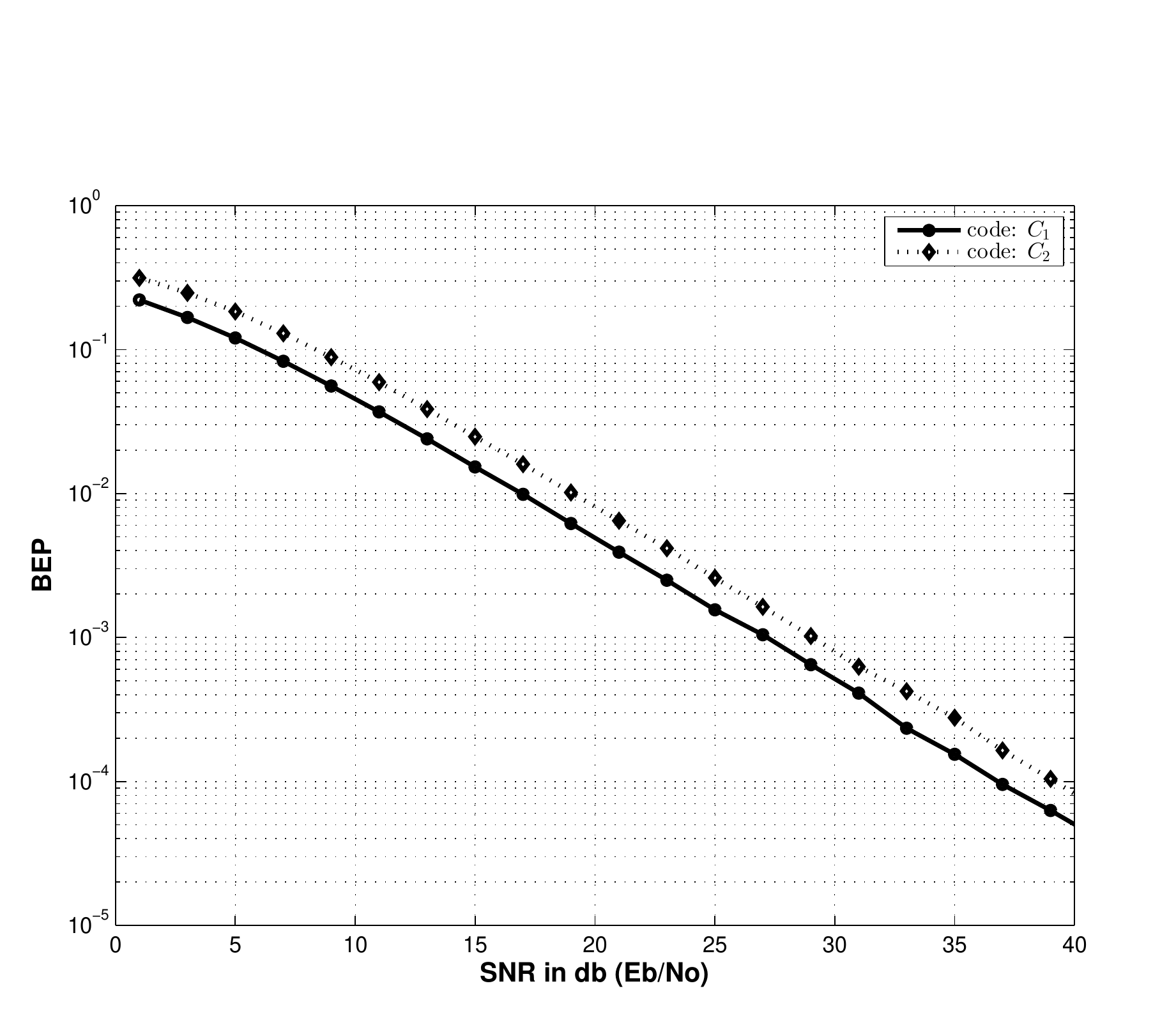}
\caption{\scriptsize SNR Vs BEP for codes $\mathfrak{C}_{1}$ and $\mathfrak{C}_{2}$ for Rayleigh fading scenario, at receiver $R_{3}$ of Example \ref{eg:simulation9receivers}. }
\label{fig:SimExample2RayleighR3}
\end{figure} 
In Fig. \ref{fig:SimExample2RayleighR1} - Fig. \ref{fig:SimExample2RayleighR9}, SNR Vs. BEP plots for all receivers other than $R_3$ are given. We can observe from Fig. \ref{fig:SimExample2RayleighR1} and Fig. \ref{fig:SimExample2RayleighR2} that code $\mathfrak{C}_{1}$ performs better than code $\mathfrak{C}_{2}$ at receivers $R_{1}$ and $R_{2}$ also. However for receiver $R_{4}$, code $\mathfrak{C}_{2}$ performs better than code $\mathfrak{C}_{1}$. The reason is that the number of transmissions used by receiver $R_{4}$ in decoding its demand is more for code $\mathfrak{C}_{1}$ than code $\mathfrak{C}_{2}$. The SNR Vs. BEP for the two codes for receiver $R_{4}$ is given in Fig. \ref{fig:SimExample2RayleighR4}. Note that the index code given by proposed Algorithm \ref{alg:optimalic}, does not guarantee better performance at all receivers. The algorithm ensures that the index code has minimum maximum error probability across all receivers.

\begin{figure}
\centering
\begin{minipage}{0.45\textwidth}
\centering
\includegraphics[width=3in,height=2in]{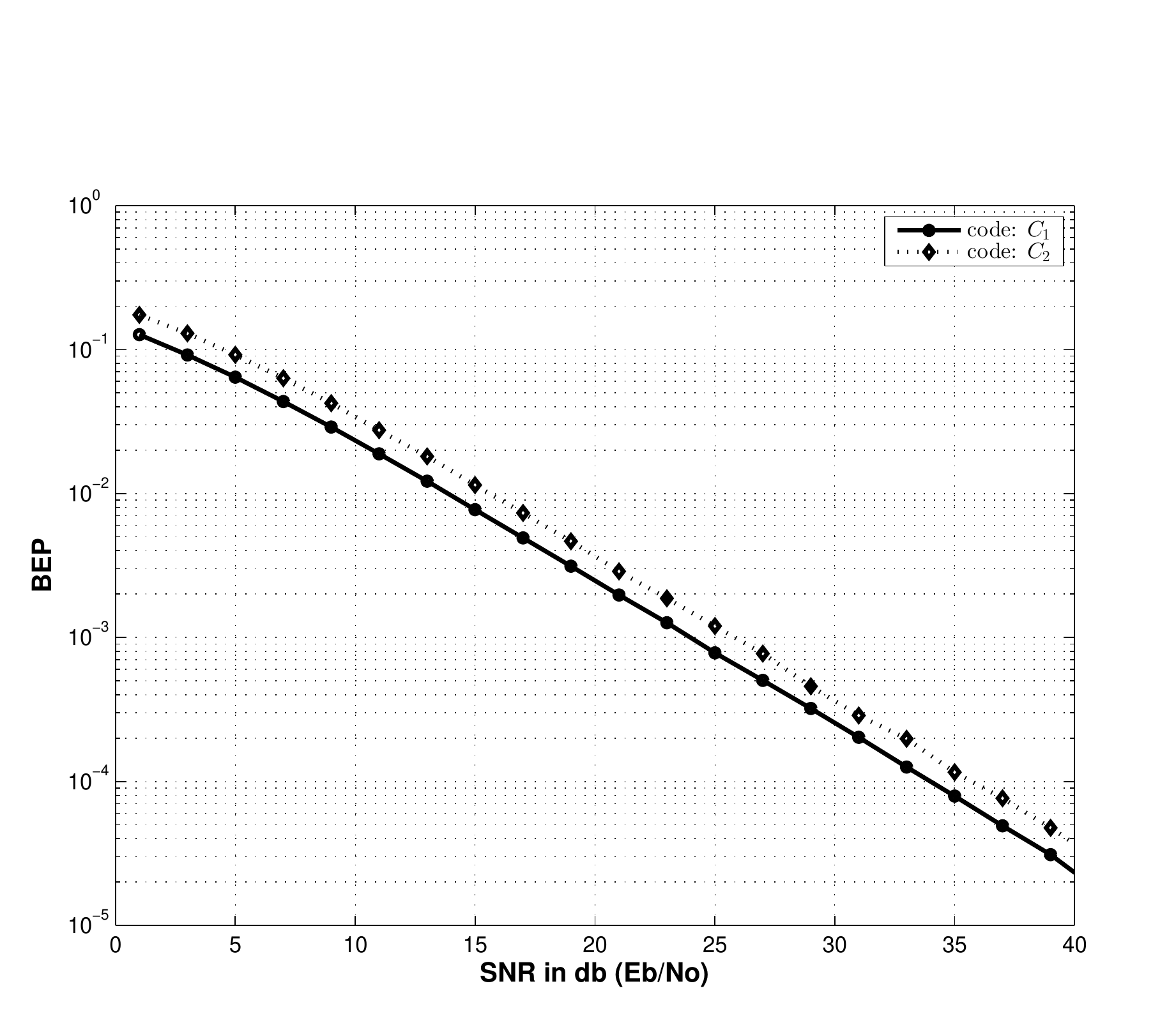}
\caption{\scriptsize SNR Vs BEP for codes $\mathfrak{C}_{1}$ and $\mathfrak{C}_{2}$ for Rayleigh fading scenario, at receiver $R_{1}$ of Example \ref{eg:simulation9receivers}. }
\label{fig:SimExample2RayleighR1}
\end{minipage}\hfill
\begin{minipage}{0.45\textwidth}
\centering
\includegraphics[width=3in,height=2in]{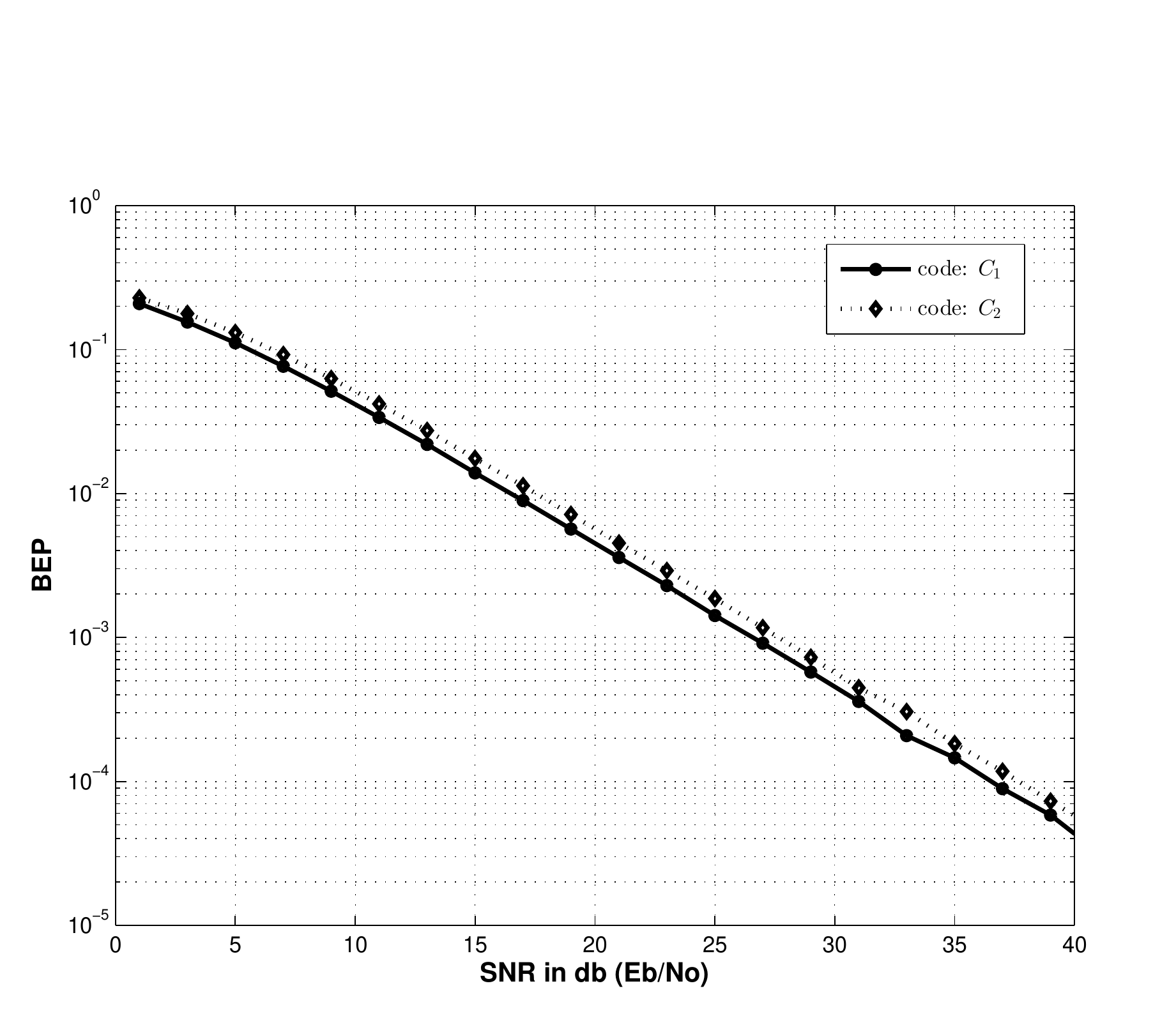}
\caption{\scriptsize SNR Vs BEP for codes $\mathfrak{C}_{1}$ and $\mathfrak{C}_{2}$ for Rayleigh fading scenario, at receiver $R_{2}$ of Example \ref{eg:simulation9receivers}. }
\label{fig:SimExample2RayleighR2}
\end{minipage}
\end{figure}
\begin{figure}
\centering
\begin{minipage}{0.45\textwidth}
\centering
\includegraphics[width=3in,height=2in]{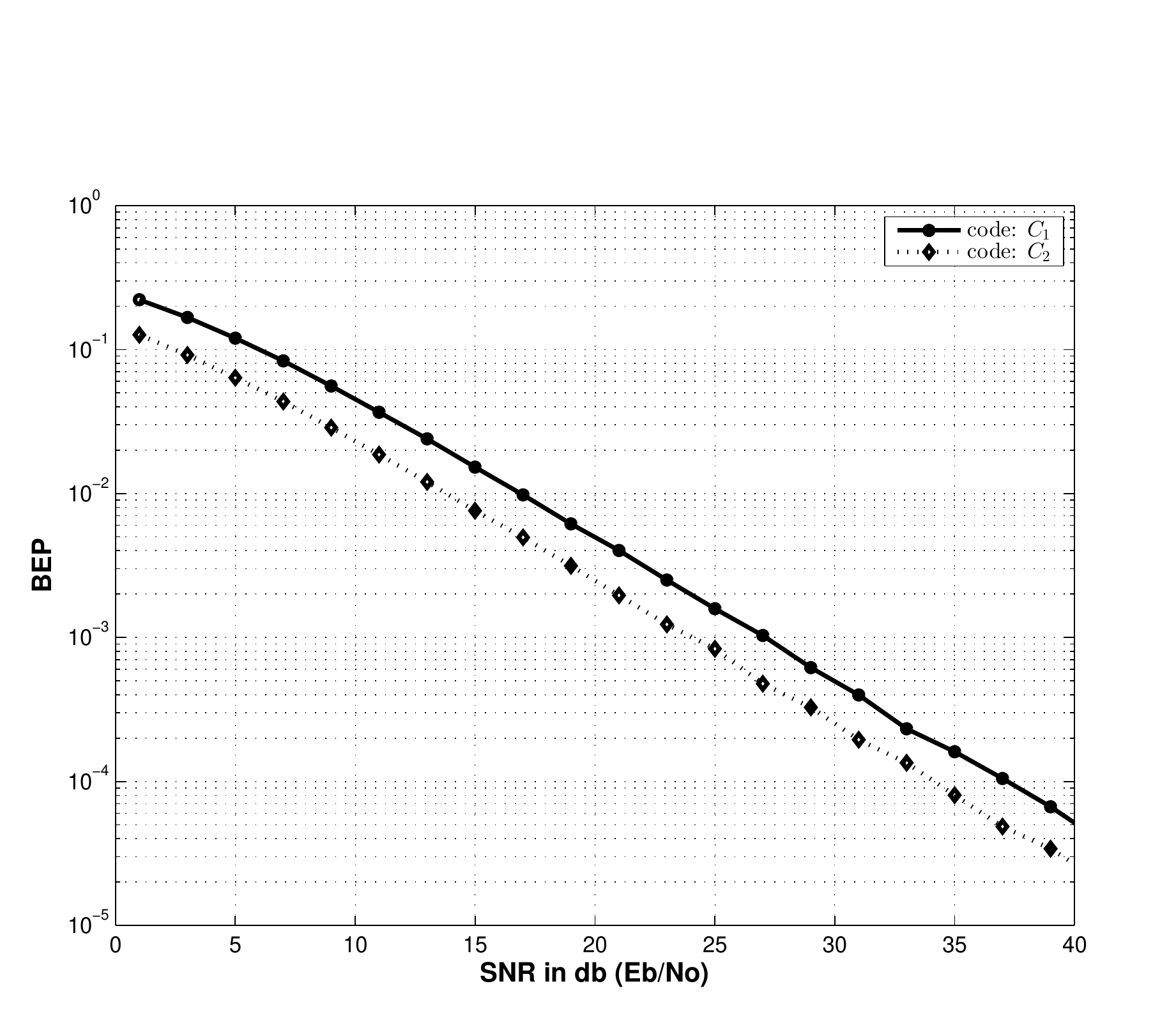}
\caption{\scriptsize SNR Vs BEP for codes $\mathfrak{C}_{1}$ and $\mathfrak{C}_{2}$ for Rayleigh fading scenario, at receiver $R_{4}$ of Example \ref{eg:simulation9receivers}. }
\label{fig:SimExample2RayleighR4}
\end{minipage}\hfill
\begin{minipage}{0.45\textwidth}
\centering
\includegraphics[width=3in,height=2in]{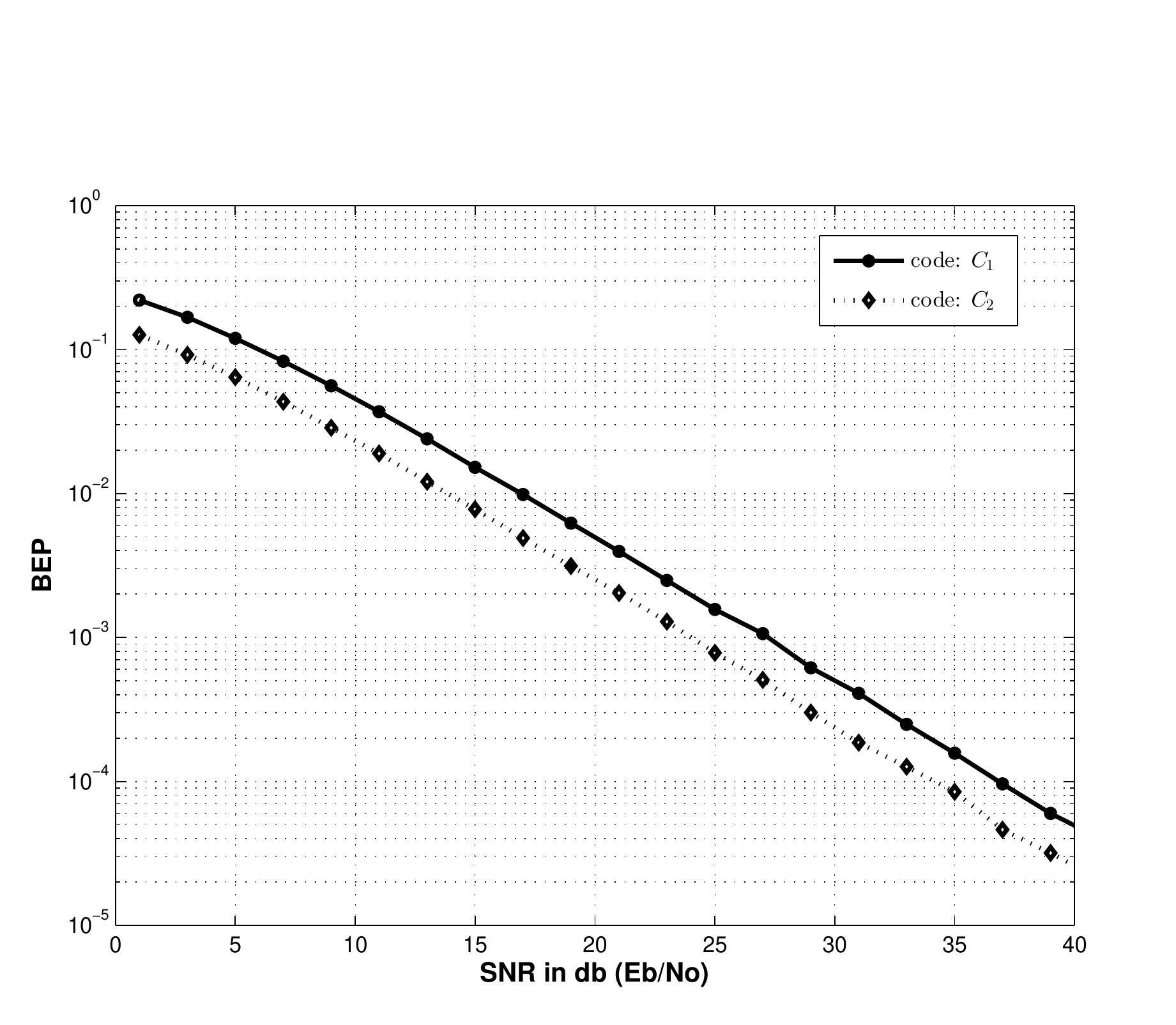}
\caption{\scriptsize SNR Vs BEP for codes $\mathfrak{C}_{1}$ and $\mathfrak{C}_{2}$ for Rayleigh fading scenario, at receiver $R_{5}$ of Example \ref{eg:simulation9receivers}. }
\label{fig:SimExample2RayleighR5}
\end{minipage}
\end{figure}
\begin{figure}
\centering
\begin{minipage}{0.45\textwidth}
\centering
\includegraphics[width=3in,height=2in]{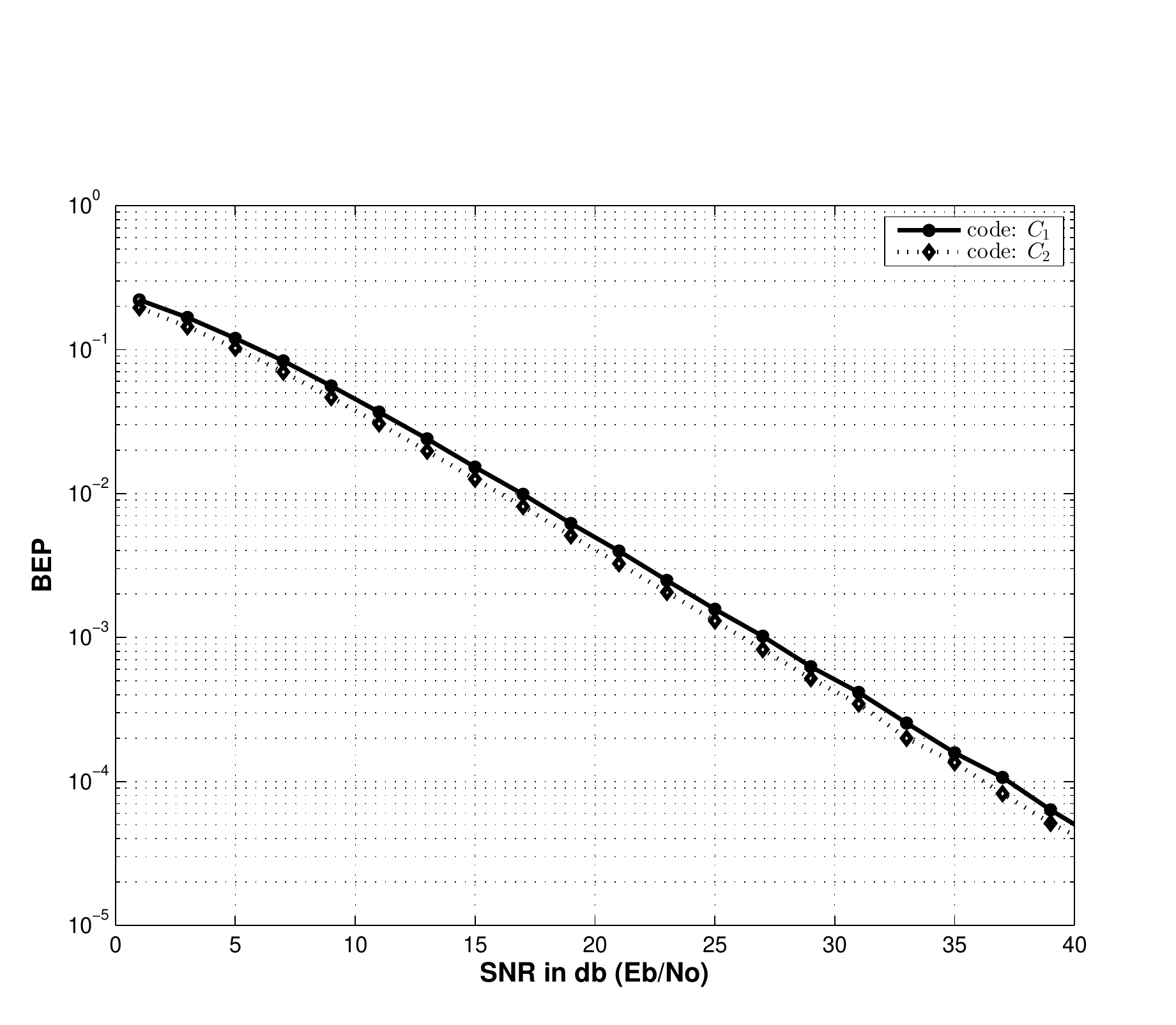}
\caption{\scriptsize SNR Vs BEP for codes $\mathfrak{C}_{1}$ and $\mathfrak{C}_{2}$ for Rayleigh fading scenario, at receiver $R_{6}$ of Example \ref{eg:simulation9receivers}. }
\label{fig:SimExample2RayleighR6}
\end{minipage}\hfill
\begin{minipage}{0.45\textwidth}
\centering
\includegraphics[width=3in,height=2in]{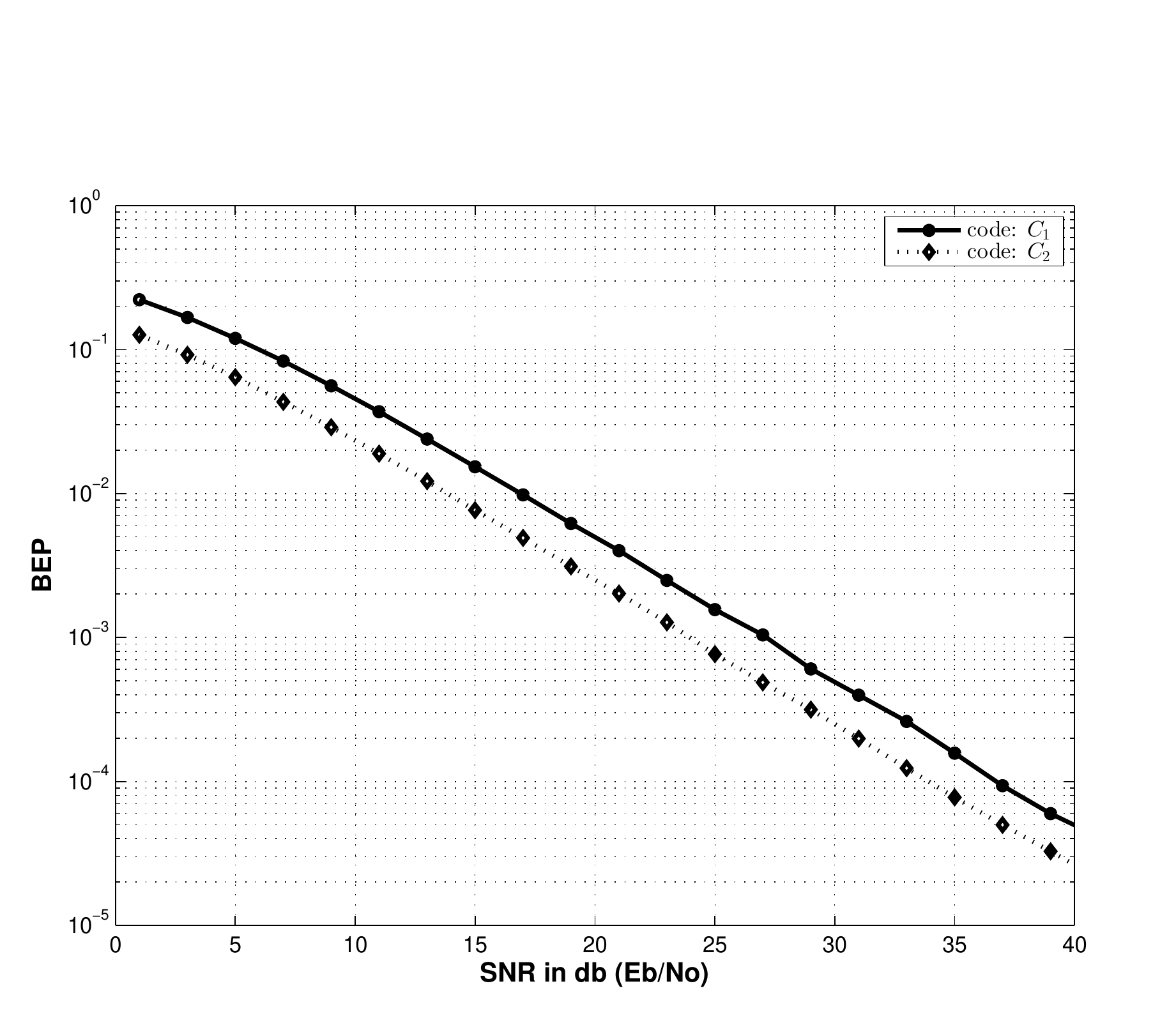}
\caption{\scriptsize SNR Vs BEP for codes $\mathfrak{C}_{1}$ and $\mathfrak{C}_{2}$ for Rayleigh fading scenario, at receiver $R_{7}$ of Example \ref{eg:simulation9receivers}. }
\label{fig:SimExample2RayleighR7}
\end{minipage}
\end{figure}

\begin{figure}
\centering
\begin{minipage}{0.45\textwidth}
\centering
\includegraphics[width=3in,height=2in]{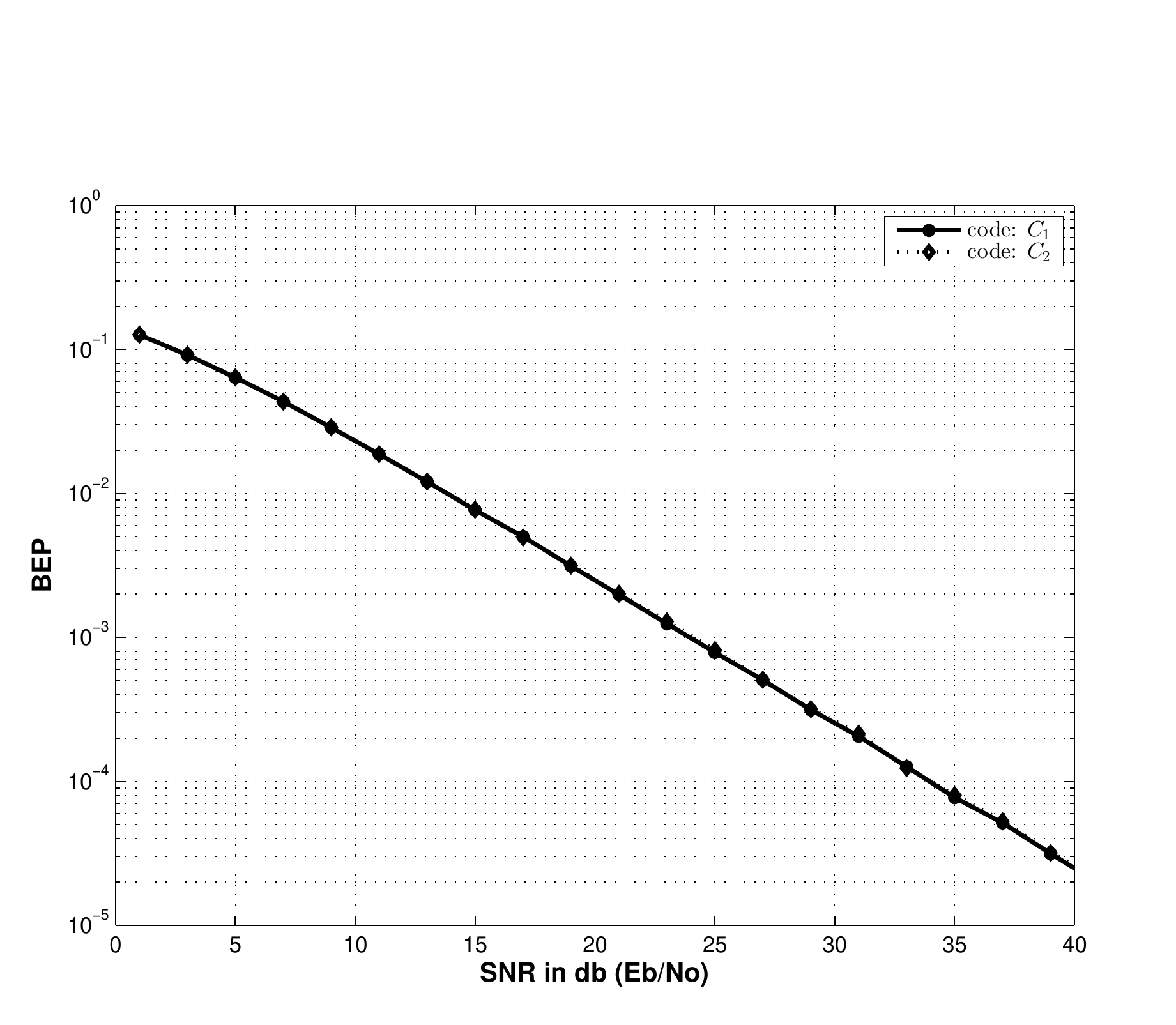}
\caption{\scriptsize SNR Vs BEP for codes $\mathfrak{C}_{1}$ and $\mathfrak{C}_{2}$ for Rayleigh fading scenario, at receiver $R_{8}$ of Example \ref{eg:simulation9receivers}. }
\label{fig:SimExample2RayleighR8}
\end{minipage}\hfill
\begin{minipage}{0.45\textwidth}
\centering
\includegraphics[width=3in,height=2in]{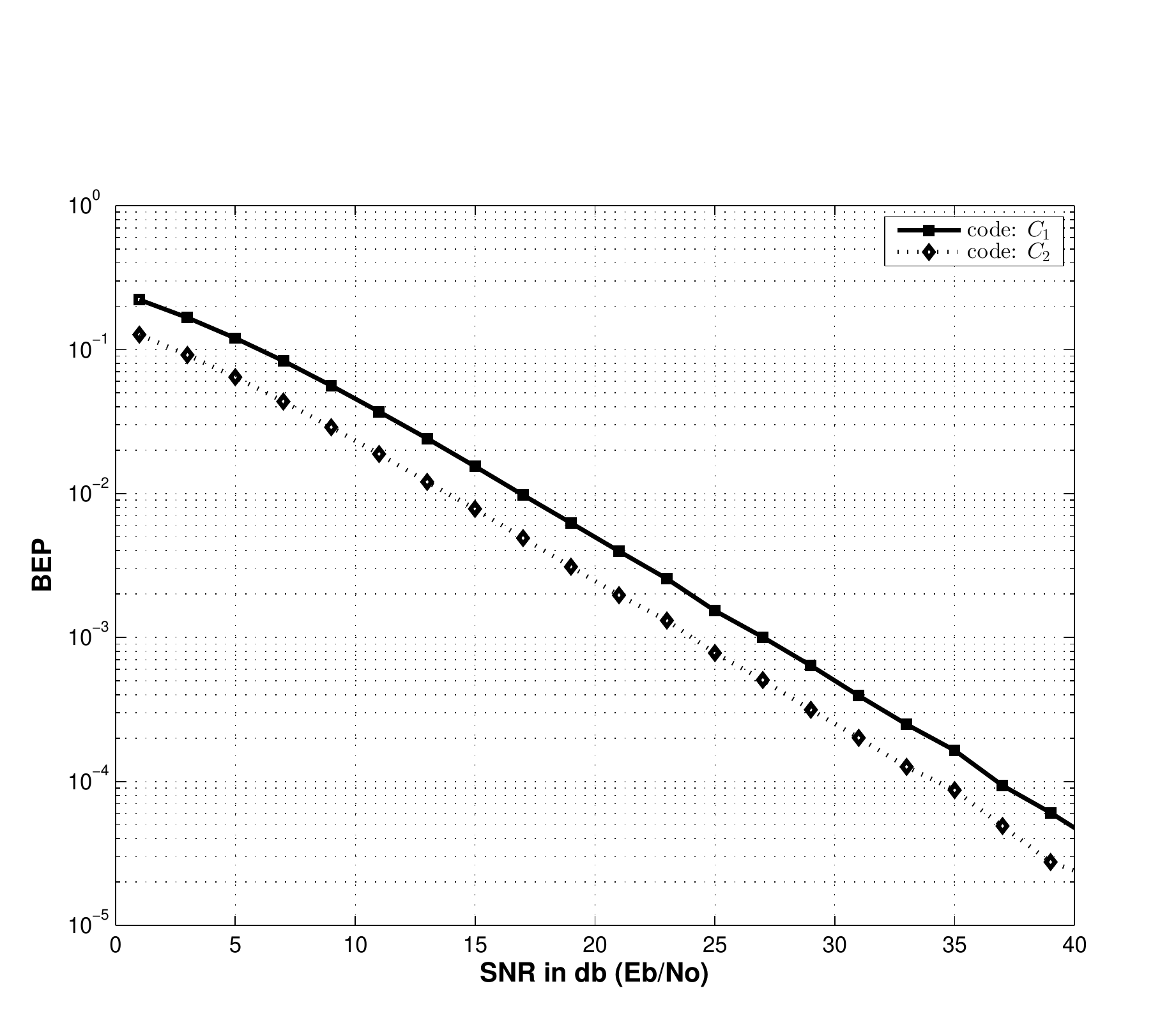}
\caption{\scriptsize SNR Vs BEP for codes $\mathfrak{C}_{1}$ and $\mathfrak{C}_{2}$ for Rayleigh fading scenario, at receiver $R_{9}$ of Example \ref{eg:simulation9receivers}. }
\label{fig:SimExample2RayleighR9}
\end{minipage}
\end{figure}


Simulations were also carried out with the channel between source and receiver $R_{j}$ modelled as a Rician fading channel. The fading coefficient $h_{j}$  is Rician with a Rician factor 2. The source uses $4$-PSK signal set along with Gray mapping. The SNR Vs. BEP curves for all receivers while using code $\mathfrak{C}_{1}$ and code $\mathfrak{C}_{2}$ is given in Fig. \ref{fig:SimExample2RicianAll1} and Fig. \ref{fig:SimExample2RicianAll2} respectively. Similar to the Rayleigh fading scenario maximum error probability was observed at receiver $R_{3}$. The SNR Vs. BEP curves for both the codes at receiver $R_{3}$ are given in Fig. \ref{fig:SimExample2RicianR3}. We can observe from Fig. \ref{fig:SimExample2RicianR3} that for the Rician fading scenario also, maximum error probability is less for code $\mathfrak{C}_{1}$. The SNR Vs. BEP plots for both the codes at receivers other than $R_{3}$ are given in Fig. \ref{fig:SimExample2RicianR1}-Fig. \ref{fig:SimExample2RicianR9}. For Rician fading also we infer the same results from the plots.  Code $\mathfrak{C}_{2}$ performs better than code $\mathfrak{C}_{1}$ for few receivers where the number of transmissions used for decoding its demand is less, but in terms of minimizing maximum error probability across all receivers code $\mathfrak{C}_{1}$ performs better.

\begin{figure}
\centering{}
\includegraphics[width=15cm,height=9cm]{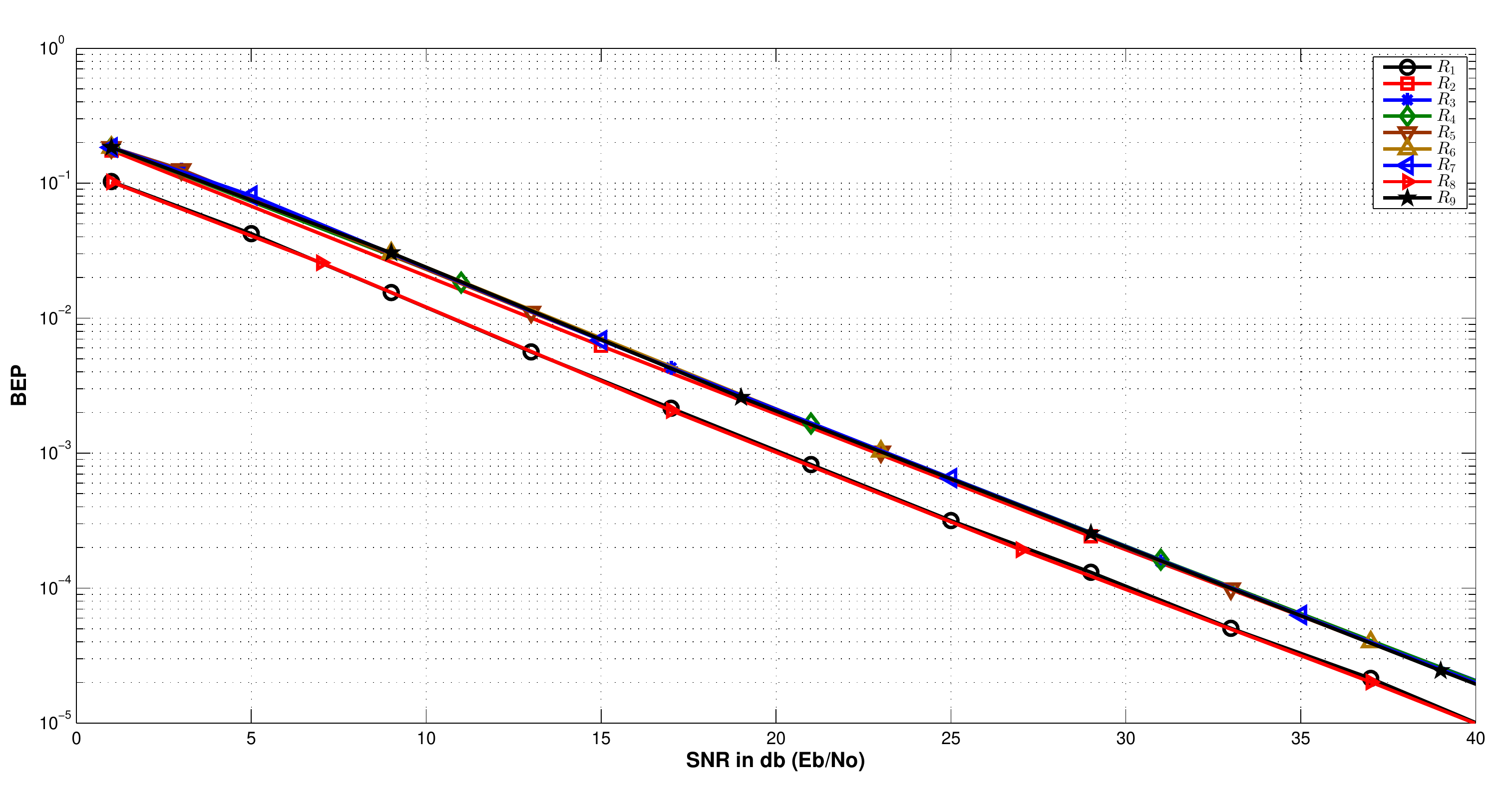}
\caption{\small  SNR Vs BEP for code $\mathfrak{C}_{1}$ for Rician fading scenario, at all receivers of Example \ref{eg:simulation9receivers}. }
\label{fig:SimExample2RicianAll1}
\end{figure}
\begin{figure}
\centering{}
\includegraphics[width=15cm,height=9cm]{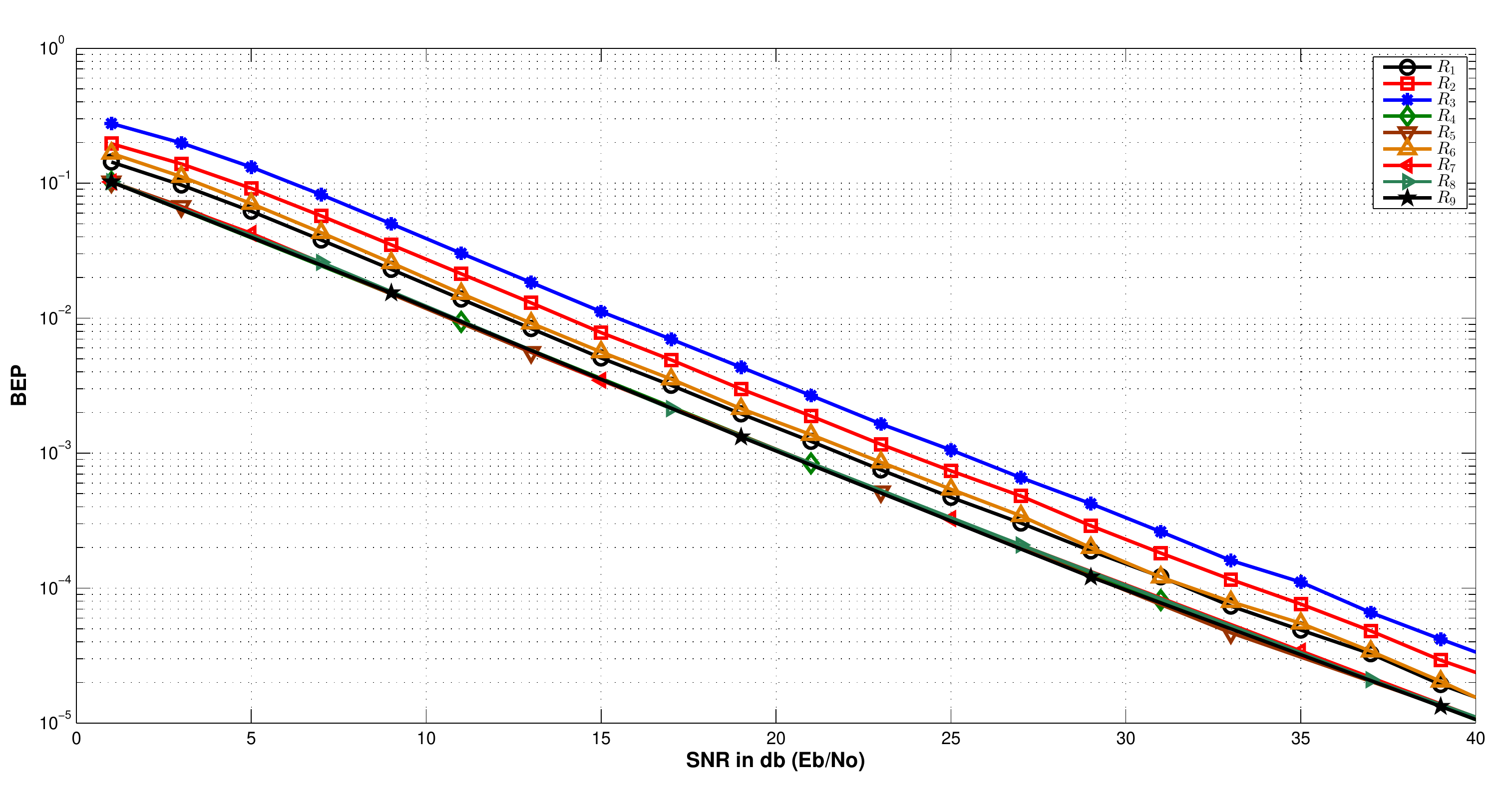}
\caption{\small  SNR Vs BEP for code $\mathfrak{C}_{2}$ for Rician fading scenario, at all receivers of Example \ref{eg:simulation9receivers}. }
\label{fig:SimExample2RicianAll2}
\end{figure}

\begin{figure}
\centering{}
\includegraphics[width=3in,height=2.25in]{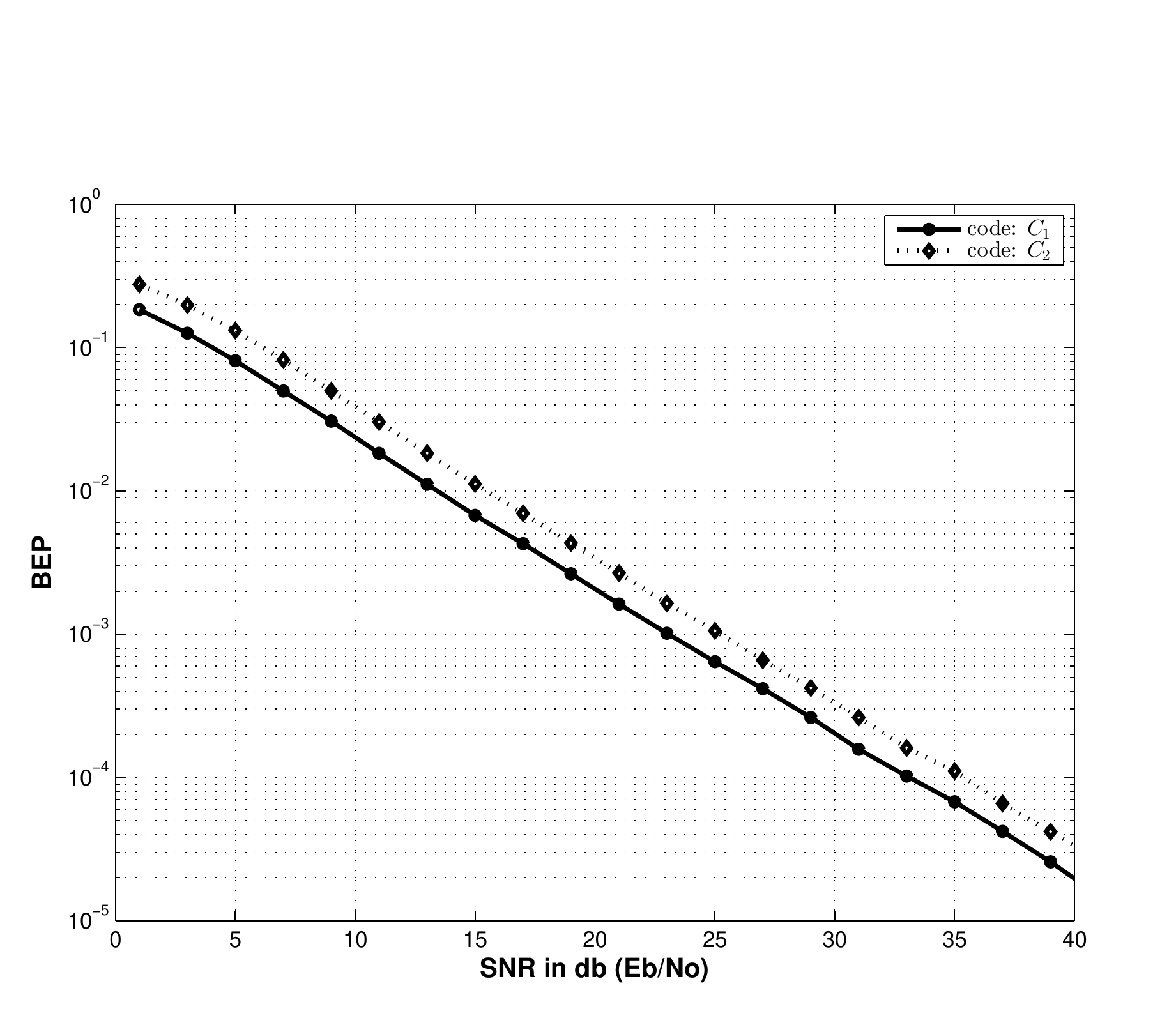}
\caption{\scriptsize SNR Vs BEP for codes $\mathfrak{C}_{1}$ and $\mathfrak{C}_{2}$ for Rician fading scenario, at receiver $R_{3}$ of Example \ref{eg:simulation9receivers}. }
\label{fig:SimExample2RicianR3}
\end{figure}

\begin{figure}
\centering
\begin{minipage}{0.45\textwidth}
\centering
\includegraphics[width=3in,height=2in]{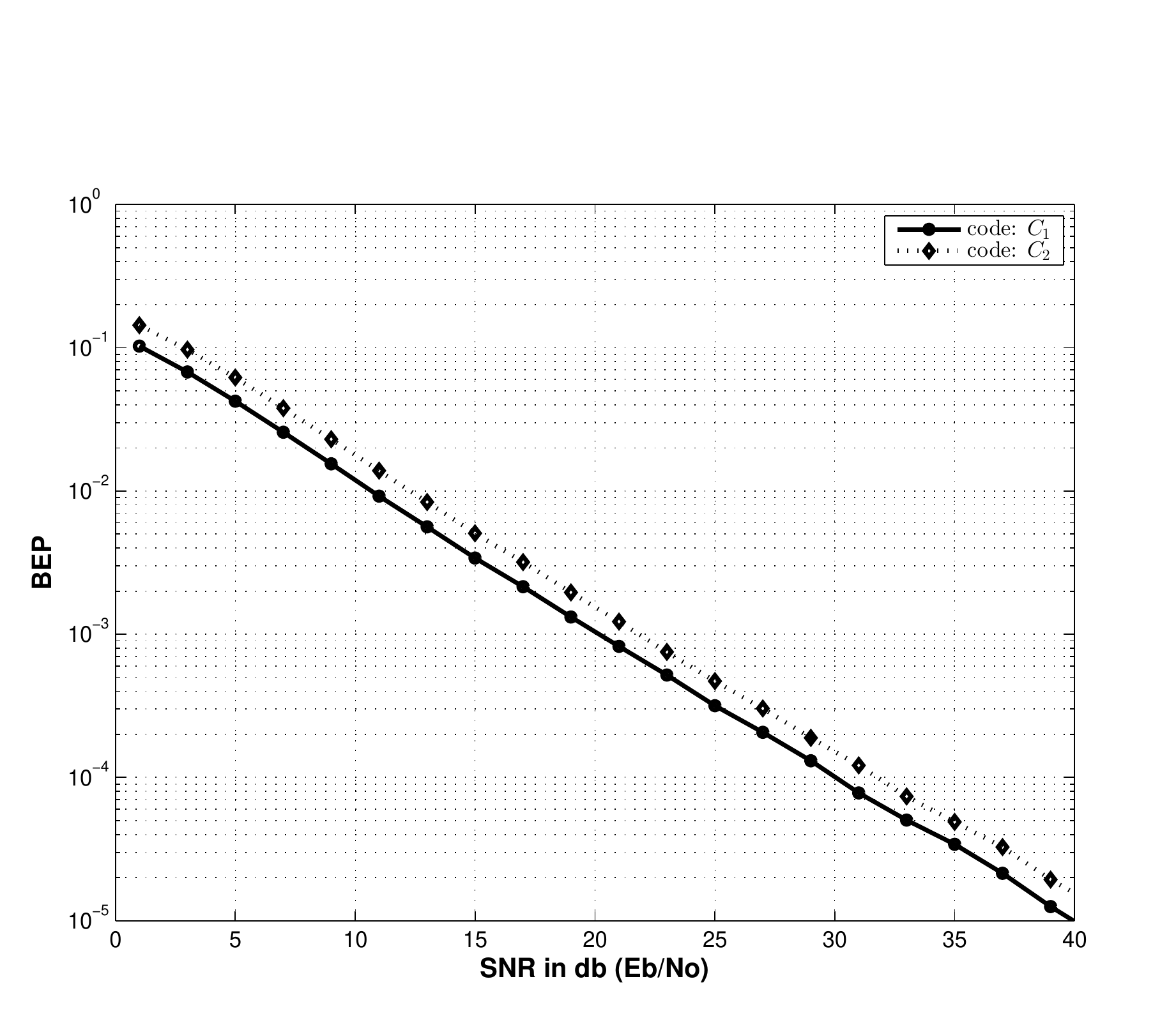}
\caption{\scriptsize SNR Vs BEP for codes $\mathfrak{C}_{1}$ and $\mathfrak{C}_{2}$ for Rician fading scenario, at receiver $R_{1}$ of Example \ref{eg:simulation9receivers}. }
\label{fig:SimExample2RicianR1}
\end{minipage}\hfill
\begin{minipage}{0.45\textwidth}
\centering
\includegraphics[width=3in,height=2in]{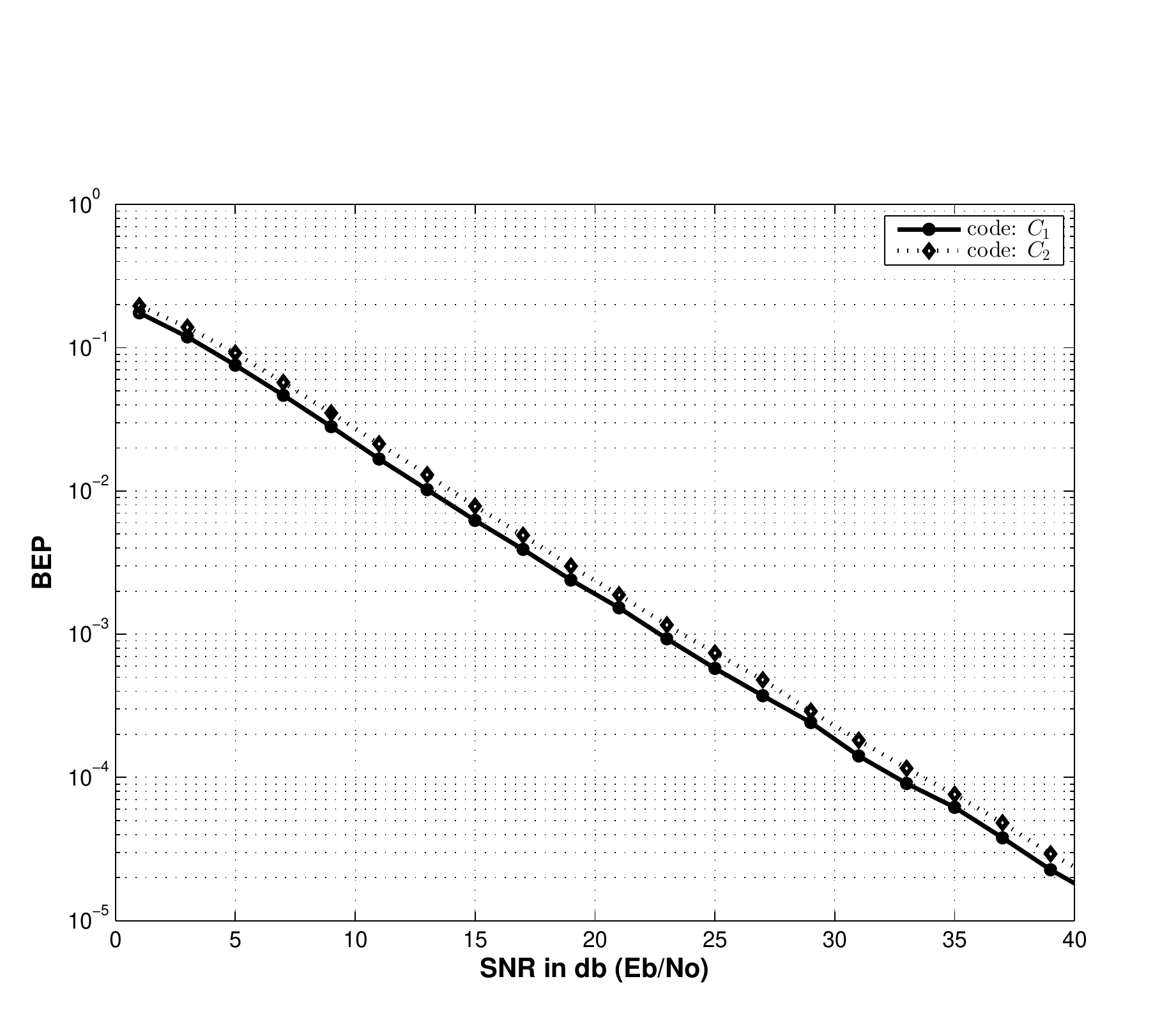}
\caption{\scriptsize SNR Vs BEP for codes $\mathfrak{C}_{1}$ and $\mathfrak{C}_{2}$ for Rician fading scenario, at receiver $R_{2}$ of Example \ref{eg:simulation9receivers}. }
\label{fig:SimExample2RicianR2}
\end{minipage}
\end{figure}
\begin{figure}
\centering
\begin{minipage}{0.45\textwidth}
\centering
\includegraphics[width=3in,height=2.25in]{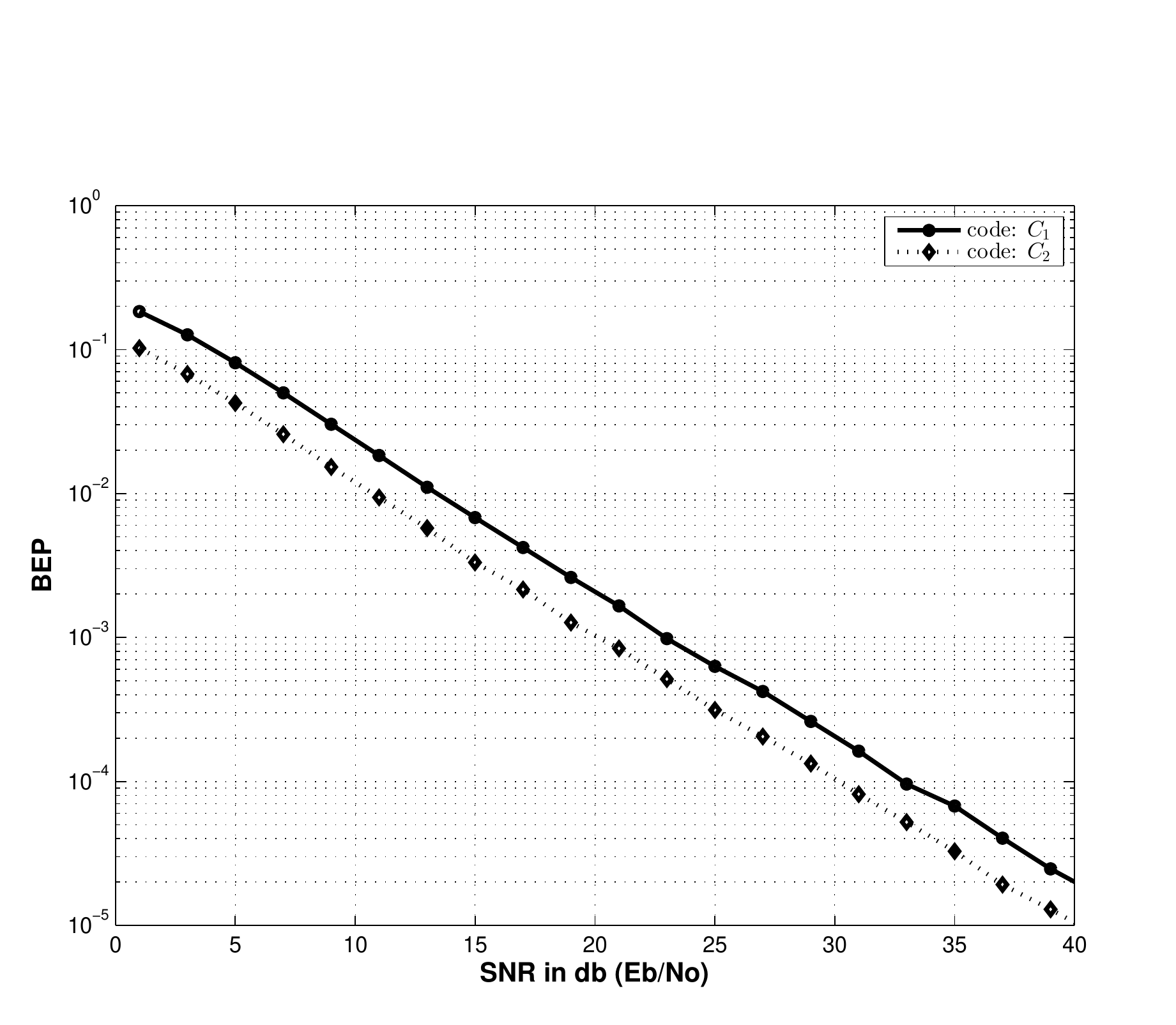}
\caption{\scriptsize SNR Vs BEP for codes $\mathfrak{C}_{1}$ and $\mathfrak{C}_{2}$ for Rician fading scenario, at receiver $R_{4}$ of Example \ref{eg:simulation9receivers}. }
\label{fig:SimExample2RicianR4}
\end{minipage}\hfill
\begin{minipage}{0.45\textwidth}
\centering
\includegraphics[width=3in,height=2.25in]{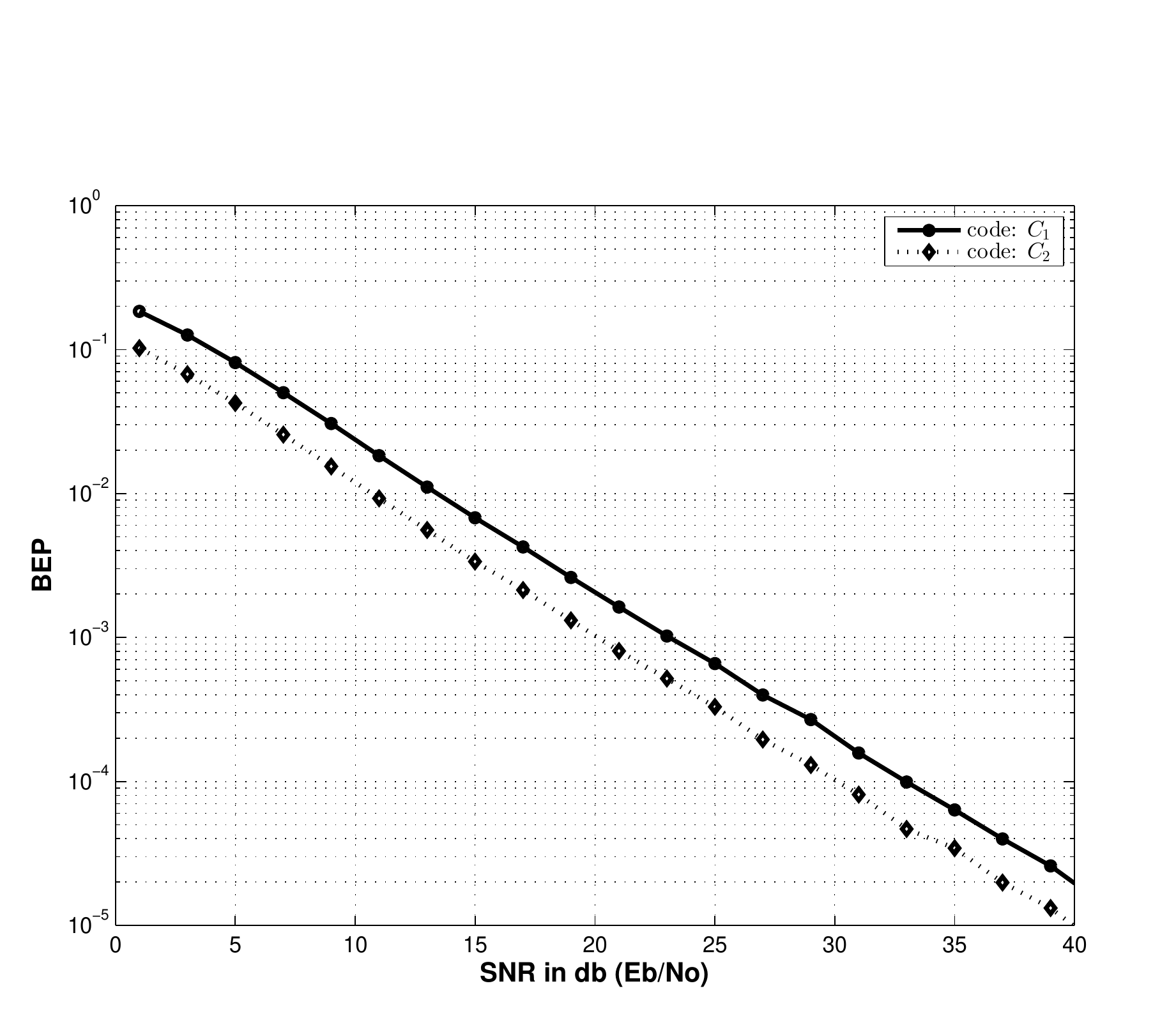}
\caption{\scriptsize SNR Vs BEP for codes $\mathfrak{C}_{1}$ and $\mathfrak{C}_{2}$ for Rician fading scenario, at receiver $R_{5}$ of Example \ref{eg:simulation9receivers}. }
\label{fig:SimExample2RicianR5}
\end{minipage}
\end{figure}
\begin{figure}
\centering
\begin{minipage}{0.45\textwidth}
\centering
\includegraphics[width=3in,height=2in]{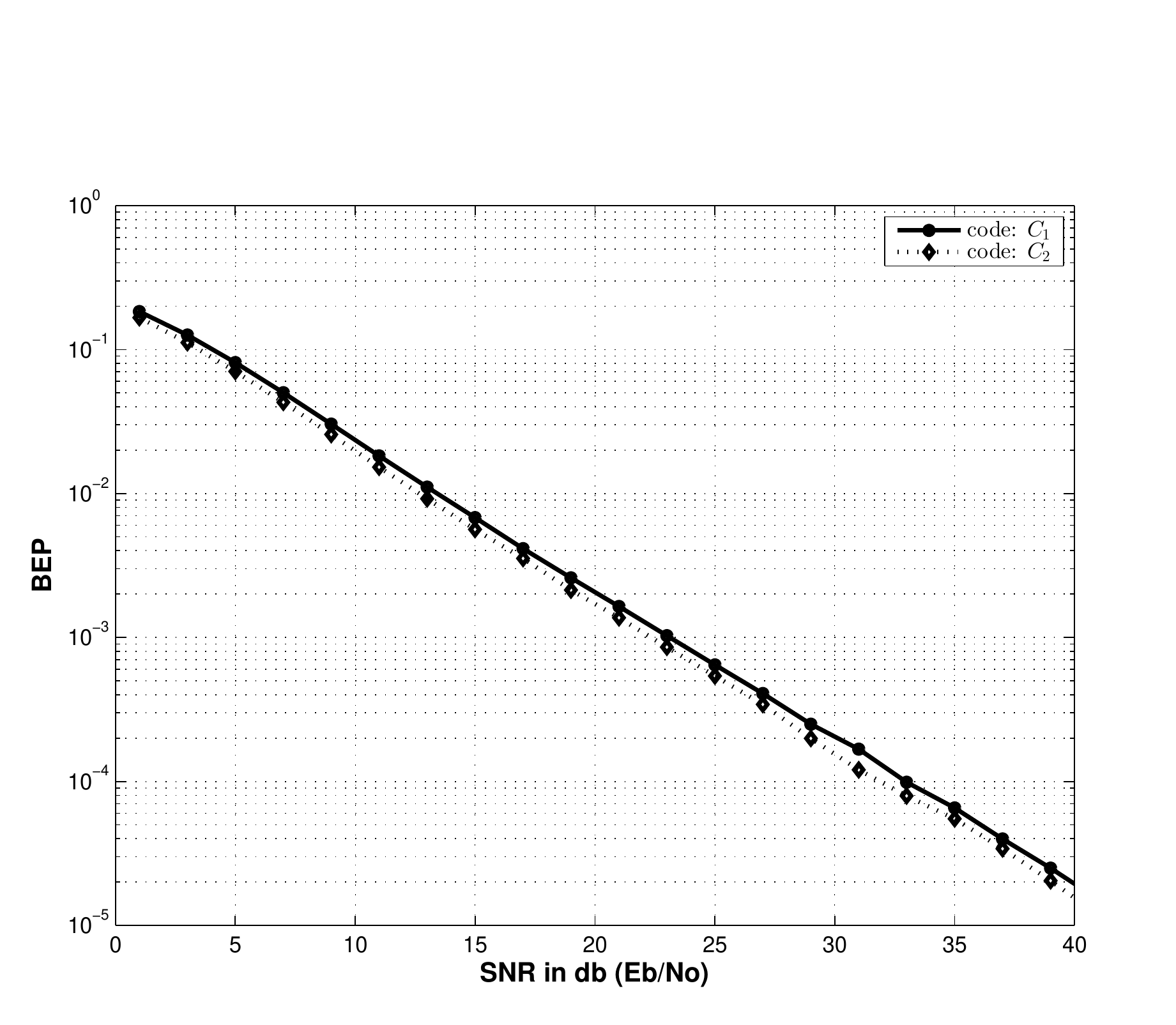}
\caption{\scriptsize SNR Vs BEP for codes $\mathfrak{C}_{1}$ and $\mathfrak{C}_{2}$ for Rician fading scenario, at receiver $R_{6}$ of Example \ref{eg:simulation9receivers}. }
\label{fig:SimExample2RicianR6}
\end{minipage}\hfill
\begin{minipage}{0.45\textwidth}
\centering
\includegraphics[width=3in,height=2in]{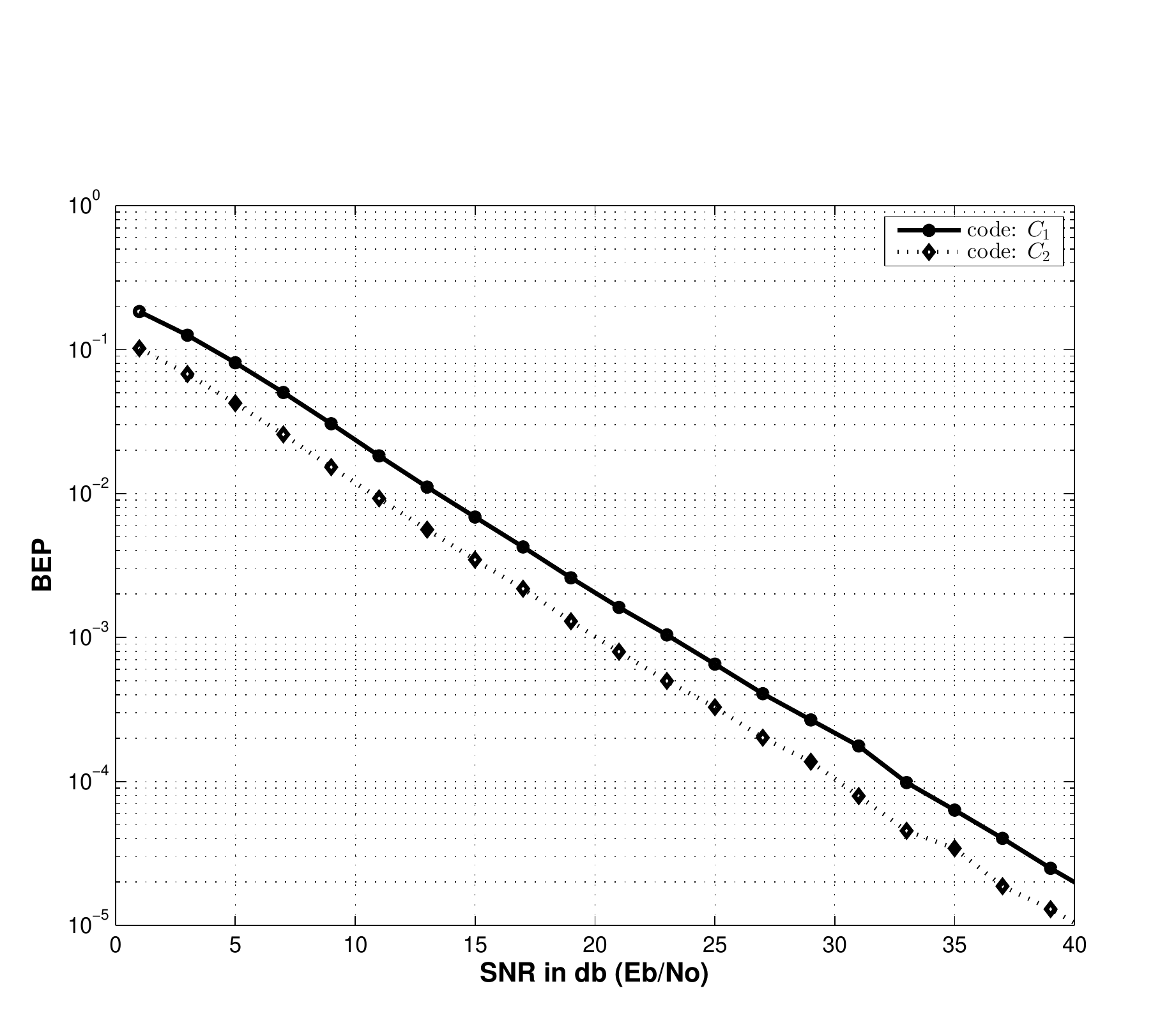}
\caption{\scriptsize SNR Vs BEP for codes $\mathfrak{C}_{1}$ and $\mathfrak{C}_{2}$ for Rician fading scenario, at receiver $R_{7}$ of Example \ref{eg:simulation9receivers}. }
\label{fig:SimExample2RicianR7}
\end{minipage}
\end{figure}
\begin{figure}
\centering
\begin{minipage}{0.45\textwidth}
\centering
\includegraphics[width=3in,height=2.25in]{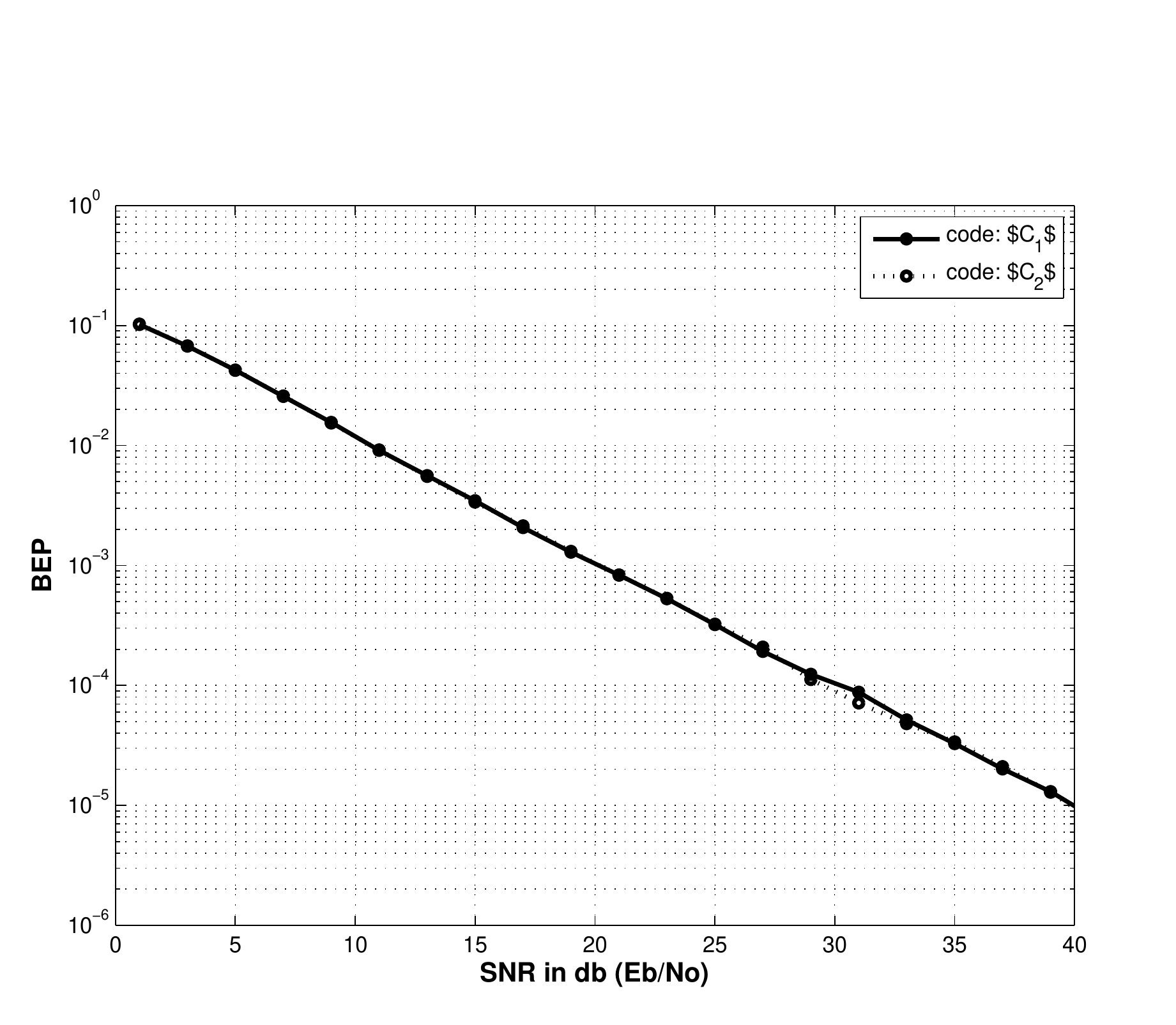}
\caption{\scriptsize SNR Vs BEP for codes $\mathfrak{C}_{1}$ and $\mathfrak{C}_{2}$ for Rician fading scenario, at receiver $R_{8}$ of Example \ref{eg:simulation9receivers}. }
\label{fig:SimExample2RicianR8}
\end{minipage}\hfill
\begin{minipage}{0.45\textwidth}
\centering
\includegraphics[width=3in,height=2.25in]{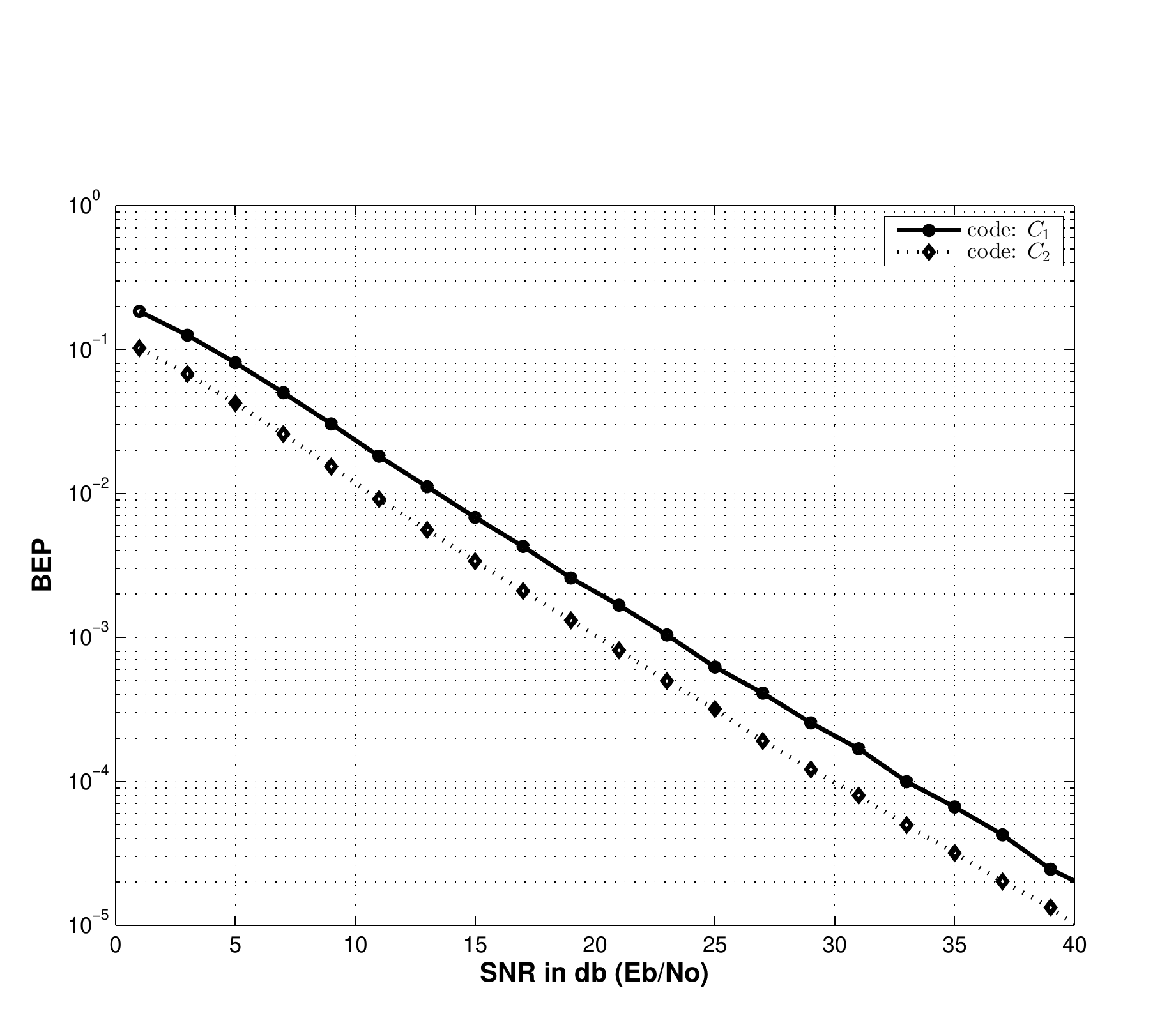}
\caption{\scriptsize SNR Vs BEP for codes $\mathfrak{C}_{1}$ and $\mathfrak{C}_{2}$ for Rician fading scenario, at receiver $R_{9}$ of Example \ref{eg:simulation9receivers}. }
\label{fig:SimExample2RicianR9}
\end{minipage}
\end{figure}

We also consider the scenario in which the source uses 16-PSK signal set for transmission. The mapping from bits to complex symbol is assumed to be Gray Mapping. Rayleigh fading scenario is considered. The SNR Vs. BEP curves for all receivers while using code $\mathfrak{C}_{1}$ and $\mathfrak{C}_{2}$ are given in Fig. \ref{fig:SimExample2Rayleigh16PSK_C1} and Fig. \ref{fig:SimExample2Rayleigh16PSK_C2} respectively. From Fig. \ref{fig:SimExample2Rayleigh16PSK_C1} and Fig. \ref{fig:SimExample2Rayleigh16PSK_C2}, we observe that the maximum error probability across different receivers occurs at receiver $R_{2}$ for code $\mathfrak{C}_{1}$ and at receiver $R_{3}$ for code $\mathfrak{C}_{2}$. Note that while using 16-PSK signal set the error probabilities of the transmissions depends on the mapping. The SNR Vs. BEP curves at receiver $R_{2}$ for code $\mathfrak{C}_{1}$ and at receiver $R_{3}$ for code $\mathfrak{C}_{2}$ is plotted in Fig. \ref{fig:SimExample2Rayleigh16PSKMax}. From Fig. \ref{fig:SimExample2Rayleigh16PSKMax}, it is evident that the maximum probability of error is less for $\mathfrak{C}_{1}$.  Thus code $\mathfrak{C}_{1}$ performs better than code $\mathfrak{C}_{2}$ in terms of reducing the maximum probability of error.

\begin{figure*}
\centering{}
\includegraphics[width=15cm,height=9cm]{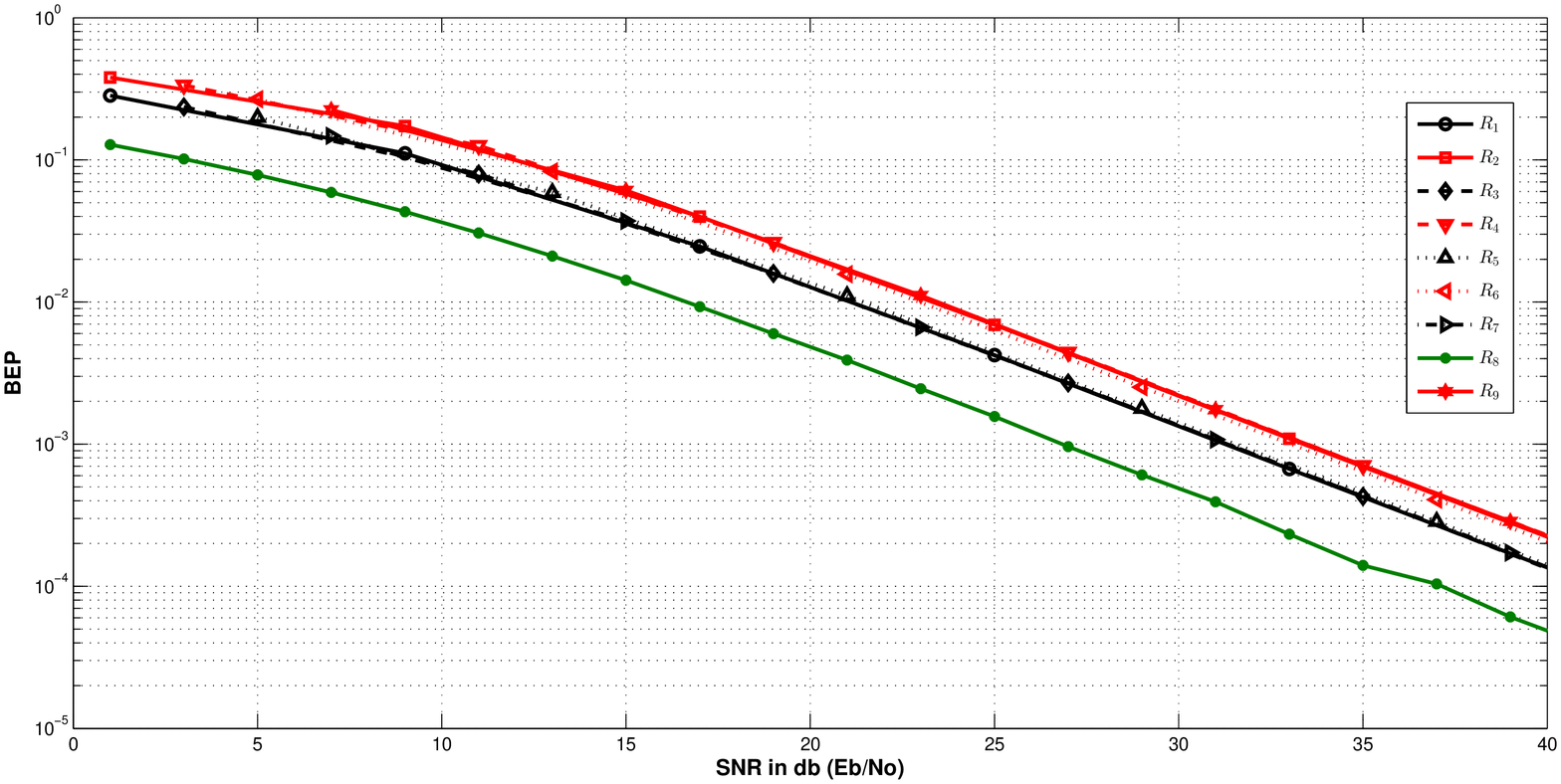}
\caption{\small SNR Vs BEP for code $\mathfrak{C}_{1}$ for Rayleigh fading scenario using 16-PSK modulation, at all receivers of Example \ref{eg:simulation9receivers}. }
\label{fig:SimExample2Rayleigh16PSK_C1}
\end{figure*} 

\begin{figure*}
\centering{}
\includegraphics[width=15cm,height=9cm]{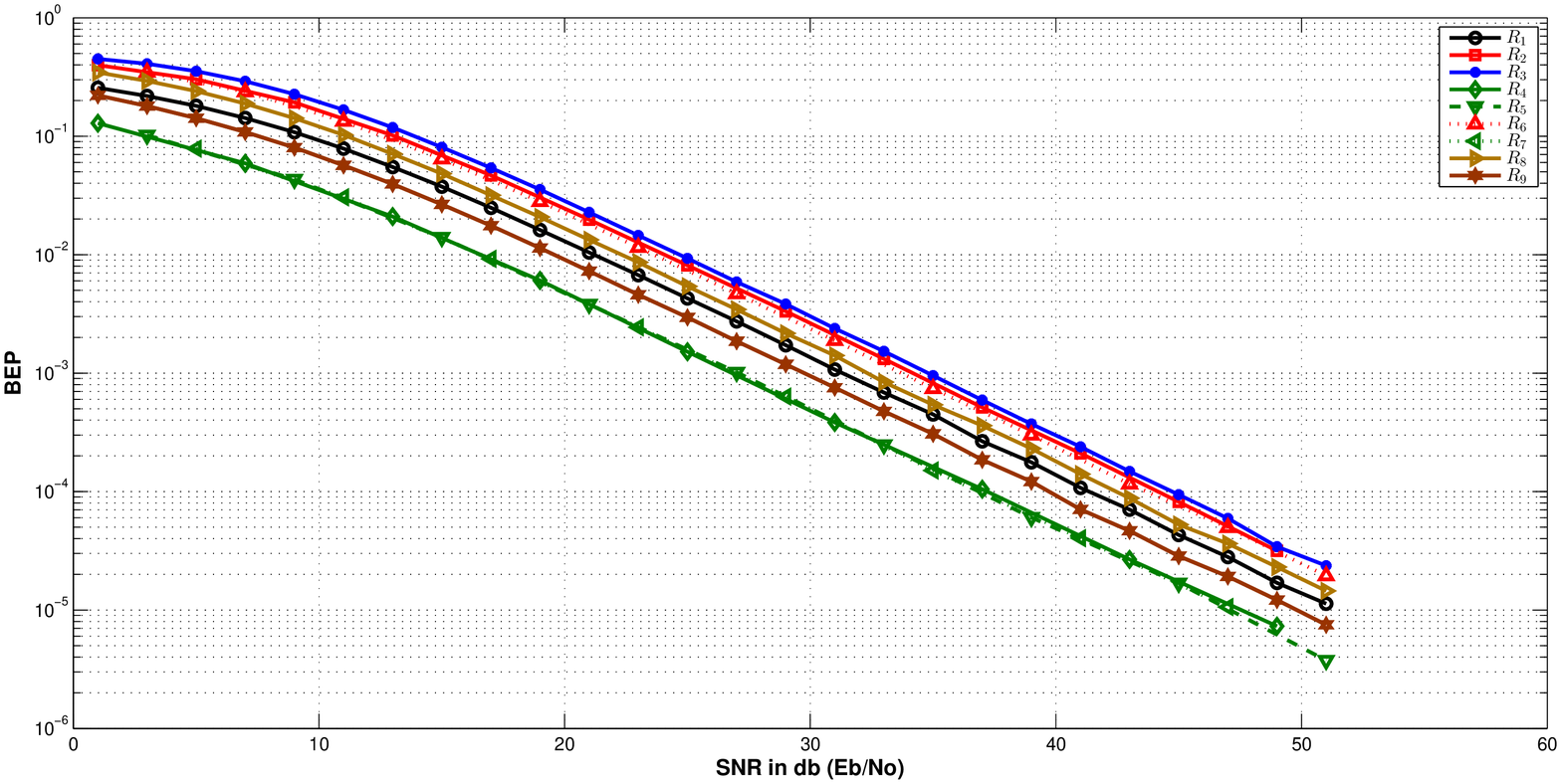}
\caption{\small SNR Vs BEP for code $\mathfrak{C}_{2}$ for Rayleigh fading scenario using 16-PSK modulation, at all receivers of Example \ref{eg:simulation9receivers}. }
\label{fig:SimExample2Rayleigh16PSK_C2}
\end{figure*} 

\begin{figure}
\centering{}
\includegraphics[scale=0.6]{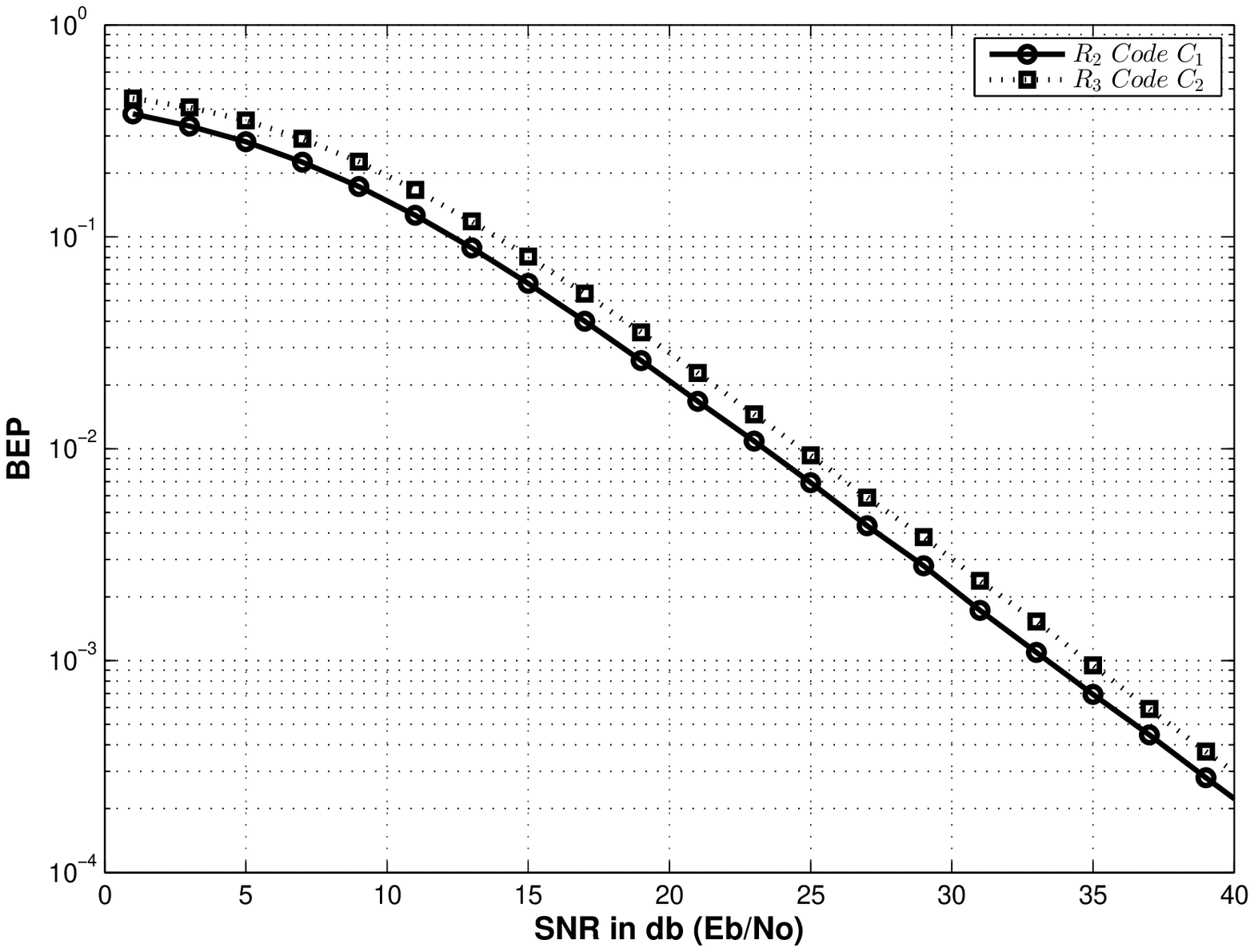}
\caption{\small SNR Vs BEP for Rayleigh fading scenario using 16PSK signal set at receiver $R_{2}$ for code $\mathfrak{C}_{1}$ and at receiver $R_{3}$ for code $\mathfrak{C}_{2}$  of Example \ref{eg:simulation9receivers}.}
\label{fig:SimExample2Rayleigh16PSKMax}
\end{figure}

\end{example}

\section{Conclusion}
\label{sec:Conclusion}
In this work, we considered a  model for index coding problem in which the transmissions are broadcasted over a wireless fading channel. To the best of our knowledge, this is the first work that considers such a model. We have described a decoding procedure in which the transmissions are decoded to obtain the index code and from the index code messages are decoded. We have shown that the probability of error increases as the number of transmissions used for decoding the message increases. This shows the significance of optimal index codes such that the number of transmissions used for decoding the message is minimized. 

For single uniprior index coding problems, we described an algorithm to identify the index code which minimizes the maximum probability of error. We showed simulation results validating our claim. The problem remains open for all other class of index codes. For other class of index coding problems the upper bound on the number of transmissions required by receivers to decode the messages is not known. Finally other methods of decoding could also be considered and this could change the criterion required in reducing the probability of error. The optimal index codes in terms of error probability and bandwidth using such a criterion could also be explored.

\end{document}